\documentclass[11pt,a4paper]{amsart}

\usepackage[left=1in,right=1in,top=1in,bottom=1in,foot=0.5in]{geometry}

\usepackage{amsmath,amsfonts,amssymb}
\usepackage[T1]{fontenc}
\usepackage{color}
\usepackage{epsfig}
\usepackage{graphicx}
\usepackage{cite,url}
\usepackage{hyperref}
\usepackage{cleveref}
\usepackage{xspace}
\usepackage{enumerate}

\usepackage[algo2e,algoruled,vlined,english]{algorithm2e} 

\newtheorem{thml}{Theorem}

\newtheorem{mytheorem}{Theorem}[section]
\newtheorem{mylemma}[mytheorem]{Lemma}
\newtheorem{myproposition}[mytheorem]{Proposition}
\newtheorem{mycorollary}[mytheorem]{Corollary}
\newtheorem{myclaim}[mytheorem]{Claim}

\newtheorem{myquestion}[mytheorem]{Question}

\theoremstyle{definition}

\newtheorem{myremark}[mytheorem]{Remark}
\newtheorem{myexample}[mytheorem]{Example}

\newcommand{\cO}{\ensuremath{\mathcal{O}}\xspace}
\newcommand{\ctO}{\ensuremath{\widetilde{\mathcal{O}}}\xspace}

\DeclareMathOperator{\UC}{UC}
\DeclareMathOperator{\WP}{WP}

\DeclareMathOperator{\St}{St}
\DeclareMathOperator{\supp}{supp}

\DeclareMathOperator*{\argmin}{arg\,min}

\DeclareMathOperator{\Fib}{\Phi}
\DeclareMathOperator{\proj}{pr}

\newcommand{\INC}{\mathrm{(INC)}}
\newcommand{\TC}{\mathrm{(TC)}}
\newcommand{\PC}{\mathrm{(PC)}}
\newcommand{\QC}{\mathrm{(QC)}}
\newcommand{\TPC}{\mathrm{(TPC)}}

\DeclareMathOperator{\diam}{diam}
\DeclareMathOperator{\rad}{rad}

\newcommand{\whp}{\alpha}
\newcommand{\hg}{Helly-gap}

\newcommand{\gd}[1]{{#1}}
\newcommand{\commentout}[1]{}

\begin{document}

\title{On $G^p$-unimodality of radius functions in graphs: structure and algorithms}

\author[J.\ Chalopin]{J\' er\'emie Chalopin}
\address{CNRS and Aix-Marseille Universit\'e, LIS, Marseille, France}
\email{jeremie.chalopin@lis-lab.fr}

\author[V.\ Chepoi]{Victor Chepoi}
\address{Aix-Marseille Universit\'e and CNRS, LIS, Marseille, France}
\email{victor.chepoi@lis-lab.fr}

\author[F.\ Dragan]{Feodor Dragan} \address{Computer Science
  Department, Kent State University, Kent, USA}
\email{dragan@cs.kent.edu}

\author[G.\ Ducoffe]{Guillaume Ducoffe} \address{National Institute
  for Research and Development in Informatics and University of
  Bucharest, Bucureşti, Rom\^{a}nia} \email{guillaume.ducoffe@ici.ro}

\author[Y.\ Vax\`{e}s]{Yann Vax\`{e}s}
\address{Aix-Marseille Universit\'e and CNRS, LIS, Marseille, France}
\email{yann.vaxes@lis-lab.fr}


\date{}

\maketitle

\begin{abstract}
    For every weight assignment $\pi$ to the vertices in a graph $G$, the \emph{radius function} $r_\pi$ maps every vertex of $G$ to its largest weighted distance to the other vertices.
    The center problem asks to find a center, i.e., a vertex  of $G$ that minimizes the radius function $r_\pi$.
    We here study some local properties of radius functions in graphs, and their algorithmic implications;
    our work is inspired by the nice property that in Euclidean spaces every local minimum of every radius function $r_\pi$ is a center.
    We study a discrete analogue of this property for graphs, which we name $G^p$-unimodality: specifically, every vertex that minimizes the radius function in its ball of radius $p$ must be a central vertex.
    While it has long been known since Dragan (1989) that graphs with $G$-unimodal  radius functions $r_\pi$ are exactly the Helly graphs, 
    the class of graphs with $G^2$-unimodal radius functions has not been studied insofar.
    We prove the latter class to be much larger than the Helly graphs, since it also comprises (weakly) bridged graphs, graphs with convex balls, and bipartite  Helly graphs (alias bipartite absolute retracts).
    These classes share basic properties with  geodesic metric spaces of $\ell_2$, $\ell_1$, and $\ell_{\infty}$-types; e.g., bridged graphs (one of the most important classes of graphs in metric graph theory)
    are considered as  the combinatorial analogs of CAT(0) spaces.
    We also show that in $\delta$-hyperbolic
    or (only for 0-1-weights) $\delta$-coarse Helly graphs, $r_\pi$ is $G^{\cO(\delta)}$-unimodal. 

    Recently, using the $G$-unimodality of radius functions $r_\pi$, a randomized $\ctO(\sqrt{n}m)$-time local search algorithm  for the center problem on Helly graphs was proposed by Ducoffe (2023). 
    Assuming the Hitting Set Conjecture (Abboud et al., 2016), we prove that a similar result for the class of graphs with $G^2$-unimodal radius functions is unlikely.
    However, we design local search algorithms (randomized or deterministic) for the center problem on many of its important subclasses:  our descent algorithm has (1) randomized complexity $\ctO(\sqrt{n}m)$ 
    if $G$ is (weakly) bridged or bipartite Helly
    with arbitrary weights, 
    (2) deterministic complexity $\cO(\sqrt{n}m)$ for the same graphs with 0-1-weights, and (3) randomized complexity $\ctO(m^{\frac{3}{2}})$ 
    if $G$ is a graph with convex balls and arbitrary weights. To overcome the $\cO(\sqrt{n}m)$ complexity barrier and using the $G^2$-unimodality of radius functions and other structural results, we
    further design a (4) divide-and-conquer algorithm with deterministic complexity $\cO(n\log^2 n)$
    for cube-free median graphs with arbitrary  weights.
    More specifically, we present a nontrivial generalization of the algorithm by Kariv and Hakimi (1979) for weighted centers in trees to another class of graphs.
    %
    Summarizing, our structural and algorithmic results contribute to a better understanding on the power of local search 
    for solving optimization problems related to distances in graphs.
\end{abstract}

\newpage
\section{Introduction}\label{intro}

\subsection{Avant-propos.} The center problem is a basic problem in graph theory, computational geometry, and optimization.
It naturally arises in facility location,  communication and transportation networks, network analysis,
robot-motion planning, and other areas. The center problem can be
defined for any metric space $(X,d)$. In the unweighted  version, given a finite set $P\subseteq X$  of points, 
it consists in computing a point $c\in X$ minimizing the largest distance from $c$  to a point of $P$. In the weighted version,
each point of $P$ has a positive multiplicative weight and one seeks to compute a point $c\in X$ minimizing the largest weighted
distance from $c$ to a point of $P$. The goal of the center problem is to design efficient algorithms for computing $c$.
The center problem has a long history, both in discrete and continuous settings. Sylvester \cite{Sy} formulated in 1857 the center problem in the Euclidean plane (the smallest
enclosing circle). 
Around the same time, Jordan \cite{Jo} defined the center of a tree and established  that it always consists of one or two adjacent vertices located in the middle of a diametral path.
The center problem was extensively studied in  Euclidean spaces, in polygons (simple or with holes), and \gd{more relevant to this work,} in graphs.  Under  plausible computational
assumptions, it was shown in \cite{AVW16} that the unweighted
center problem in graphs  cannot be solved in subquadratic time.
However, faster algorithms can be designed if the radius function has additional properties.
Unimodality of a function (i.e., the property
that its local minima are global minima) is at the heart of descent
methods of optimization. These methods have numerous variations and specifications and, more importantly,  numerous practical applications in
optimization, combinatorial optimization, operations research, machine
learning, statistics, and discrete and metric geometry.
Convex functions constitute the most common setting of unimodality. The distance function of a normed space is an example of a convex function;
the objective functions in facility location problems (center, median, and $p$-moments) 
in normed spaces are thus convex and can be minimized by descent algorithms.
The radius functions in trees and in simple polygons with geodesic  metric are also convex (see \cite{TaFrLo} and \cite{PoShRo}); \cite{PoShRo} also
establishes that simple polygons satisfy other properties of CAT(0) metric spaces (e.g., convexity of balls); convexity of balls also holds in
simple polygons with link metric \cite{LePoSaSeShSuTouWhYa}.
In fact, metric trees and simple polygons are CAT(0) and the convexity of the distance function is among  the most important and
useful properties of CAT(0) geometries \cite{BrHa,Pa}, an important type of metric geometries extensively investigated in geometric group theory and metric graph theory.

In this paper, we \gd{introduce} the $G^p$-unimodality of the (weighted and unweighted) radius functions $r_\pi$ in graphs $G$
and use it to design  subquadratic-time local search algorithms for the center problem in several classes of graphs, among which are (weakly) bridged graphs, graphs with convex balls, bipartite Helly graphs, and cube-free median graphs.
Our results can be viewed as a contribution in the case of radius functions
to the general question ``How does the geometry of the graph influence the complexity of local search?'', recently raised in \cite{BrChRe}.

\subsection{Related work.} There is a vast literature on the center problem. We mostly review the results related to unimodality and to the classes of graphs occurring in this paper. 
By \cite{Jo}, the unweighted center of a tree $T$ is located in the middle of a diametral path of $T$. It is folklore that a diametral pair of vertices can be found in $\cO(n)$ time
by picking an arbitrary vertex $u$ of $T$, computing a furthest from $u$ vertex $v\in T$, and then computing a furthest from $v$ vertex $w\in T$; then the pair $v,w$ is  diametral.
The weighted center problem in trees is much
more difficult: Kariv and Hakimi \cite{KaHa} presented an $\cO(n\log n)$-time algorithm combining divide-and-conquer,  convexity, and centroid computation.
At each step, a centroid $v$ of a subtree $T'$ of $T$ and a neighbor $v'$ of $v$ with $r_\pi(v')<r_\pi(v)$ are computed. The radius function $r_\pi$ of $T$ is convex,
thus either $v'$ does not exist and $v$ is central or the subtree of $T'$ rooted at $v'$ contains the center and the recursion  is performed on it.
We refer to this  method as \emph{cut-on-best-neighbor}.
Meggido \cite{Me2} refined this approach into an optimal $\cO(n)$-time algorithm by significantly refining the approach of \cite{KaHa}
and using the same ideas as his linear-time algorithm for linear programming in ${\mathbb R}^2$ and ${\mathbb R}^3$. 
In \cite{Me2}, he also presented the first $\cO(n)$-time algorithm for the smallest enclosing circle (disproving the conjecture
that $\cO(n\log n)$ is its best complexity). 
Both algorithms are based on the same ideas as
the linear-time algorithm for linear programming in ${\mathbb R}^2$ and ${\mathbb R}^3$ \cite{Me2}. 
Linear time algorithms for linear programming and the unweighted center problem in any fixed dimension were given in \cite{Me3};
see \cite{Co,Dy2} for an $\cO(n)$-time algorithm for the weighted version.
For the geodesic (unweighted) center in simple polygons,  an $\cO(n\log n)$-time algorithm
was proposed in \cite{PoShRo} and an optimal $\cO(n)$-time algorithm was given in \cite{AhBaBoCaKoOh}. 
The center problem for simple polygons with the link-metric
was investigated in \cite{LePoSaSeShSuTouWhYa}, which proposed an $\cO(n^2)$-time algorithm (the best known algorithm has complexity $\cO(n\log n)$ and was given in \cite{DjLiSa}).

One computation of a furthest vertex is called a \emph{FP scan} \cite{MaBo}.
Thus the unweighted center of a tree is computed
via two FP scans.  
This  FP scans paradigm was successful in several cases, e.g., \cite{ChDr} uses it to compute an unweighted center of a chordal
graph $G$ in $\cO(m)$ time. In this case, a pair $v,w$ returned after two FP scans is almost diametral and
there is a center of $G$ located in the neighborhood of  the middle of $(v,w)$-geodesics.  The algorithm of \cite{LePoSaSeShSuTouWhYa} is based on the fact
that the link-radius of a simple polygon (which can be viewed as an infinite connected graph) is approximatively equal to half of the link-diameter. Knowing the diameter (the link-diameter is $\cO(n)$-time computable),
one can locate the link-center around the middle of a diametral pair. 
The graphs investigated in \cite{ChDr} and \cite{LePoSaSeShSuTouWhYa} have small hyperbolicity.
Chepoi et al. \cite{ChDrEsHaVa} proved that similar results  hold for the center problem in $\delta$-hyperbolic graphs $G$: two FP scans return
a pair $v,w$ such that $d(v, w)\ge \diam(G)-2\delta$ and the center of $G$ is contained in the ball of radius $5\delta+1$ centered at a middle vertex of any $(v,w)$-geodesic.
This provides a general framework for the unweighted center problem (parameterized by hyperbolicity). 

Another approach to the center problem
is based on unimodality and local search. Dragan \cite{Dr_thesis,Dr_Helly} proved that  Helly graphs
(the graphs in which the family of balls satisfies the Helly property) are exactly
the graphs in which all
radius functions are unimodal. 
%
Therefore local search can be used to solve the center problem in Helly graphs, however the number of steps until the minimum cannot be easily bounded.
Ducoffe \cite{Du_Helly} suggested to  pick a random set $U$
of size $\ctO(\sqrt{n})$ and run local search starting from a vertex $u\in U$ with smallest $r_\pi(u)$ (see also Aldous \cite{Al} for a similar method for unimodal functions on graphs).
This leads to an $\ctO(\sqrt{n}m)$-time algorithm for computing a weighted center of a Helly graph \cite{Du_Helly,DuDr}, where $\ctO(\sqrt{n})$ is the number
of steps and $\cO(m)$ is the complexity of each improvement step. This algorithmic paradigm (which we call  \emph{sample-select-descent}) was later applied to other
classes of graphs 
\cite{DrDu_alpha,Du_abs}. 
A main difficulty when dealing with these classes is to efficiently implement an improvement step of local search (the solution proposed in \cite{Du_Helly} heavily relies on the Helly property).

There are also several papers which present efficient algorithms for the unweighted center problem, which do not use local search or FP-scans.
The unweighted center problem in median graphs was considered in \cite{BeHa}, \cite{BeDuHa1}, and \cite{BeDuHa2}: \cite{BeHa} presents an algorithm with  complexity $\cO(2^{\cO(d\log d)}n)$ for computing the center of
a median graph with constant largest cube-dimension $d$ and \cite{BeDuHa1} presents an algorithm with complexity $\cO(n^{1.6408}\log^{\cO(1)}n)$ for  median graphs with $n$ vertices.
This result was recently improved by \cite{BeDuHa2} that provides an $\cO(n\log^{4}n)$-time algorithm for computing all unweighted eccentricities of a median graph.
Finally, \cite{ChDrVa} presents linear-time algorithms for squaregraphs and trigraphs, planar instances of median and bridged graphs. There are  also subquadratic-time algorithms for the center problem in planar graphs~\cite{Cab18,GKHM+18} and in graphs in which the VC-dimension of the family of all balls is constant~\cite{DuHaVi}. 


Buseman \cite{Bu}  introduced peakless functions by replacing the quantitative 
inequality $f(z)\le \lambda\cdot f(x)+(1-\lambda)\cdot f(y)$
by the qualitative one $f(z)\le \max \{ f(x), f(y)\}$, where equality holds only if $f(x)=f(z)=f(y)$.
Buseman and Phadke \cite{BuPh} investigated peakless functions in Busemann spaces
(which generalize CAT(0) spaces and where each pair of points can be connected by a unique geodesic \cite{Pa}). 
In the case of discrete metric spaces, where there exist
several geodesics between the same points, Arkhipova and Sergienko \cite{ArSe} introduced the notion of $r$-convex functions (see also \cite{LeSeSo}).
A function $f$ is $r$-convex  if for each pair of points there exists a geodesic on which $f$ is convex. Bandelt and Chepoi \cite{BaCh_med} combined these notions and defined a function $f$
to be weakly peakless if for each pair of points there is a geodesic on which $f$ is peakless. More recently, B\'en\'eteau et al. \cite{BeChChVa_G2} introduced the notions of $p$-weakly peakless
functions by requiring that each pair of vertices is connected by a step-$p$ geodesic 
on which the function is peakless. While peakless, $r$-convex, and weakly peakless functions are unimodal on  $G$, $p$-weakly peakless functions
are unimodal on the $p$th power $G^p$ of $G$. 
%
All these classes of functions were suitable for investigating the unimodality of the median and radius functions on graphs. Namely, it was shown in \cite{BaCh_med} that the graphs in which all median functions are unimodal
are exactly the graphs in which all median functions are weakly peakless. 
This bijection was extended in \cite{BeChChVa_G2} to $G^p$-unimodality of median functions
and $p$-weakly peaklessness  of all median functions. The results of \cite{BaCh_med} and \cite{BeChChVa_G2} show that numerous important classes of graphs have $G$- or $G^2$-unimodal median functions.
As we mentioned before, Dragan \cite{Dr_thesis,Dr_Helly} established that the graphs in which all radius functions are $G$-unimodal  are exactly the Helly graphs. 
Properties close to relaxed unimodality (called locality) were established and used in the case of $\delta$-hyperbolic graphs \cite{DrGu_hyp} and graphs with $\alpha_i$-metrics \cite{DrDu_alpha}.

\subsection{Our results.} 
One can ask to what extent the  two results about the center problem in Helly graphs 
are generalizable. In this paper,
we investigate the unimodality of the radius function $r_\pi$ of a graph $G$ not in $G$ but in some its power $G^p$. We show that
several important classes of graphs 
have $G^p$-unimodal radius functions, i.e., every vertex that minimizes the radius function in its ball of radius $p$ is a central vertex:


\begin{thml}\label{Gpunimodality} (Propositions \ref{thm:wb},\ref{bar-wp},\ref{thm:cb},\ref{hyperbolic-wp},\ref{cor:cHelly-almost-unimodal},\ref{thm:cb})
For bridged and weakly bridged graphs, graphs with convex balls, bipartite Helly graphs, and cube-free median graphs, all radius functions $r_\pi$ are $G^2$-unimodal. For
$\delta$-hyperbolic graphs, all radius functions $r_\pi$ are $G^{4\delta+1}$-unimodal and for  $\delta$-coarse Helly graphs all 0-1 radius functions $r_\pi$
are $G^{\cO(\delta)}$-unimodal.
\end{thml}

To prove Theorem \ref{Gpunimodality}, 
a stronger property,
$p$-weakly peaklessness of the radius function $r_\pi$ is established; we recall, as mentioned earlier, that it is a  relaxation of convexity, which originates from Busemann/CAT(0) geometry \cite{Bu,BuPh},  and
was used before in \cite{BaCh_med,BeChChVa_G2}. 

The recognition problem for graphs with $G^p$-unimodal radius functions is polynomial:

\begin{thml}\label{th:recognition} (Theorem \ref{thm:recognition}) Given a graph $G$ with $n$ vertices and a positive integer $p$, one can decide in
$\cO(n^4)$ time whether all radius functions $r_\pi$ of $G$ are $G^p$-unimodal.
\end{thml}

To implement efficiently local search 
on graphs with $G^p$-unimodal radius functions, we need an efficient algorithm that for any vertex $v$ outputs an improving $p$-neighbor of $v$ if it exists.
In general, this cannot be done in subquadratic time even for $p=2$:

\begin{thml}\label{th:localminimum} (Theorem \ref{thm:localminimum})
    Assuming the Hitting Set Conjecture, for any $\varepsilon > 0$, deciding if $v$ is a local minimum in $G^2$ of
    the radius function $r_\pi$ for a graph $G$ cannot be solved in $\cO(n^{2-\varepsilon})$ time, even if $G$ has only $\ctO(n)$ edges, and every radius function $r_\pi$  of $G$ is $2$-weakly-peakless.
\end{thml}

Nevertheless, we show (and this is our first main result) that for the
classes of graphs from Theorem \ref{Gpunimodality} (except hyperbolic
and coarse Helly graphs), getting an improving 2-neighbor or a
certificate that it does not exist can be done efficiently (e.g., in
$\cO(m)$ time for bridged graphs). This leads to the first subquadratic-time
algorithms for the center problem for all these classes of graphs. Furthermore, in
the case of graphs with 2-weakly peakless radius functions and
0-1-weights, 
we present a deterministic algorithm with the same
complexity as the sample-select-method. Consequently, using all this
and local and global metric and geometric properties of those classes of
graphs, we obtain the following algorithmic results: 

\begin{thml}\label{main} (Theorems \ref{thm:main-bridged},\ref{thm:main-bar},\ref{thm:center-cb},\ref{center-hyp})
    If $G$ is a bridged, weakly bridged, or bipartite Helly graph with arbitrary weights, then a central vertex of $G$ can be computed in randomized time $\ctO(\sqrt{n}m)$.
    For 0-1-weights,  a central vertex of $G$ can be computed in deterministic time $\cO(\sqrt{n}m)$ (and in deterministic time  $\cO(\delta m)$, if $G$ is $\delta$-hyperbolic).  More generally, if $G$ is a graph with convex balls and arbitrary weights, then
    a central vertex of $G$ can be computed in randomized time $\ctO(m^{\frac{3}{2}})$. 
\end{thml}

The sample-select-descent algorithm cannot get better than $\cO(\sqrt{n}m)$ even for trees. To overcome this $\cO(\sqrt{n}m)$ barrier, we need to use  $G^p$-unimodality
differently.  For cube-free median graphs (a subclass of bipartite Helly graphs), we do this by adapting the cut-on-best-neighbor approach from \cite{KaHa,Me2} (this is our second main result):

\begin{thml} \label{thml:cube-free-median} (Theorem \ref{thm:cube-free-median})
    If $G$ is a cube-free median graph with arbitrary weights, then a central vertex of $G$ can be computed in deterministic time $\cO(n\log^2 n)$.
\end{thml}

\subsection{Bridged, median,  and (bipartite) Helly graphs.} Together with hyperbolic graphs, median graphs, bridged graphs and Helly graphs constitute
the most important classes of graphs in metric graph theory \cite{BaCh_survey,CCHO}. All these classes have been also
extensively studied in geometric group theory.

Median graphs are exactly the 1-skeleta of CAT(0) cube complexes \cite{Ch_CAT}. Gromov \cite{Gr} characterized
CAT(0) cube complexes as the simply connected cube complexes in which links  are flag  complexes. Median graphs can also be
characterized as the domains of event structures \cite{BaCo} from concurrency theory and
as the  solution sets of 2-SAT formulas \cite{Schaefer}. For characterizations and applications of median graphs to discrete mathematics
and geometric group theory, see the book \cite{Bo} and the surveys \cite{BaCh_survey} and \cite{Sa_survey}.

Bridged graphs are
the graphs in which balls around convex sets are convex (which is
one of the basic properties of Euclidean convexity and CAT(0) spaces) and they have been characterized
in \cite{SoCh,FaJa} as the graphs in which all isometric cycles have length 3. Bridged graphs generalize 
chordal graphs.
Chepoi \cite{Ch_CAT} presented a local-to-global characterization of bridged graphs as the 1-skeleta of simply connected flag simplicial complexes with 6-large links
(i.e., they do not have induced 4- and 5-cycles).  Those simplicial complexes have been rediscovered by Januszkiewicz and \'{S}wia\c{t}kowski \cite{JaSw}
and dubbed systolic complexes.  Systolic complexes are contractible \cite{Ch_CAT,JaSw} and they are considered as combinatorial analogs of
simplicial nonpositive curvature.  The  groups acting  on systolic complexes/bridged graphs have been studied in numerous papers;
\cite{HuOs_Artin,JaSw,OsPr,Soe} is a small sample. 
Graphs with convex balls generalize bridged graphs. They have been introduced and characterized in \cite{SoCh,FaJa}
and thoroughly studied in \cite{ChChGi}. Weakly bridged graphs constitute a subclass of graphs with convex balls; they have been introduced in \cite{Os} and
characterized in \cite{ChOs}. 

Helly graphs are the graphs in which the balls satisfy the Helly property, i.e., any collection
of pairwise intersecting balls has a nonempty intersection.  The bipartite Helly graphs (alias bipartite absolute retracts) are the bipartite analogs of Helly graphs. 
Namely, bipartite Helly graphs are the bipartite graphs in which the family of half-balls
(the intersections of balls with the halves of the bipartition) satisfies the Helly property. Helly graphs and bipartite Helly graphs are the discrete analogues of hyperconvex spaces \cite{ArPa}  (complete geodesic metric
spaces in which the set of closed balls satisfies the Helly property).
%
%
In perfect analogy with hyperconvexity, there is a close relationship between Helly graphs and absolute retracts.
A graph is an absolute retract exactly when it is a retract of any larger graph into which it embeds isometrically. Then absolute retracts and Helly graphs are the same.
Analogously, the bipartite Helly graphs are the absolute retracts in the category of bipartite graphs \cite{BaDaSc}, which is why they are also called bipartite absolute retracts.
Helly graphs have been characterized in  the papers \cite{BaPe,BaPr} metrically and in \cite{CCHO} topologically, in a local-to-global way. The study of groups acting
on Helly graphs was initiated in the paper \cite{CCGHO} and continued in many other papers, e.g. \cite{HaHoPe,Ho,HuOs_Helly,OsVa}. Numerous examples of such groups are also
known: hyperbolic groups and groups acting on CAT(0) cube complexes are Helly \cite{CCGHO}.
The bipartite Helly graphs have been introduced and metrically characterized in \cite{BaDaSc}.  Bandelt, Hell, and Farber \cite{BaHeFa} established a correspondence between Helly graphs
and bipartite Helly graphs, using natural transformations between reflexive graphs and bipartite graphs. An important subclass of bipartite Helly graphs is the class of hereditary modular graphs \cite{Ba_hmg}.
Since the hereditary modular graphs are exactly the graphs in which all isometric cycles have length 4 \cite{Ba_hmg}, they can be viewed as bipartite analogs of bridged graphs. Hoda \cite{Ho_quadric} associated
cell complexes to hereditary modular graphs and investigated  their topological properties as well as  the groups acting on them.

\smallskip
From the point of view of graph theory, median, bridged, and
(bipartite) Helly graphs are quite general and universal. First, any
connected graph can be isometrically embedded into a Helly graph and
any connected bipartite graph can be isometrically embedded into a
bipartite Helly graph. Furthermore, in analogy to injective
hulls/tight spans of general metric spaces, for any connected graph
(respectively, connected bipartite graph) $H$ there exists a unique
minimal Helly graph (respectively, bipartite Helly graph) $G$
containing $H$ as an isometric subgraph; this graph $G$ is called the
Hellyfication (respectively, the bipartite Hellyfication) of
$H$. Second, any connected graph $H$ is a minor of a cube-free median
or $K_4$-free bridged
graph. 
Furthermore, any graph $H$ not containing any induced $C_4$ and $C_5$ is
an induced subgraph of a bridged
graph. 
The local structure of median and bridged graphs is also universal
because arbitrary graphs may occur in their stars. 
For any graph $H$ (respectively, any triangle-free graph $H$) its
simplex graph 
(the covering graph of the poset of cliques of $H$ ordered by
inclusion) is a median graph (respectively, a cube-free median graph)
\cite{BaVdV}. The stars (the union of all cubes containing a given
vertex) of any median graph $G$ are simplex
graphs. 
In the same vein, Weetman \cite{Weetman} described a nice construction
$W(H)$ of (infinite) graphs in which the links (neighborhoods) of all
vertices are isomorphic to a prescribed finite graph $H$ of girth
$\ge 6$.  For example, if $H = C_6$, then $W(H)$ is the triangular
grid. If $H = C_n$ for $n \ge 7$, then $W(H)$ is the graph of a tiling
of the (hyperbolic) plane into triangles in which each vertex has $n$
neighbors. For general $H$, $W(H)$ are always bridged and $K_4$-free.

Nonetheless, median, bridged, hyperbolic, and (bipartite) Helly graphs  have a rich and versatile metric structure. Median graphs can be viewed as discrete  $\ell_1$-geometries, bridged graphs and graphs with convex balls as discrete  $\ell_2$-geometries, and
Helly graphs as discrete $\ell_\infty$-geometries.
This structure  has  been  thoroughly studied in metric graph theory, see \cite{BaCh_survey} for a survey. 
On the other hand, the algorithmic results about
median, bridged, and (bipartite) Helly graphs are still rare. For median graphs, we can mention the polynomial-time  algorithm for computing the geodesic between any two points
in  the associated CAT(0) cube complex \cite{Ha}, the linear time algorithm for computing the median \cite{BeChChVa_med}, and the subquadratic-time algorithms
for computing the 0-1-centers \cite{BeDuHa1,BeDuHa2,BeHa}. For bridged graphs,  our algorithm for computing  a central vertex
is, to the best of our knowledge, the first such algorithmic result. Previously, it was only known how to compute the 0-1-centers in linear time for chordal graphs~\cite{ChDr}
and planar bridged graphs \cite{ChDrVa}.

\subsection{The structure of the paper.} The rest of the paper is organized as follows. In Section \ref{prelim} we present the main notions and definitions used in all other sections. More specific definitions are given
in respective sections. In Section~\ref{s:technical} we present the main ideas of the algorithm sample-select-descent for bridged graphs and bipartite Helly graphs, and of the algorithm cut-on-best-neighbor for cube-free median graphs.
Section~\ref{descent} shows how to derandomize the sample-select-descent algorithm for 0-1-profiles. Section~\ref{s:bridged} presents our results about the center problem in bridged and weakly bridged graphs, namely: the $G^2$-unimodality of their radius functions,
the sample-select-descent algorithm, and its correctness and complexity. Similar results for bipartite Helly graphs are presented in Section~\ref{sec:bar-alg}. Section~\ref{sec:median} presents the cut-on-best-neighbor algorithm
for cube-free median graphs and establishes its correctness and complexity. Section~\ref{sec:hyperb-cHelly} establishes the $G^{\cO(\delta)}$ unimodality of all radius functions in $\delta$-hyperbolic graphs and (for 0-1-profiles) in $\delta$-coarse
Helly graphs. In Section~\ref{sec:lower-recogn} we prove that the graphs with $G^p$-unimodal radius functions can be recognized in polynomial time; we also establish super-linear lower bounds for deciding if a vertex is a local
minimum of the radius function in a graph, which holds even if all radius functions are $G^2$-unimodal. Section \ref{sec:perspectives} presents some perspectives of our work and open questions (other open questions are locally presented in previous sections). The final
Section~\ref{sec:cb-alg} generalizes the results of Section~\ref{s:bridged} to all graphs with convex balls. Since the resulting complexity is worse than for (weakly) bridged graphs and the technical details are much
more involved we decided to present Section~\ref{sec:cb-alg} as an appendix of the paper.

\section{Preliminaries}\label{prelim}

\subsection{Graphs.} \label{ss:notions}
All graphs $G=(V,E)$ in this paper are
undirected, connected, and simple. If $G$ is a bipartite graph, then we denote
by $V_0$ and $V_1$ the sets of the bipartition of $V$.
Our non-algorithmic results hold for infinite graphs; in algorithms, $G$ has $n$ vertices and $m$ edges.
We write $u \sim v$ if $u, v \in V$ are adjacent and $u\nsim v$ otherwise. 
As usual, $K_n$ is a complete graph, $K_{p,q}$ a complete bipartite graph, and $C_n$ a  cycle; sometimes, $C_3$ is called a \emph{triangle}, $C_4$ a \emph{square} and $C_5$ a \emph{pentagon}.
The \emph{wheel}
$W_n$ consists of $C_n$ and a vertex adjacent to all vertices of
$C_n$. The \emph{3-cube} $Q_3$ is the 1-skeleton of the 3-dimensional unit cube. 

A $(u,v)$-\emph{geodesic}
is a $(u,v)$-path with a minimum number of edges.
The \emph{distance} $d(u,v):=d_G(u,v)$ between  $u$ and $v$ is the length of a $(u,v)$-geodesic.
The \emph{interval} between $u$ and $v$ is the set of all vertices on $(u,v)$-geodesics, i.e.,
$I(u,v)=\{w\in V: d(u,w)+d(w,v)=d(u,v)\}$. Let $I^{\circ}(u,v)=I(u,v)\setminus \{ u,v\}$.  The \emph{ball} of radius $r$ centered at a vertex $v$ of $G$ is the set
$B_r(v)=\{ x\in V: d(v,x)\le r\}$. As usual, $N(v)=B_1(v)\setminus\{ v\}$ and
$N(A)=B_1(A)\setminus A$. More generally, $N_r(v)=B_r(v)\setminus B_{r-1}(v)=\{ x\in V: d(v,x)=r\}$. A \emph{half-ball} of a bipartite graph $G=(V_0\bigcup V_1,E)$ is the intersection of any ball $B_r(v)$ of $G$ with $V_0$ or $V_1$.
More generally, the $r$--{\it ball  around a set} $A\subseteq V$
is the set 
$B_r(A)=\{ v\in V: d(v,A)\le r\},$ where $d(v,A)=\mbox{min} \{ d(v,x): x\in A\}$.
Let also $\proj_z(A) = \{a \in A: d(z,a) = d(z,A)\}$ \gd{be the metric projection of vertex $z$ on $A$}.
For an integer $p\ge 1$, the \emph{$p$th power} of 
$G$ 
is a graph $G^p$ with 
the same vertex-set $V$ as $G$ and $u$, $v$ are adjacent in $G^p$ if and only if $u \ne v$ and
$d_G(u,v)\le p$. 

A subgraph $H$  of $G$ is (1)  \emph{isometric}  if $d_H(u,v)=d_G(u,v)$
for any $u,v\in V(H)$, (2)  {\it  convex}
if $I(u,v)\subset H$ for any $u,v\in V(H)$, (3) {\it locally-convex} if $H$ is connected and $I(u,v)\subset H$ for any $u,v\in V(H)$ with $d(u,v)=2$, 
and (3) {\it gated}  if for every $x$ outside $H$ there
is  $x'$ in $H$ (the {\it  gate} of $x$)
such that  $x'\in I(x,y)$ for any $y\in V(H)$. 
An {\em outergate} of $z$ with respect to $S\subset V$ is a vertex $z^*$ such that $d(z,z^*)=d(z,S)-1$ and $z^*$ is adjacent to all vertices
of the projection $\proj_z(S)$ of $z$ on $S$. The set   $S$ is {\em outergated} if any $z\in V \setminus S$ has an outergate.

A graph $G=(V,E)$
is {\it isometrically embeddable} into a graph $G'=(W,F)$ if there exists a mapping $\varphi : V\rightarrow W$
such that $d_{G'}(\varphi (u),\varphi(v))=d_G(u,v)$ for $u,v\in V$. A retraction $\varphi$ of a graph $G$ is an idempotent nonexpansive mapping of $G$ into itself, that is,
$\varphi^2=\varphi: V(G)\rightarrow V(G)$ with $d(\varphi(x),\varphi(y))\le d(x,y)$ for all $x,y\in V(G)$.  The subgraph $H$ of $G$ induced by the image of $G$
under $\varphi$ is referred to as a retract of $G$. A graph $H$ is called an \emph{absolute retract} if $H$ is a retract of any graph $G$ into which $H$ is isometrically embedded.
Analogously, a bipartite graph $H$ is  a \emph{bipartite absolute retract} if $H$ is a retract of any bipartite graph $G$ into which $H$ is isometrically embedded.
A family of sets $\mathcal F$ on a universe $V$ satisfies the \emph{Helly property} if any subfamily ${\mathcal F}'$ of ${\mathcal F}$ has a nonempty intersection whenever
each pair of sets of ${\mathcal F}'$ intersects.

Vertices $v_1,v_2,v_3$ form a {\it metric triangle}
$v_1v_2v_3$ if $I(v_1,v_2),I(v_2,v_3),$ and
$I(v_3,v_1)$ pairwise intersect only in 
common end-vertices, i.e.,
$I(v_i, v_j) \cap I(v_i,v_k) = \{v_i\}$ for any $1 \leq i, j, k \leq 3$.
If $d(v_1,v_2)=d(v_2,v_3)=d(v_3,v_1)=k,$
then this metric triangle is called {\it equilateral} of {\it size} $k.$ An equilateral metric triangle
$v_1v_2v_3$ of size $k$ is called \emph{strongly equilateral} if $d(v_i,x)=k$ for any vertex $x\in I(v_j,v_k)$, where $\{ i,j,k\}=\{ 1,2,3\}$.
Obviously, all metric triangles of size 0 or of size 1 are strongly equilateral.
A metric triangle $v_1v_2v_3$ of $G$ is a {\it quasi-median} of the triplet $x,y,z$
if the following metric equalities are satisfied: $d(x,y)=d(x,v_1)+d(v_1,v_2)+d(v_2,y), d(y,z)=d(y,v_2)+d(v_2,v_3)+d(v_3,z),$ and
$d(z,x)=d(z,v_3)+d(v_3,v_1)+d(v_1,x)$ (such a quasi-median always exists). A quasi-median $v_1v_2v_3$ of $x,y,z$ that is a metric triangle of size 0 (\textit{i.e.}, $v_1=v_2=v_3$) is called a \emph{median}.

\subsection{Weakly modular graphs.} \label{ss:wmg}
A graph $G$ is \emph{weakly modular} \cite{Ch_metric,CCHO}
if $G$ satisfies the triangle $\TC$ and quadrangle $\QC$ conditions:
\begin{enumerate}
    \item[$\TC$] \emph{Triangle Condition}: for  $v,x,y\in V$ with $d(v,x)=d(v,y)=k$ and $x\sim y$, $\exists z\in B_{k-1}(v)$, $z\sim x,y$. 
    \item[$\QC$] \emph{Quadrangle Condition}: for $v,x,y,u\in V$ with $d(v,x)=d(v,y)=k=d(v,u)-1$ and $u\sim x,y$, $x\nsim y$, $\exists z\in B_{k-1}(v)$, $z\sim x,y$.
\end{enumerate}

By \cite{Ch_metric}, the weakly modular graphs are exactly the graphs in which all metric triangles are strongly equilateral. A graph $G$ is called
\emph{modular} if all metric triangles have size 0, i.e., the quasi-medians of all triplets are medians. \emph{Median graphs} are the modular graphs
in which each triplet has a unique median. \emph{Cube-free median graphs} are the median graphs not containing the 3-cube as a subgraph.

\emph{Helly graphs} are the graphs in which the family of balls satisfies the Helly property. Helly graphs are weakly modular, namely all quasi-medians in
Helly graphs are metric triangles of size 0 or 1. \emph{Bipartite Helly graphs} are the bipartite graphs in which the family of half-balls satisfies the Helly property.
Bipartite Helly graphs are modular. We stress here the fact that \emph{the class of bipartite Helly graphs is not a subclass of Helly graphs}, because in Helly graphs and in bipartite Helly
graphs the Helly property holds for different families of sets: balls in the first case and half-balls in the second case.

A graph $G$ is \emph{bridged} if any isometric cycle of $G$ has size 3. By \cite{FaJa,SoCh}, the bridged graphs are exactly the
graphs in which the balls $B_r(A)$ around convex sets $A$ are convex. Bridged graphs are weakly modular, in fact they are exactly the weakly modular graphs not containing
induced $C_4$ and $C_5$ \cite{Ch_metric,Ch_CAT}. A \emph{graph with convex balls} (a \emph{CB-graph}) is a graph in which all balls are convex. Obviously, bridged graphs are
CB-graphs. CB-graphs are not weakly modular but they satisfy $\QC$ and the \emph{Triangle-Pentagon Condition} $\TPC$ asserting that for
any $v,x,y\in V$ such that $d(v,x)=d(v,y)=k$ and $x\sim y$ either $\TC$ or the following $\PC$ holds \cite{ChChGi}:

\begin{enumerate}
\item[$\PC$] \emph{Pentagon Condition}:  there exist  $z,w,w'\in V$ such that $z\in B_{k-2}(v)$ and  $xwzw'y$ is a pentagon of $G$.
\end{enumerate}

The \emph{weakly bridged graphs} are the weakly modular graphs with convex balls. By \cite{ChOs} they are exactly the weakly modular graphs not containing induced $C_4$.

\subsection{Radius function and $G^p$-unimodality.}  A {\it profile} (or a {\it weight function})  is a map $\pi: V\rightarrow
{\mathbb{R}^+\cup \{ 0\}}$ with finite support $\supp(\pi)=\{ v\in V: \pi(v)>0\}$.
The \emph{radius function} 
is given by $r_{\pi}(v)=\max_{u\in \supp(\pi)} \pi(u)d(u,v).$
A vertex $v$ minimizing $r_{\pi}(v)$ is called a {\it center} of $G$ with respect to  $\pi$ and  the set of all centers is the {\it center}
$C_{\pi}(G)$. 
The value $r_\pi(v)$ for any $v\in C_\pi(G)$ is  the \emph{radius} of $G$ with respect to $\pi$ and is denoted by
$\rad_\pi(G)$. For $v\in V$, let $F_\pi(v) = \{z \in \supp(\pi): \pi(z)d(v,z) = r_\pi(v)\}$ be  the set of \emph{furthest vertices} of $v$. 
If 
$\pi(v)\in \{ 0,1\}$ for any  $v$, then
we call $\pi$ a \emph{0-1-profile} and $r_\pi$ a \emph{0-1-radius function}. 
%
%

Let $f:V\rightarrow {\mathbb R}$ be a function defined on the vertex-set $V$  of $G$. The \emph{global minima}
of $f$ is the set $\argmin(f)$ of vertices of $G$  minimizing $f$ on $G$. By a {\it local minimum} one means a vertex $v$ 
such that $f(v)\le f(w)$ for any $w$ in  $B_1(v)$. 
A function $f$ is called {\it unimodal} on $G$ if every local minimum of $f$ is global, that is, if the inequality $f(v)\le f(y)$ holds for all neighbors of $y$ of a vertex $v$ of $G$ then
it holds for all vertices $y$ of $G.$ Weakly peakless functions introduced in \cite{BaCh_med} is an important subclass of unimodal functions:
$f:V\rightarrow {\mathbb R}$ is \emph{weakly peakless} if for any two non-adjacent vertices $u,v$ it satisfies the following condition $\WP(u,v)$:
\begin{enumerate}[$\WP(u,v)$:]
\item
\emph{there exists $w \in I^{\circ}(u,v)$ such that
 $f(w) \leq \max \left\{f(u),f(v)\right\}$, and  equality holds only if $f(u)=f(w)=f(v)$.}
\end{enumerate}
In \cite{BeChChVa_G2} the notions of unimodal and weakly peakless functions have been generalized to $G^p$-unimodal and $p$-weakly peakless functions, with $p$ a positive integer.
A function $f$  is  {\it $G^p$-unimodal} if every
local minimum of $f$ in the $p$th power $G^p$ of $G$ is a global minimum of $f$. 
A function $f$  is called \emph{$p$-weakly peakless} if for any two vertices $u,v$ such that $d(u,v)\ge p+1$ it satisfies the condition $\WP(u,v)$.
It is shown in \cite{BeChChVa_G2} that $p$-weakly peakless functions are $G^p$-unimodal. 
A function $f$ is \emph{locally-$p$-weakly peakless} on $G$
if it satisfies $\WP(u,v)$  for any $u,v$ with $p+1 \leq d(u,v) \leq 2p$. By \cite{BeChChVa_G2}, $f$ is $p$-weakly peakless
iff  $f$ is locally-$p$-weakly peakless. Note that if $f$ is $p$-weakly peakless, then  any pair of vertices $u,v$ of $G$
can be connected by a $p$-geodesic (a subset of a geodesic in which consecutive vertices have distance $\le p$) along which $f$ is peakless.
Using $p$-weakly peaklessness is more handy when proving $G^p$-unimodality because we can better use the metric structure of graphs.
Note that any radius function $r_\pi$ is defined with respect to the distance function of  $G$, but its unimodality
is considered in $G^p$. 

\subsection{Cartesian trees.}
We often need to compute $r_\pi$ for  vertices of a set $S$, 
by running a  BFS from $S$ and computing
for each $s\in S$ the maximal weighted distance $\kappa(s)$ from $s$ to a vertex from $\supp(\pi)$ in the branch of the BFS-tree rooted at $s$.  For this
we will use  the following lemma:

\begin{mylemma}\label{lem:max-nonneighbor}
    Let $G=(V,E)$ be a graph 
    and  $\kappa : V \mapsto \mathbb{R}^+ \cup \{0\}$. 
    There is an $\cO(n+m)$-time algorithm that computes, for every $u\in V$, $c_{\kappa}(u) := \max\{ \kappa(v) : v \in V \setminus N(u) \}$.
\end{mylemma}
\begin{proof}
    Denote the $i^{th}$ vertex of $G$ by $v_i$.
    We construct an array $A_{\kappa}$ 
    such that $A_{\kappa}[i] = \kappa(v_i)$ for   $1 \leq i \leq n$.
    We pre-process it in $\cO(n)$ time so that  
    any $q(i,j) := \max\{ A_{\kappa}[k] : i \leq k \leq j  \}$ can be computed in $\cO(1)$ time.
    This is done using Cartesian trees~\cite{GaBeTa}.
    Thereafter, we reorder in $\cO(n+m)$ time the adjacency lists $N(u)$ of every $u$ 
    by increasing index.
    Doing so, 
    we can scan every ordered neighborhood $N(u)$ once and compute
    the pairs  $(i_1,j_1),
    \ldots,(i_r,j_r)$, for some $r \leq \deg(u)+1$, such that $V \setminus N(u) = \bigcup_{t=1}^r \{v_{i_t},
    \ldots,v_{j_t}\}$.
    Note that $c_{\kappa}(u) = \max\{ q(i_t,j_t): 1 \leq t \leq r \}$.
    Thus we can compute $c_{\kappa}(u)$ in $\cO(r) = \cO(\deg(u))$ time, giving $\cO(n+m)$ total time.
\end{proof}

\section{Technical overview} \label{s:technical}

In this section, we present the technical contributions of this paper at a high level. We begin by recalling the sample-select-descent method on which most of our algorithmic results are based. This allows us to highlight the main difficulty in implementing this method on a given graph class. We then explain how to overcome this difficulty in the case of bridged  and bipartite Helly graphs (which is one the main contributions of this work). Kariv and Hakimi introduced the cut-on-best-neighbour method, which enabled to obtain an $\cO(n\log n)$-time algorithm for the weighted center problem in trees \cite{KaHa}. We end this section with an overview of our second main contribution, which consists in extending to cube-free median graphs the use of the cut-on-best-neighbour method.


\subsection{Sample-select-descent method.} In the case of (weakly) bridged graphs, graphs with convex balls, and
bipartite Helly graphs, we use the sample-select-descent method. For the special case of 0-1-profiles, we present a deterministic version of this method (see Section \ref{descent}).  Recall that sample-select-descent
picks a random set $U$ of size $\ctO(\sqrt{n})$, computes their eccentricities, and runs local search in $G^2$ starting from a vertex $u$ of $U$ with smallest value  $r_\pi(u)$.
Local search constructs
a minimizing path $(u_0=u,\ldots,u_\ell)$ of $G^2$ such that $r_\pi(u_0)>\ldots>r_\pi(u_\ell)$,
and $u_\ell$ is a local (and thus  global) minimum of $r_\pi$ in $G^2$. At step $i+1$, a vertex $u_{i+1}\in B_2(u_i)$ such that $r_\pi(u_i)>r_\pi(u_{i+1})$ is computed.
Algorithm \ref{alg:sample-select-descent} provides the pseudo-code of the generic sample-select-descent.
Two complexity issues arise. First, given $u_i$, how to compute  $u_{i+1}$, i.e., how to efficiently
implement {\sf ImproveEccentricity}$(G,\pi,u_i)$? 
Second, how to bound the length $\ell$ of the improving path $(u_0,\ldots, u_\ell)$?   (This second question was treated in the case of quasi $M^{\#}$- and
$L^{\#}$-convex functions from discrete convex analysis \cite{MuSh1,MuSh2} and in the computation of $L_2$-geodesics in CAT(0) cube
complexes \cite{Ha}).  In the case of radius functions, the second issue can be solved  by applying  to $G^2$ the following result
of \cite{Du_Helly} (see also \cite{Al}):

\begin{mylemma}[\!\!{\cite[Lemma~3]{Du_Helly}}]\label{lem:k-unimodal} For any graph $G=(V,E)$ and for any profile $\pi$, for any random subset $U$ of $V$ of size  $\ctO(\sqrt{n})$, if $u$ minimizes $r_\pi$ on  $U$, then with high probability  any minimizing path \gd{in $G^2$} between $u$ and any vertex of $C_\pi(G)$ has length at most  $\ctO(\sqrt{n})$.
\end{mylemma}

In the case of 0-1-profiles, we proceed in the following deterministic way. By  \cite{MeMo}, any graph $G$ can be covered in $\cO(m)$ time with $\Theta(\sqrt{n})$ balls of radius $2\sqrt{n}$ and this covering can be computed in $\cO(m)$ time.
Let $U$ be the set of centers of these balls. Compute $r_\pi$ on the vertices of $U$ and pick $u\in U$ minimizing $r_\pi$ on $U$, i.e., $f(u)=\min \{ f(u'): u'\in U\}$.
Then we run local search on $G^2$ starting from $u$ until we find a local minimum $v^*$. We prove that $v^*$ is reached in at most $\sqrt{n}+1$ improvement steps.


\begin{algorithm2e}[h]
   \label{alg:sample-select-descent}
   \SetKwInOut{Input}{Input}
   \SetKwInOut{Output}{Output}
   \Input{A graph $G=(V,E),$ a profile $\pi:V\rightarrow R^+$}
   \Output{A central vertex of $G$}

   Pick a random set $U\subseteq V$ of size $\ctO(\sqrt{n})$\;
   Compute $r_\pi(u)$ for all $u\in U$\;
   Select $u_0\in U$ with minimum $r_\pi(u_0)$\;
   $i\leftarrow 0$ \;
   $u_1 \leftarrow$ {\sf ImproveEccentricity}$(G,\pi,u_0)$\;
   \While{$u_{i+1} \neq u_i$}
             {
               $i\leftarrow i+1$\;
               $u_{i+1}\leftarrow$ {\sf ImproveEccentricity}$(G,\pi,u_i)$\;
             }
 \KwRet{$u_i$}\;

 \caption{{\sf SampleSelectDescent}$(G,\pi)$}
 \end{algorithm2e}

 For radius function $r_\pi$, the issue of computing an improving
 neighbor in $G^2$ may be a bottleneck. Theorem \ref{th:localminimum}
 shows that deciding if $v$ is a local minimum of $r_{\pi}$ in $G$ or
 in $G^2$ is as hard as the global minimization of $r_{\pi}$ even if
 $r_\pi$ is $G^2$-unimodal.  However, for (weakly) bridged graphs,
 graphs with convex balls, and bipartite Helly graphs, we show (and
 this is the main technical content of the paper) that given a vertex
 $v$, deciding if $v$ is a local minimum of $r_{\pi}$ in $G^2$ or
 computing a vertex $u\in B_2(v)$ with $r_\pi(u)<r_\pi(v)$ otherwise,
 can be done in $\cO(m)$ time.  Consequently, for those classes of
 graphs, each call of {\sf ImproveEccentricity} requires $\cO(m)$
 time. Combining with Lemma \ref{lem:k-unimodal} this yields
 algorithms with complexity $\ctO(\sqrt{n}m)$ ($\cO(\sqrt{n}m)$ for
 0-1-profiles) for the center problem on such graphs.

\subsection{The case of bridged graphs.}

We describe how to implement {\sf ImproveEccentricity}$(G,\pi,v)$
(i.e., to compute a vertex $u\in B_2(v)$ with $r_\pi(u)<r_\pi(v)$) in
the case of bridged graphs. For graphs with convex balls this is done
analogously but some technical extra work is needed for the last step
(i.e., to assert that $v$ is a local minimum of $r_{\pi}$ in $G^2$),
details are given in the appendix (Section~\ref{sec:cb-alg}).

\begin{figure}
  \begin{center}
    \includegraphics[width=0.45\textwidth]{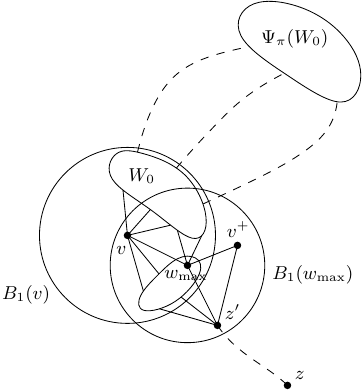}
  \end{center}
  \caption{ {\sf ImproveEccentricity}$(G,\pi,v)$ for bridged graphs.}\label{fig-algos1}
\end{figure}

In bridged graphs $G$, the balls around convex sets are convex. This
property is crucial for {\sf ImproveEccentricity}$(G,\pi,v)$  and the proof of its
correctness. In particular, using it, we show that any 1-ball $B$ of
$G$ is outergated (see Section \ref{ss:notions}) and we can compute in
$\cO(m)$ time the outergates on $B$.
Using Lemma~\ref{lem:max-nonneighbor}, we compute $u \in B_1(v)$
minimizing $r_\pi$ in $\cO(m)$ time. If $r_\pi(u)<r_\pi(v)$, then we are
done.
%
%
Otherwise, $v$
is a local minimum of $r_\pi$ in $G$ and we have to decide if $v$ is a
local minimum of $r_\pi$ in $G^2$. The key step of the algorithm is to
identify a neighbor $w_{\max}$ of $v$ such that $w_{\max} \sim u$ if
there exists a vertex $u \in B_2(v)$ with $r_\pi(u) < r_\pi(v)$ (see Figure~\ref{fig-algos1} for an illustration).
For this, consider the set $F_\pi(v)$ of all furthest from $v$
vertices of the profile.  For every $w \in N(v)$, the \emph{shadow}
$\Psi_\pi(w)$ of $w$ is the set of all $z \in F_\pi(v)$ such that
$w \in I(v,z)$, i.e., $w$ lies on a shortest $(v,z)$-path.
Next, we partition the neighborhood $N(v)$ of $v$ into equivalence
classes $W_0, W_1, \ldots, W_k$, where two vertices $w, w' \in N(v)$
are equivalent if they have the same shadows
$\Psi_\pi(w) = \Psi_\pi(w')$. Using outergates and partition
refinement techniques, this partition of $N(v)$ can be computed in
$\cO(m)$ time. Doing so, we also compute $|\Psi_\pi(W_i)|$ for every
equivalence class $W_i$ (where $\Psi_\pi(W_i) = \Psi_\pi(w)$ for any
$w \in W_i$). Assume that $W_0$ is the equivalence class maximizing
the size $|\Psi_\pi(W_0)|$ of its shadow. Since $v$ is a local minimum
of $r_\pi$ in $G$, $F_\pi(v) \neq \Psi_\pi(W_0)$.
We pick any vertex $z \in F_\pi(v) \setminus \Psi_\pi(W_0)$ and
compute in $\cO(m)$ time the intersection
$N(v) \cap I(v,z) = N(v) \cap N(z')$, where $z'$ is any outergate of
$z$ on the 1-ball $B_1(v)$.  Let $w_{\max} \in N(v) \cap I(v,z)$ be a
vertex maximizing $|N(w_{\max}) \cap W_0|$.
Using the convexity of balls, we prove that if $v$ is not a local
minimum of $r_\pi$ in $G^2$, then $w_{\max}$ has a neighbor $v^+$ such
that $r_\pi(v^+)<r_\pi(v)$ (see Proposition~\ref{thm:loc-search}).
Using Lemma~\ref{lem:max-nonneighbor}, the algorithm selects in $\cO(m)$
time a vertex $v^+$ minimizing $r_\pi$ in $B_1(w_{\max})$. If
$r_\pi(v^+) = r_\pi(v)$, then $v \in C_\pi(G)$ and otherwise we return
$v^+$ as an improving vertex $u \in B_2(v)$.




\subsection{The case of bipartite Helly graphs}
We describe how to implement the function {\sf ImproveEccentricity}$(G,\pi,v)$ in the case of a bipartite Helly graph $G=(V_0\cup V_1,E)$.
The computation is different but follows
the same lines as for bridged graphs (see Section \ref{sec:bar-alg}).
Suppose without loss of generality that $v\in V_0$.

\begin{figure}
    \begin{center}
      \includegraphics{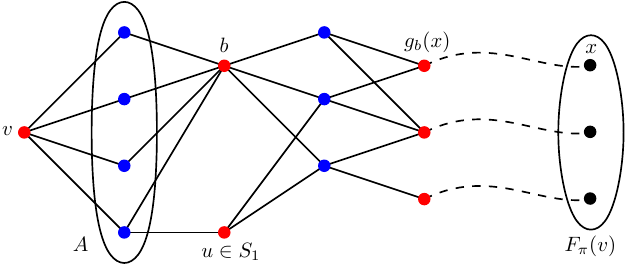}
    \end{center}
    \caption{Computing $S_1$ in the case of bipartite Helly
      graphs.}\label{fig-bipHelly}
\end{figure}

First, using the Helly property for half-balls, we show that any
$1$-ball of $G$ is outergated. Using this property, we show how to
compute in $\cO(m)$ time a vertex $u\in N(v)$ with
$r_\pi(u)<r_\pi(v),$ if $u$ exists.  Now, suppose that $v$ is a local
minimum of $r_\pi.$ Then we have to find a vertex $u\in N_2(v)$ with
$r_\pi(u)<r_\pi(v),$ if such $u$ exists. When $u$ exists, since $G$ is
bipartite, necessarily $u\in I(v,x)$ for any $x\in F_\pi(v)$, i.e.,
$u$ belongs to the set
$S_1=\bigcap\{ N_2(v)\cap I(v,x): x\in F_\pi(v)\}$.  Let also
$A=\bigcap\{ N(v)\cap I(v,x): x\in F_\pi(v)\}$. If
$S_1\ne \varnothing$, then any common neighbor $a$ of $v$ and any
vertex of $S_1$ belongs to $A$, thus $A\ne \varnothing$.  The set
$S_1$ can be computed in the following way (see
Figure~\ref{fig-bipHelly} for an illustration).  Since $B_1(v)$ is
outergated, the set $A$ can be computed in $\cO(m)$ time. If
$A=\varnothing$, then $S_1$ is also empty. Otherwise, the computation
of $S_1$ is done in two steps. First, we decide if
$S_1\ne \varnothing$ and if this is the case, we compute a vertex $b$
of $S_1$ such that $A \subseteq N(b)$.  Without loss of generality,
assume that $v \in V_0$. Note that if $S_1\ne \varnothing$, then the
half-balls $B_1(a)\cap V_0, a\in A$ and
$B_{d(x,v)-2}(x)\cap V_0, x\in F_\pi(G)$ pairwise intersect.  By the
Helly property for half-balls, this implies that there exists a common
vertex $b$ that belongs to all these half-balls, i.e., such that
$b \in S_1$ and $A \subseteq N(b)$. Consequently, $S_1$ is non-empty
if and only if there exists such a vertex $b$.
To test the existence of such a vertex in $\cO(m)$ time, we transform
the profile $\pi$ to a new profile $\pi'$ and reduce the existence of
$b\in S_1$ to testing if an arbitrary vertex $a$ of $A$ is a local
minimum of the new radius function $r_{\pi'}$.  Now, with the vertex
$b\in S_1$ at hand, for each $x \in F_\pi(v)$ we compute the outergate
$g_b(x)$ of $x$ with respect to $B_1(b)$. We show that $z \in N_2(v)$
belongs to $S_1$ if and only if $z$ is at distance $2$ from $g_b(x)$
for all $x \in F_\pi(v)$. Based on this local characterization, $S_1$
can be computed in $\cO(m)$ time since the intersection of the balls
$B_2(y), y\in Y$ for any set $Y$ can be computed in $\cO(m)$
time. 

Since $G$ is bipartite, if $u \in  N_2(v)$ and $d(u,x) > d(v,x)$, then
$d(u,x) = d(v,x)+2$. Consequently, if $u\in N_2(v)$ and
$r_\pi(u)<r_\pi(v)$, necessarily for any $x$ such that
$\pi(x) (d(v,x)+2) \geq r_\pi(v)$, we should have
$d(u,x) \leq d(v,x)$. Let $X = \{x: \pi(x) (d(v,x)+2) \geq r_\pi(v)\}$
and note that if there exists $u\in N_2(v)$ such that
$r_\pi(u)<r_\pi(v)$, then $d(u,x) \leq d(v,x)$ for all $x \in X$.  We
show that $u \in N_2(v)$ satisfies this condition if $u$ is at
distance $2$ from the outergate $g_v(x)$  with respect to $B_1(v)$ for all $x \in X$. 
Using the same subroutine
as before, we can compute the set $S_2$ of vertices $u \in N_2(v)$
such that $d(u,x) \leq d(v,x)$ for all $x \in X$. Finally, if $S_1 \cap S_2 \neq \varnothing$,
we output any vertex of
$S_1 \cap S_2$ as the improving vertex $u\in N_2(v)$ and $v$ otherwise. The correctness of this procedure is
obtained by proving that together the two necessary conditions are also sufficient.

\subsection{Cut-on-best-neighbor for cube-free median graphs.}
Cube-free median graphs are bipartite Helly graphs and the previous
algorithm performs a single step of local search for $r_\pi$ in
$\cO(m)$ time. The complexity of the sample-select-descent also
depends on the number of descent steps and the size of the initial set
$U$.  Therefore sample-select-descent cannot be used to get an
algorithm faster than $\ctO(\sqrt{n}m)$.

For cube-free median graphs, we design an $\ctO(m)$-time
algorithm. To do so, we use the $G^2$-unimodality of $r_\pi$
differently: we generalize in a non-trivial way the
cut-on-best-neighbor method by Kariv and Hakimi \cite{KaHa} for
weighted centers in trees.  Their algorithm computes a centroid $v$ of
$T$ and a neighbor $v'$ of $v$ with $r_\pi(v')<r_\pi(v).$ Since the
radius function $r_\pi$ of $T$ is convex, if $v'$ does not exist then
$v$ is central.  Otherwise, the center belongs to the subtree $T'$ of
$T$ containing $v'$ and the recursion is performed on it.  For median
graphs $G$, the role of subtree $T'$ is played by \emph{halfspaces},
i.e., convex sets with convex complements.  The halfspaces of $G$ are
exactly the sets of the form $H(x,y)=\{ z\in V: d(z,x)<d(z,y)\}$ where
$xy$ is an edge of $G$ and can be elegantly constructed in linear time
by LexBFS \cite{BeChChVa_med}.

Let $G=(V,E)$ be a cube-free median graph; then $G$ contains at most $2n$
edges and $r_\pi$ is $G^2$-unimodal. We start with $G$ as the search
region $R$. At each iteration, the search region $R$ is convex (and
thus gated) and contains the center $C_\pi(G)$. Furthermore, $R$
is the intersection of the search region $R'$ of the previous iteration with a
certain halfspace $H$ of $G$ and has size at most one half of the size of
$R'$: $R=R'\cap H$ and $|R|\le |R'|/2$. Consequently, the algorithm has $\log_2 n$ iterations.
Now, we show how,  given the search region $R$,  to compute in $\cO(n\log n)$ time a halfspace $H$
such that $C_\pi(G)\subseteq H$ and $|R\cap H|\le |R|/2$.
%
First, a median/centroid $v$ of $R$ is computed in $\cO(|R|)$ time using
\cite{BeChChVa_med}. Then we compute the star
$\St(v) \subseteq B_2(v)$ of $v$ in $R$ (the union of all edges and
squares containing $v$) and $r_\pi(z)$ for all $z\in \St(v)$ (this can
be done in linear time because $\St(v)$ is gated).  If $v$ minimizes
$r_\pi(v)$ in the star $\St(v)$, using the $G^2$-unimodality of $r_\pi$,
we show that $v$ is a central vertex of $G$ and we return
it. Otherwise, we compute a halfspace $H$ of $R$ separating $v$ from
$C_\pi(G)$, i.e., $C_\pi(G)\subseteq H$ and $v\in R\setminus
H$. 
Since $v$ is a median of $R$ and $C_\pi(G)\subseteq H$,
$|R\cap H|\le |R|/2$ holds. 
%

\begin{figure}
    \begin{center}
        \includegraphics[width=.8\textwidth]{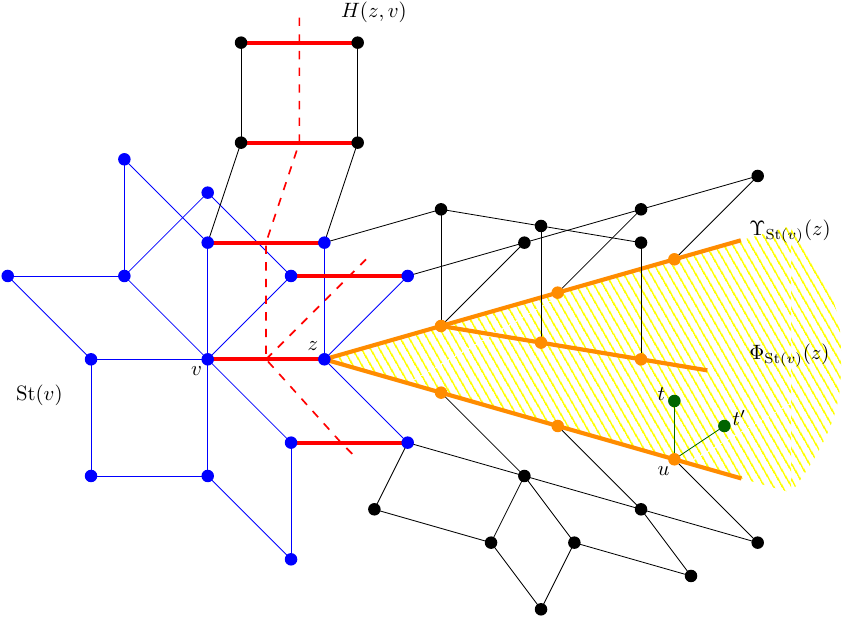}
    \end{center}
    \caption{{\sf ReduceConvexRegion}  for cube-free median graphs.
      Starting from a vertex $v$, its star $\St(v)$ is in blue and
      given an improving neighbor $z$ of $v$, the parallel with $vz$ edges defining the halfspace $H(z,v)$ are drawn
      in bold red, the fiber $\Psi_{\St(v)}(z)$ is drawn in stripped yellow, and its
      boundary $\Psi_{\St(v)}(z)$ in orange.}\label{fig-algos2}
\end{figure}

The computation of the halfspace $H$ 
is the main technical issue of the algorithm. This is done in $\cO(n\log n)$ time by using the function {\sf ReduceConvexRegion}, which is
described  in Section \ref{sec:median} and which we now outline (see Figure~\ref{fig-algos2} for an illustration).
A vertex $z\in \St(v)$ with $r_\pi(z)<r_\pi(v)$ is called
an \emph{improving neighbor} of $v$ if $z\sim v$ and an \emph{improving second neighbor} if $z\nsim v$.
First we show
that $v$ has at most two improving neighbors. 
If $v$ does not have improving neighbors but has an improving second neighbor $w\in R$, then as $H$
we consider a halfspace $H(z,v)$ where $z\sim v,w$. 
Otherwise, we pick an improving neighbor $z$ and we construct the fiber
$\Fib_{\St(v)}(z)$ of $z$ (\textit{i.e.}, the set of vertices having $z$ as their
gates in $\St(v)$).  The boundary $\Upsilon_{\St(v)}(z)$ is the set of
vertices of $\Fib_{\St(v)}(z)$ that have a neighbor outside of
$\Fib_{\St(v)}(z)$.  For cube-free median graphs, this boundary is a tree with gated branches.
Using the  cut-on-best-neighbor approach  for trees, we find in $\cO(n \log n)$
time a local-minimum $u$ of the radius function $r_\pi$ on $\Upsilon_{\St(v)}(z)$
(i.e., a vertex $u \in \Upsilon_{\St(v)}(z)$ such that $r_\pi(u)\le r_\pi(u')$ for all
neighbors $u'$ of $u$ in $\Upsilon_{\St(v)}(z)$). Then we compute in linear time the star
$\St(u)$ of $u$ and $r_\pi(t)$ for all vertices $t \in \St(u)$. If $u$ is a local minimum in
$\St(u)$, we return $u$ as a central vertex. If $u$ has  one
improving neighbor $t$, then we select the halfspace $H(t,u)$ as $H$.
Analogously, if $u$ does not have an improving neighbor but has an improving second neighbor $w$, then
we take as $H$ a halfspace $H(t,u)$, where $t\sim u,w$ and $t$ does not belong to the $(z,u)$-path of $\Upsilon_{\St(v)}(z)$.
Finally,   if $u$ has two improving neighbors, we prove that they
belong to $\Phi_{\St(v)}(z)$ and that the halfspace $H(z,v)$ defined by the edge
$vz$ separates $v$ from $C_\pi(G)$ (and thus can be taken as $H$).
Since  {\sf ReduceConvexRegion} is called $\cO(\log n)$ times, we obtain an algorithm with complexity $\cO(n\log^2 n)$.

\section{Deterministic sample-select-descent}\label{descent} Let $f: V\rightarrow {\mathbb Z}^+\cup \{ 0\}$  be a 2-weakly peakless function on $G$. \emph{Local search} (the \emph{descent algorithm})  constructs
a minimizing path $(u_0,\ldots,u_\ell)$ of $G^2$ such that $f(u_0)>\cdots>f(u_\ell)$,
and $u_\ell$
is a local (and thus  global) minimum of $f$. At step $i+1$,  a vertex $u_{i+1}\in B_2(u_i)$ such that $f(u_i)>f(u_{i+1})$ is computed.
We say that $f$ has
\emph{descent} $p$ if for any $v$ and  $x\in B_2(v)$,  $|f(v)-f(x)|\le p$. Any 0-1-radius function $r_{\pi}$  has descent $2$. 
For a vertex $u$, denote by $\delta(u)$ 
the number of descent steps necessary to  achieve a vertex from $\argmin(f)$ starting from $u$. 

\begin{mylemma} \label{norm} Let $f$ be a 2-weakly peakless function on $G$  with descent $p$. For any $u\in V$ and $v\in \argmin(f)$ closest to $u$,  $f(u)\le f(v)+p\tau$ and $\delta(u)\le p\tau$, where $\tau=\frac{d(u,v)}{2}+1$.
\end{mylemma}

\begin{proof} Since $f$ is \gd{weakly} peakless in $G^2$, there is a  $(u,v)$-geodesic $P'=(u=v'_0,v'_1,\ldots, v'_k=v)$  in $G^2$
    such that $f(v'_0)>f(v'_1)>\ldots>f(v'_k)$, where $k=d_{G^2}(u,v)\le \tau$. 
    At each  step, $f$ decreases by $\le p$ units, thus $f(u)\le f(v)+ p\tau$. 
    Pick any longest minimizing path $P''=(v_0=u,v_1\ldots,v_{k'}=v')$ of $G^2$ such that $f(v_0)>f(v_1)>\ldots>f(v_{k'})$ and $v'\in \argmin(f)$. 
    By definition, $\delta(u)=k'$. 
    Since $d(u,v)\le d(u,v')$, $k'\ge k$. Since at each step $f$ decreases  by $\ge 1$ and $f(u)\le f(v)+ p\tau=f(v')+p\tau$, we get $k'\le p\tau$.
\end{proof}

By  \cite{MeMo}, any $G$ can be covered in $\cO(m)$ time with $\Theta(\sqrt{n})$ balls of radius $2\sqrt{n}$. 
Let $U$ be the set of centers of these balls. Compute $f$ on the vertices of $U$ and let $u$ minimize $f$ on $U$. 
Run local search on $G^2$ starting from $u$ until we find a local minimum $v^*$. 

\begin{myproposition} \label{number-of-steps} If $f$ is a 2-weakly peakless function with descent $p$, then the number of descent steps to compute $v^*\in \argmin(f)$ starting from $u\in U$ is at most $p(\sqrt{n}+1)$.
\end{myproposition}

\begin{proof} Let $u'\in U$ so that $d(u',v^*)\le 2\sqrt{n}$.
Set $\tau = \frac{d(u',v^*)}2+1 \le \frac{2\sqrt{n}}2+1=\sqrt{n}+1$.
By Lemma~\ref{norm},  $f(u')\le f(v^*)+p\tau$.
    From the choice of $u$,  $f(u)\le f(u')$, thus  $f(u)\le f(u')\le f(v^*)+p\tau$. Since each  step decreases $f$ by $\ge 1$ and  $d(u',v^*)\le 2\sqrt{n}$, we need $\le p\cdot(\sqrt{n}+1)$  steps to reach $v^*$ from $u$.
\end{proof}

\section{The center problem in (weakly) bridged graphs}\label{s:bridged}

In this section, we present an algorithm for the center problem in weakly  bridged graphs based on $G^2$-unimodality of the radius function.

\subsection{Bridged and weakly bridged graphs.}
Recall that a graph $G$ is  \emph{bridged} if  balls $B_r(C)$ around convex sets $C$ are convex \cite{SoCh,FaJa}. 
A graph $G$ is \emph{weakly bridged} (\emph{WB-graph} for short) if $G$ is a graph with convex balls in which each $C_5$ is included in a 5-wheel \cite{ChOs}.
Topological characterization of bridged graphs, respectively of weakly bridged graphs, via their clique complexes were given in \cite{Ch_CAT} and \cite{ChOs}. Notice that the clique complexes of bridged graphs are exactly the systolic simplicial complexes
introduced in \cite{JaSw} and which are considered as combinatorial analogs of simplicial nonpositive curvature.

In this paper, we only need a few metric
characterizations and properties of these graphs, which we now recall.
%
%
%
As we noted already (see Section~\ref{ss:wmg}), bridged graphs and weakly bridged graphs are weakly modular. 
Furthermore, weakly bridged graphs  $G$ satisfy the  \emph{Interval Neighborhood Condition} $\INC$: for any $u,v\in V$, the neighbors of $v$ in $I(u,v)$ form a clique.
%
%
Recall now the metric characterization of bridged and weakly bridged graphs: 

\begin{mytheorem}\label{thm:B-graphs}
  For a graph $G$, the following are equivalent:
  \begin{enumerate}
  \item $G$ is bridged;
  \item any isometric cycle of $G$ has length 3~\cite{SoCh,FaJa};
  \item $G$ is a weakly modular graph without induced $C_4$ and
    $C_5$~\cite{Ch_metric,Ch_CAT}.
  \end{enumerate}
\end{mytheorem}

\begin{mytheorem}[\!\!\cite{ChOs}]\label{thm:WB-graphs}
  A graph $G$ is weakly bridged \gd{iff} $G$ is weakly modular without
  induced $C_4$.
\end{mytheorem}


A graph $G$ satisfies the \emph{interval-outergate property} if for any $u,v\in V$ with $d(u,v)=k \geq 2$, there exists  $w\in I(u,v)$ (called the \emph{interval-outergate} of $u$ with respect to $v$)
at distance $k-2$ from $u$ and adjacent to all vertices of $I(v,u)\cap N(v)$.

\begin{mylemma}[\!\!\cite{ChOs}]\label{lem:wb-outergate}
    Any  weakly bridged graph $G$ satisfies the interval-outergate property.
\end{mylemma}

\begin{mylemma}\label{lem:wm-wx} If $G$ is weakly modular and $u,v,x\in V$ are such that $d(x,u)<d(x,v)$, then there exists $w_x\in N(v)\cap I(v,u)$ such that $w_x\in I(v,x)$. 
\end{mylemma}

\begin{proof} Let $u'v'x'$ be a metric triangle such that  $d(u,v)=d(u,u')+d(u',v')+d(v',v)$, $d(u,x)=d(u,u')+d(u',x')+d(x',x)$, and $d(v,x)=d(v,v')+d(v',x')+d(x',x)$. If $v'\ne v$, then any neighbor of $v$ in $I(v,v')\subset I(v,u)\cap I(v,x)$ can be selected as $w_x$. Now suppose that $v'=v$.
    Since $d(x,u)<d(x,v)$, the vertices $u',v',x'$ are pairwise distinct and the metric triangle $u'v'x'$ cannot be equilateral, which is impossible since $G$ is weakly modular.
\end{proof}

\subsection{$G^2$-unimodality of the radius function.}
The following result establishes the statement of Theorem \ref{Gpunimodality} for (weakly) bridged graphs: 

\begin{myproposition}\label{thm:wb} All radius functions of a weakly bridged graph $G$ are $2$-weakly peakless. 
\end{myproposition}

\begin{proof}    Let $u,v$ be any two vertices of $G=(V,E)$ with $d(u,v)\ge 3$ and suppose that $r_\pi(u)\le r_\pi(v)$.  If $r_\pi(u)=r_\pi(v)$, then for any vertex $x\in \supp(\pi)$
    and any vertex $w\in I(u,v), w\ne u,v$, by convexity of balls we have
$\pi(x)d(x,w)\le \max\{ \pi(x)d(x,u),\pi(x)d(x,v)\}\le r_\pi(u)=r_\pi(v),$  yielding $r_{\pi}(w)\le r_\pi(u)=r_\pi(v)$. Thus, suppose that $r_\pi(u) < r_\pi(v)$. To establish
$\WP(u,v)$, we will prove the following assertion: \emph{For any  $w\in I(v,u)$ such that $d(v,w)=2$ and $N(v)\cap I(v,u)\subseteq N(w)$, we have  $r_\pi(w) < r_\pi(v)$}
    (such a vertex exists by Lemma~\ref{lem:wb-outergate}).
    %
    %
    %
    Recall that $F_\pi(v) = \{ x \in \supp(\pi): \pi(x)d(x,v)=r_\pi(v) \}$. Since $G$ has convex balls,  for any $z \in I(v,u)\setminus \{ v,u\}$ and any $y \in \supp(\pi) \setminus F_\pi(v)$, the following inequality holds:
$\pi(y) \cdot d(z,y) \leq \max\{\pi(y) \cdot d(v,y), \pi(y) \cdot d(u,y)\} \leq \max\{\pi(y) \cdot d(v,y), r_\pi(u)\} < r_\pi(v)$.
    Therefore, in order to prove $r_\pi(w) < r_\pi(v)$, it suffices to prove that $d(w,x) < d(v,x)$ for every $x \in F_\pi(v)$.
    By Lemma~\ref{lem:wm-wx}, for every $x \in F_\pi(v)$, there is some $w_x \in N(v) \cap I(u,v)$ such that $d(w_x,x) < d(v,x)$.
    Set $W = \{w_x: x \in F_\pi(v)\}$. By definition $W \subseteq N(v) \cap I(u,v) \subseteq N(w)$.
    Finally, for any $x \in F_\pi(v)$ we have $w_x,u\in B_{d(x,v)-1}(x)$.  Since $w\in I(w_x,u)$, the convexity of $B_{d(x,v)-1}(x)$ yields $d(w,x) \leq \max\{d(w_x,x),d(u,x)\}<d(x,v)$.
    Thus, $r_\pi$ is $2$-weakly peakless. 
\end{proof}

%

\subsection{The algorithm.}
We consider the algorithmic implications of Proposition~\ref{thm:wb} and local search; namely,
we show how for weakly bridged graphs to implement each iteration of the descent algorithm in $\cO(m)$ time.  


\subsubsection{Computing $r_\pi$ for a clique.} We start by recalling the following result: 

\begin{mylemma}[\!\!{\cite[Lemma 6]{Du_Helly}}]\label{lem:outergate}
    For a set $S$ of a graph $G$, in $\cO(m)$ time one can map every  $z \notin S$
    to $z^* \in B_{d(v,S)-1}(z)\cap B_1(S)$ maximizing $|N(z^*) \cap S|$.
    Furthermore, if $S$ is outergated, then $z^*$ is an outergate of $z$.
\end{mylemma}



\begin{mylemma}\label{lem:cb-clique-1}
    Any clique $K$ of $G$ is outergated.  
\end{mylemma}

\begin{proof} Let $z \notin K$ be arbitrary, let $K'=\proj_z(K)$ and $u\in V$ with $d(z,u)=d(z,K)-1$ maximizing $|N(u)\cap K'|$. Suppose for the sake of contradiction that $u$ is not an outergate of $z$. 
    Let $u\nsim w$ for $w\in K'$. Pick $v\in K, v\sim u$. By $\TC$, there exists
$u'\sim v,w$ with $d(z,u')=d(z,K)-1$. Since $u\nsim w$, $u'\ne u.$ By $\INC$, $u\sim u'. $ Pick $x\in K', x \sim u$.  Since $G$ is $C_4$-free, 
$u'\sim x$. Thus $|N(u')\cap K'|>|N(u)\cap K'|$, 
    a contradiction. 
    %
\end{proof}

\begin{mylemma}\label{thm:wb-cliques}
    For any clique $K$, in $\cO(m)$ time, we can compute $r_\pi(w)$ for every $w \in K$.
\end{mylemma}

\begin{proof}
    By Lemma~\ref{lem:cb-clique-1}, $K$ is outergated.
    Then, by Lemma~\ref{lem:outergate}, in $\cO(m)$ time we can compute an outergate $g_K(y)$ for every $y \notin K$.
    For every $u \in N(K)$, we compute the weights $\alpha(u) = \max\{ \pi(y)d(y,K): y \notin K  \mbox{ and }  g_K(y)=u \}$ and $\beta(u) = \max\{ \pi(y)(d(y,K)+1): y \notin K  \mbox{ and }  g_K(y) = u  \}$.
    Pick any $w \in K$.
    Then
    \begin{flalign*}
    r_\pi(w) = \max \Big\{&\{ \pi(w'): w' \in K \setminus \{w\} \} \cup \{ \pi(x)d(x,K): x \in X \} \\
    &\cup \{ \alpha(u): u \in N(K) \cap N(w) \} \cup \{ \beta(u): u \in N(K) \setminus N(w) \}\Big\}.
    \end{flalign*}
    We assert that we can compute in $\cO(m)$ time $\max\{ \beta(u): u \in N(K) \setminus N(w) \}$, for every $w \in K$.
    For that, we apply Lemma~\ref{lem:max-nonneighbor} to the subgraph induced by $N[K]$ and to the weight function $\kappa$, where $\kappa(w) = 0$ if $w \in K$ and $\kappa(u) = \beta(u)$ if $u \in N(K)$.
    Thus, for each $w \in K$, we compute $r_\pi(w)$ in additional $\cO(\deg(w))$ time. The total runtime is in $\cO(m)$.
\end{proof}

\subsubsection{Local minima of $r_\pi$ in $G$ and in $G^2$.} First we show how to minimize $r_\pi$ on $B_1(v)$.   

\begin{mylemma}\label{thm:local-min}
    For $v \in V$, we can compute $u$ in $B_1(v)$ minimizing $r_\pi$ in  $\cO(m)$ time. 
\end{mylemma}

\begin{proof}
    Let $F_\pi(v) \cap N(v) = \varnothing$, otherwise $|F_\pi(v) \cap N(v)| > 1$ and $v$ is a local minimum of $r_\pi$
    or $F_\pi(v) \cap N(v) = \{u\}$ and we can output a vertex amongst $v,u$ \gd{that minimizes $r_\pi$}.
    By Lemma~\ref{lem:wb-outergate}, every $z \in F_\pi(v)$ has an outergate $g_v(z)$ with respect to $B_1(v)$.
    Furthermore, by Lemma~\ref{lem:outergate} we can compute $g_v(z)$, for any $z \in F_\pi(v)$, in $\cO(m)$ time.
    Let $K = \bigcap\{ N(v) \cap I(v,z): z \in F_\pi(v) \}$.
    Note that $K= \bigcap\{N(v) \cap N(g_v(z)): z \in F_\pi(v)\}$, thus one can compute $K$ in $\cO(m)$ time.
    For any $w\in N(v)\setminus K$ there exists $z\in F_\pi(v)$ such that
    $w\notin I(v,z)$. Hence $d(w,z)\ge d(v,z)$, yielding $r_\pi(w)\ge r_\pi(v)$.  Therefore, if $K=\varnothing$, $v$
    is a local minimum of $r_\pi$. If $K \neq \varnothing$,  since $K$ is the intersection of the cliques $N(v) \cap I(v,z), z \in F_\pi(v)$,
    $K$ is a clique. By Lemma~\ref{thm:wb-cliques}, we can compute $u\in K$ minimizing $r_\pi$ 
    in $\cO(m)$ time.
    Since $r_\pi(w)\ge r_\pi(v)$ for any $w\in N(v)\setminus K$, either $v$ or $u$ is a local minimum of $r_\pi$ in $B_1(v)$.
\end{proof}

Now, we describe a local-search technique, allowing to find a vertex in $B_2(v)$ with smaller eccentricity than that of $v$ (if one exists).
First, we show that Lemma~\ref{lem:wm-wx} can be generalized as follows:

\begin{mylemma}\label{lem:wm-wX-2}
    Let $X\subset V$ be nonempty, and  $u,v,w \in V$ be such that (1) for every $x \in X$, $d(u,x) < d(v,x)$ and
    (2) $w \in \bigcap\{N(v) \cap I(v,x):  x \in X\}$.
    Then there exists  $w_X \in N(v) \cap I(v,u)$ such that $w_X \in \bigcap\{N(v) \cap I(v,x):  x \in X\}$. 
\end{mylemma}

\begin{proof}
    Assume $w \notin I(u,v)$, otherwise we can set $w_X = w$.  If $d(u,w)>d(u,v)$, then $v \in I(u,w)$ and $w,u\in B_{d(v,x)-1}(x)$ for any $x \in X$,
    contradicting the convexity of the ball $B_{d(v,x)-1}(x)$. Thus $d(u,w) = d(u,v)$.
    Pick any $y \in X$. By Lemma~\ref{lem:wm-wx}, there exists  $w_y \in N(v) \cap I(v,u)$ such that $d(y,w_y) < d(y,v)$.
    Since $w,w_y\in N(v)\cap I(v,y)$ and the ball $B_{d(y,v)-1}(y)$ is convex, the vertices $w_y,w$ are adjacent.
    Consequently, $w_y \in I(w,u)$. Since $u,w\in B_{d(x,v)-1}(x)$ for any $x\in X$, the convexity of the ball $B_{d(x,v)-1}(x)$ implies that  $d(x,w_y) \leq d(x,v)-1$.
    Thus  $w_y \in \bigcap\{N(v) \cap I(v,x):  x \in X\}$ and we can set $w_X = w_y$.
\end{proof}

%


\begin{myproposition}\label{thm:loc-search} If $v\notin C_\pi(G)$, in $\cO(m)$ time one can find $u' \in B_2(v)$ with $r_\pi(u') < r_\pi(v)$.
\end{myproposition}

\begin{proof}
    Suppose $v$ is not central. If there is $v'\sim v$ with $r_\pi(v') < r_\pi(v)$, then Lemma~\ref{thm:local-min} will return  $u\in B_1(v)$ such that $r_\pi(u)\le r_\pi(v')<r_\pi(v)$.
    So, let $v$ be a local minimum of $r_\pi$. Let $F_\pi(v)=\{z \in V: \pi(z) \cdot d(v,z) = r_\pi(v)\}$. 
    Since $v$ is a local but not a global minimum, 
    $d(v,z)>1$ for any $z\in F_\pi(v)$.
    For every  $w \in N(v)$, let $\Psi_\pi(w) = \{ z \in F_\pi(v): w\in N(v) \cap I(v,z) \}$.
    We define an equivalence relation $\equiv$ on $N(v)$ as follows: $w \equiv w' $ iff  $\Psi_\pi(w) = \Psi_\pi(w')$.
    We partition $N(v)$ according to  $\equiv$. For that, we use outergates with respect to $N(v)$,
    which exist for any $z \in F_\pi(v)$ by Lemma~\ref{lem:wb-outergate}. 
    By Lemma~\ref{lem:outergate} in $\cO(m)$ time we compute
    an outergate $g_v(z)$ for every $z \in F_\pi(v)$.
    Then, for any $w,w' \in N(v)$, we have  $w \equiv w'$ iff $N(w) \cap \{g_v(z):  z \in F_\pi(v)\} = N(w') \cap \{g_v(z) :  z \in F_\pi(v)\}.$
    Consequently, we can compute in $\cO(m)$ time the equivalence classes on $N(v)$ with respect to $\equiv$ by using partition refinement techniques.
    Next, in $\cO(n)$ time we compute,  for every vertex $u \in B_2(v) \setminus B_1(v)$, the number of vertices $z \in F_\pi(v)$ such that $g_v(z) = u$.
    Doing so, in total $\cO(m)$ time, we also compute $|\Psi_\pi(W_i)|$ for every equivalence class $W_i$ of $\equiv$
    (where we set $\Psi_\pi(W_i) = \Psi_\pi(w)$ for arbitrary $w \in W_i$).
    Let $W_0$ be an equivalence class maximizing $|\Psi_\pi(W_0)|$. If $|\Psi_\pi(W_0)| = |F_\pi(v)|$, then we abort.
    From now on, we assume that $|\Psi_\pi(W_0)| < |F_\pi(v)|$.  Let $z \in F_\pi(v) \setminus \Psi_\pi(W_0)$ be arbitrary.
    Then in $\cO(m)$ time we compute $N(v) \cap I(v,z) = N(v) \cap N(g_v(z))$.  Let $w_{\max} \in N(v) \cap I(v,z)$ be a vertex maximizing $|N(w_{\max}) \cap W_0 |$.
    Finally, we apply Lemma~\ref{thm:local-min} to compute a vertex $v^+$ minimizing $r_\pi$ within $N(w_{\max})$ and the algorithm returns $v^+$.

    Now, we prove the correctness of this algorithm. Recall that $v$ is a local minimum of $r_\pi$ and  $r_\pi(v')<r_\pi(v)$ for some $v'$.  
    By Lemma~\ref{lem:wb-outergate}, there exists $u \in I(v,v')$ such that $d(u,v)=2$ and $N(v) \cap I(v,v') \subseteq N(u)$.
    Furthermore, by the proof of Proposition~\ref{thm:wb}, $r_\pi(u) < r_\pi(v)$.  By Lemma~\ref{lem:wm-wX-2}, there exists  $w' \in N(v) \cap I(v,v') \subseteq N(u)$ such that $\Psi_\pi(W_0) \subseteq \Psi_\pi(w')$.
    By maximality of $|\Psi_\pi(W_0)|$, necessarily  $w' \in W_0$. To prove the correctness of the algorithm, we show that (1) $|\Psi_\pi(W_0)| = |F_\pi(v)|$ is impossible and
    that (2) $u$ is adjacent to $w_{\max}$. This will show that the algorithm returns a vertex $v^+\in N(w_{\max})\subset B_2(v)$ with $r_\pi(v^+)\le r_\pi(u)<r_\pi(v)$.
    First, suppose   $|\Psi_\pi(W_0)| = |F_\pi(v)|$, i.e., $\Psi_\pi(W_0)= F_\pi(v)$ holds. Since $w'\in W_0$, we have $\pi(y)d(w',y)<\pi(y)d(v,y)$ for any vertex $y\in \Psi_\pi(W_0)$.
    Since $v$ is a local
    minimum, the profile $\pi$ must contain  $y'\notin \Psi_\pi(W_0)= F_\pi(v)$ such that
    $\pi(y')d(w',y') = r_\pi(w') \ge r_\pi(v)>\pi(y')d(v,y').$  This implies $d(w',y')=d(v,y')+1$. Since $\pi(y')d(u,y')\le r_\pi(u)<r_\pi(w')=\pi(y')(d(v,y')+1)$, we also conclude that
    $d(u,y')\le d(v,y')$. Consequently, $u$ and $v$ belongs to the ball $B_{d(v,y')}(y')$ and $w'\in I(u,v)$ does not belong to this ball, a contradiction with convexity of balls.  This establishes that
    $\Psi_\pi(W_0)$ is strictly included in $F_\pi(v)$ and proves that the vertex $z$ chosen by the algorithm exists.
    %
    To prove that the vertex  $w_{\max}$ computed by the algorithm is adjacent to $u$, first we prove that $w_{\max}$ is adjacent to every  $w\in N(u)\cap W_0$ (this set is nonempty since it contains
    $w'$). 
    By Lemma~\ref{lem:wm-wx}, there exists  $w_z\in N(v) \cap I(v,z) \cap I(v,v') \subseteq N(u)$.
    Since $w,w_z \in I(u,v)$, by $\INC$  the vertices  $w,w_z$ must be adjacent.
    Furthermore, since $w_z,w_{\max} \in N(v) \cap I(v,z)$, and so they are adjacent, the sets $N(w_z) \cap W_0$ and $N(w_{\max})\cap W_0$ must be comparable for inclusion (otherwise, we get a forbidden induced $C_4$).
    By maximality choice of $w_{\max}$, we obtain $w\in N(w_z) \cap W_0\subseteq N(w_{\max})\cap W_0$, thus $w_{\max}$ and $w$ are adjacent.
    Then, $w_{\max},u \in N(w) \cap I(w,z)$, and by $\INC$  $w_{\max}\sim u$. This concludes the proof.
\end{proof}

Consequently, Algorithm~\ref{alg:improveeccentricity} implements the
{\sf ImproveEccentricity} procedure in the case of weakly bridged graphs.


\begin{algorithm2e}[h]
    \label{alg:improveeccentricity}
    \SetKwInOut{Input}{Input}
    \SetKwInOut{Output}{Output}
   \Input{A weakly bridged graph $G=(V,E),$ a profile $\pi:V\rightarrow R^+,$ and a vertex $v\in V$}
   \Output{A vertex $u\in B_2(v)$ with $r_\pi(u)<r_\pi(v)$ if it exists and $v$ otherwise}
   Compute a vertex $u$ minimizing $r_\pi$ in $B_1(v)$ \hfill \tcp{$\cO(m)$ by Lemma~\ref{thm:local-min}}
   \If{$r_\pi(u)<r_\pi(v)$}{\KwRet{$u$}}
   Compute the outergates in $B_1(v)$ of every vertex $z\in V$\;
   Partition $N(v)$ into the equivalence classes $W_0,W_1,\hdots,W_k$ of the relation $\equiv$\;
   Assume $W_0$ is an equivalence class with the largest shadow  $\Psi_\pi(W_0)$\;
   Pick a vertex $z\in F_\pi(v)\setminus \Psi_\pi(W_0)$\;
   Let $w_{\max}$ be a vertex of $N(v)\cap I(v,z)$ having a maximum number of neighbors in $W_0$\;
   Compute a vertex $v^+$ minimizing $r_\pi$ in $B_1(w_{\max})$ \hfill \tcp{$\cO(m)$ by Lemma~\ref{thm:local-min}}
   \eIf{$r_\pi(v^+)=r_\pi(v)$}{\KwRet{$v$}}{\KwRet{$v^+$}}
  \caption{{\sf ImproveEccentricity}$(G,\pi,v)$ for weakly bridged graphs}
\end{algorithm2e}

\subsubsection{Computing a central vertex.} 
Local search will construct a path $(v_0,v_1,\ldots, v_\ell)$
of $G^2$ such that $r_\pi(v_0)>r_\pi(v_1)>\ldots>r_\pi(v_\ell)$ by
applying {\sf ImproveEccentricity}$(G,\pi,v)$ at each
step, until it reaches a vertex $v^*:=v_\ell$ where the algorithm
aborts. By Proposition~\ref{thm:loc-search}, $v^*$ is a central vertex
of $G$.  By Lemma~\ref{lem:k-unimodal}, the length of
$(v_0,v_1,\ldots, v_\ell)$ is at most
$\ctO(\sqrt{n})$. In the case of 0-1-profiles, the
length of this path will be $\cO(\sqrt{n})$
(Proposition~\ref{number-of-steps}). Consequently, we obtain the following main result of this section (which is Theorem \ref{main} in the case of weakly bridged graphs):

\begin{mytheorem}\label{thm:main-bridged}
    If $G$ is a weakly bridged graph  with arbitrary weights, then a central vertex of $G$ can be computed in randomized time $\ctO(\sqrt{n}m)$.
    For 0-1-weights,  a central vertex of $G$ can be computed in deterministic time $\cO(\sqrt{n}m)$.
\end{mytheorem}

\begin{myremark}
  In the appendix (Section~\ref{sec:cb-alg}), we extend the
  algorithm for weakly bridged graphs to graphs with convex balls. In
  this case, both the implementation and the correctness proof of {\sf
    ImproveEccentricity}$(G,\pi,v)$ are technically more involved.
    We also need to design a final procedure that either asserts that the
  last vertex $v^*$, obtained with local search, is central or finds a center inside
  $B_2(v^*)\setminus B_1(v^*)$.
\end{myremark}

\section{The center problem in bipartite Helly graphs}\label{sec:bar-alg}

In this section, we present a subquadratic-time algorithm for the center problem in bipartite
Helly graphs, based on $G^2$-unimodality of the radius function. The main technical contribution
is an $\cO(m)$-time implementation of the function {\sf ImproveEccentricity}.

\subsection{Bipartite Helly graphs.}  Recall that a graph $G$ is a \emph{Helly graph} if 
the set of  balls of $G$ satisfies the Helly property: any collection of pairwise intersecting balls has a nonempty intersection.
Analogously, a graph $G$ is a \emph{bipartite Helly graph}  if $G$ is bipartite and the family of half-balls of $G$ satisfies the Helly property.
As we noticed above (see Section~\ref{ss:wmg}), bipartite Helly graphs are modular. A topological local-to-global characterization of Helly graphs was given in \cite{CCHO}, however such a
characterization  for bipartite Helly graphs is missing.

Bipartite Helly graphs have been introduced and characterized in several nice
ways by Bandelt, D\"{a}hlmann, and Sch\"{u}tte~\cite{BaDaSc}, some of which we recall next.
The graph $B_n$ is the bipartite complete graph $K_{n,n}$ minus a
perfect matching, i.e., the bipartition of $B_n$ is defined by
$a_1,\ldots,a_n$ and $b_1,\ldots,b_n$ and $a_i$ is adjacent to $b_j$
iff $i\ne j$. The graph $\widehat{B}_n$ is obtained from
$B_n$ by adding two adjacent new vertices $a$ and $b$ such that $a$ is
adjacent to $b_1,\ldots,b_n$ and $b$ is adjacent to $a_1,\ldots,a_n$.  

\begin{mytheorem}[\!\!\cite{BaDaSc}]\label{absolute-retracts}
    For a bipartite graph $G=(V_0\cup V_1,E)$, the following conditions are equivalent:
    \begin{enumerate}[(1)]
        \item\label{th-bar-1} $G$ is a bipartite Helly graph;
        \item\label{th-bar-3} $G$ is a modular graph such that every induced
        $B_n$ ($n\ge 4$) extends to $\widehat{B}_n$ in $G$;
        \item\label{th-bar-4} $G$ satisfies the following  \emph{interval-outergate}
        property: for any vertices $u$ and $v$ with $d(u,v)\ge 3$, the
        neighbors of $v$ in $I(u,v)$ have a second common neighbor $x$ in
        $I(u,v)$.
    \end{enumerate}
\end{mytheorem}

\subsection{$G^2$-unimodality of the radius function.}

The following result establishes the statement of Theorem \ref{Gpunimodality} related to bipartite Helly graphs:

\begin{myproposition} \label{bar-wp}  All radius functions $r_\pi$ of a bipartite Helly graph $G$ are 2-weakly peakless.
\end{myproposition}
\begin{proof}
    By the characterization of 2-weakly peakless functions of \cite{BeChChVa_G2}, it suffices
    to establish the property $\WP(u,v)$ for any two vertices $u,v$ with $3\le d(u,v)\le 4$.
    By Theorem~\ref{absolute-retracts}, the vertices $u$ and $v$ have interval-outergates $a$ and  $b$,
    i.e., there exist vertices  $a,b\in I^{\circ}(u,v)$ such that $d(u,a)=d(v,b)=2$ and $a$ is adjacent to all
    vertices of $A=N(u)\cap I(u,v),$ $b$ is adjacent to all vertices of $B=N(v)\cap I(u,v)$. Let $C=I^{\circ}(u,v)-(A\cup B).$ The set $C$ is empty when $d(u,v)=3$ and contains all middle vertices of shortest $uv$-paths when $d(u,v)=4.$
    Suppose without loss of generality that $r_{\pi}(u)\le r_{\pi}(v)$.
    %
    %
    Since any bipartite Helly graph is modular, every triplet of
    vertices has a median.  Let $z$ be any vertex and $z'$ be a median
    of the triplet $u,v,z$. Then $z'$ either belongs to
    $A\cup B\cup C$ or coincides with $u$ or $v$. If $z'\in A$, then
    $d(a,z)\le d(a,z')+d(z',z)\le 1+d(z',z)=d(u,z)$.  Since any vertex of
    $C$ has a neighbor in $A$ and $a$ is adjacent to every vertex in
    $A,$ $d(a,w)\le 2$ for any $w\in C.$ Hence, if $z'\in C$ then
    $d(a,z')\le d(u,z').$ Since $d(a,b)\le 2$ when $b\in C$ and
    $d(a,b)=1$ when $b\in A$, $d(a,b)\le d(u,b).$ Additionally,
    $b\in I(u,w)$ for any $w\in B\cup \{v\}.$ Hence, if
    $z'\in B\cup \{v\}$ then $d(a,z')\le d(u,z').$ We conclude that
    $d(a,z)\le d(u,z)$ except when $z'=u.$
    Let $x$ be any furthest from $a$ vertex   of the profile $\pi$, i.e., $r_\pi(a)=\pi(x)d(x,a).$ Let $x'$ be any median of the triplet
$u,v,x$. If $x'\ne u$, then from the previous conclusion we deduce that $r_{\pi}(u)\ge \pi(x)d(x,u)\ge \pi(x)d(x,a)=r_{\pi}(a)$.
    Since $r_{\pi}(u)\le r_\pi(v)$,  $a$ verifies the condition $\WP(u,v)$. Now, let $x'=u$.
    Then $r_{\pi}(v)\ge \pi(x)d(v,x)>\pi(x)d(a,x)=r_\pi(a)$.
    We conclude that
$r_\pi(a)<\max\{ r_\pi(u),r_\pi(v) \}.$  
    %
    %
\end{proof}

Since cube-free median graphs are hereditary modular graphs \cite{Ba_hmg}, and thus bipartite Helly graphs, the statement of Theorem \ref{Gpunimodality} related to cube-free median graphs is a corollary of Proposition \ref{bar-wp} :
\begin{mycorollary} \label{wp-cube-free-median} All radius functions $r_\pi$ of a cube-free median graph $G$ are 2-weakly peakless. 
\end{mycorollary}

\subsection{The algorithm.}
In this section, we consider the algorithmic implications of Theorem~\ref{bar-wp}; namely,
we show how for bipartite Helly graphs to implement each iteration of the descent algorithm in $\cO(m)$ time. From now on,  $G=(V_0\cup V_1,E)$ is a bipartite Helly graph and $\pi$ is any profile on $G$.
We also set $V=V_0\cup V_1$.

We first show the following proposition.
\begin{myproposition}\label{prop:abs-rectract-1}
    For  $v \in V$,  we can compute $u \in N(v)$ such that $r_\pi(u) < r_\pi(v)$  in $\cO(m)$ time or assert that $v$ is a local minimum of $r_\pi$.
\end{myproposition}
\begin{proof}
    Let $A=\{ z \in V: \pi(z)\cdot(d(v,z)+1) \geq r_\pi(v) \}$.
    Note that $F_\pi(v) \subseteq A$. 
    First assume $A \cap B_1(v) \neq \varnothing$.
    If $v \in A$, or $|A \cap N(v)| > 1$, then $v$ is a local minimum of $r_\pi$.
    Otherwise, let $u$ be the unique vertex of $A \cap N(v)$.
    Then,  $u$ or $v$ minimizes $r_\pi$ within $B_1(v)$.
    From now on, we assume $A \cap B_1(v) = \varnothing$. Pick any $u \in N(v)$.
    We claim that $r_\pi(u) < r_\pi(v)$ if and only if $u \in \bigcap\{ I(v,z) : z \in A \}$.
    For that, we use that since $G$ is bipartite, $|d(u,w)-d(v,w)| = 1$ for every $w \in V$.
    In one direction, suppose $r_\pi(u) < r_\pi(v)$ but there exists some $z \in A$ such that $u \notin I(v,z)$.
    Then $d(u,z) = d(v,z)+1$, and therefore, $r_\pi(u) \ge \pi(z) \cdot (d(v,z)+1) \ge r_\pi(v)$, a contradiction.
    Conversely, suppose $u \in \bigcap\{ I(v,z): z \in A \}.$
    Pick any $w \in F_\pi(u)$.
    If $w \notin A$ then $r_\pi(u) \leq \pi(w) \cdot (d(v,w)+1) < r_\pi(v)$.
    Otherwise, $w\in A,$ and $r_\pi(u) = \pi(w) \cdot d(u,w) < \pi(w) \cdot d(v,w) \leq r_\pi(v).$ In both cases, $r_\pi(u)<r_\pi(v)$ as required.
    This establishes the claim. We are left computing $\bigcap\{ N(v) \cap I(v,z) : z \in A \}$.
    By Theorem~\ref{absolute-retracts}, $G$ satisfies the interval-outergate property.
    Furthermore, by Lemma~\ref{lem:outergate} we can compute an outergate $g_v(z)$, for every $z \in A$, in $\cO(m)$ total time.
    We end up observing that $\bigcap\{ N(v) \cap I(v,z) : z \in A \} = \bigcap\{N(v) \cap N(g_v(z)): z \in A\}$, and therefore we can compute this subset in $\cO(m)$ time.
\end{proof}

Before proving the more general Proposition~\ref{prop:abs-rectract-2} (building on our Proposition~\ref{prop:abs-rectract-1},
but extending the local search to the ball of radius two), we need two technical lemmas.
Our proof strategy for Lemma~\ref{lem:abs-rectract-1} below is the same as for Lemma $3$ in~\cite{Du_abs}.

\begin{mylemma}\label{lem:abs-rectract-1}
    If $\pi$ is a 0-1-profile on $G$ and $k$ is a positive integer, then we can compute in $\cO(km)$ time the set of all vertices $v$ such that $r_\pi(v) \le k$
    and their corresponding values $r_\pi(v)$.
\end{mylemma}
\begin{proof}
    Since we can consider the restrictions of $\pi$ to the partite sets $V_0,V_1$ separately, keeping the maximum of both radius functions computed for every vertex, we may assume, without loss of generality, that $\supp(\pi) \subseteq V_0$.
    Then, following~\cite{Du_abs}, we consider a more general problem where for every $i$ such that $0 \le i \le k$, the goal is to compute both a partition ${\mathcal P}_i = (X_1^i,X_2^i,\ldots,X_{\ell_i}^i)$ of $\supp(\pi)$, and a partition ${\mathcal Q}_i = (Y_1^i,Y_2^i,\ldots,Y_{\ell_i}^i)$ of some subset of $V_{i \pmod 2}$, with the following property:
    \emph{for every $j$ such that $1 \le j \le \ell_i$, we must have $\bigcap\{ B_i(x) \cap V_{i \pmod 2} : x \in X_j^i\} = Y_j^i \neq \varnothing$.}
    We stress that all subsets $Y_1^i,Y_2^i,\ldots,Y_{\ell_i}^i$ must be nonempty and pairwise disjoint.
    In particular, the least $i \geq 1$ such that $\ell_i = 1$ must equal $\rad_\pi(G)$.
    Furthermore, for every vertex $v$, the least $i \ge 1$ such that $\ell_i = 1$ and $v \in X_1^i$ must equal $r_\pi(v)$.
    Let $\supp(\pi) = \{x_0,x_1,x_2,\ldots,x_{\ell_0}\}$.
    If $i = 0$, then we set ${\mathcal P}_0 = {\mathcal Q}_0 = (\{x_0\},\{x_1\},\ldots,\{x_{\ell_0}\})$.
    Otherwise, when $i > 0$, we compute ${\mathcal P}_i,{\mathcal Q}_i$ from $G$ and ${\mathcal P}_{i-1},{\mathcal Q}_{i-1}$ in $\cO(m)$ time.
    For that, let us define an intermediate family of subsets ${\mathcal R}_{i-1} = (Z_1^{i-1},Z_2^{i-1},\ldots,Z_{\ell_{i-1}}^{i-1})$
    where, for every $j$ such that $1 \le j \le \ell_{i-1}$, $Z_j^{i-1} = N(Y_j^{i-1})$. Since the
    subsets of ${\mathcal Q}_{i-1}$ are pairwise disjoint, we can compute ${\mathcal R}_{i-1}$ in $\cO(m)$ time.

    \begin{myclaim}
        For every $j$ such that $1 \le j \le \ell_{i-1}$, $Z_j^{i-1} = \bigcap\{B_i(x) \cap V_{i \pmod 2}: x \in X_j^{i-1}\}$.
    \end{myclaim}
    By construction,  $Z_j^{i-1} \subseteq \bigcap\{B_i(x) \cap V_{i \pmod 2}: x \in X_j^{i-1}\}$.
    Conversely, let $z \in \bigcap\{B_i(x) \cap V_{i \pmod 2}: x \in X_j^{i-1}\}$ be arbitrary.
    Then, the half-balls $B_1(z) \cap V_{i-1 \pmod 2}$ and $B_{i-1}(x) \cap V_{i -1 \pmod 2}$ for every $x \in X_j^{i-1}$ pairwise intersect.
    By the Helly property for half-balls, $z$ must have a neighbor in $\bigcap\{B_{i-1}(x) \cap V_{i-1 \pmod 2}: x \in X_j^{i-1}\} = Y_j^{i-1}$, which implies $z \in Z_j^{i-1}$.

    In general, we do not have ${\mathcal Q}_i = {\mathcal R}_{i-1}$ because the subsets of ${\mathcal R}_{i-1}$ may not be disjoint.
    We compute ${\mathcal P}_i,{\mathcal Q}_i$ from ${\mathcal P}_{i-1}, {\mathcal R}_{i-1}$ by using the following algorithm.
    We initialize $t := 0$.
    While ${\mathcal R}_{i-1}$  is nonempty, we select any vertex $z_t$ that appears in a maximum number of the remaining subsets in ${\mathcal R}_{i-1}$.
    Let $J_t = \{j : z_t \in Z_j^{i-1}\}$.
    We construct the subsets $X_t^i := \bigcup\{ X_j^{i-1} : j \in J_t \}$ and $Y_t^i := \bigcap\{ Z_j^{i-1} : j \in J_t \}$.
    Then, we add $X_t^i, Y_t^i$ to ${\mathcal P}_i,{\mathcal Q}_i$.
    After that, we remove all subsets $Z_j^{i-1}$ such that $j \in J_t$ from ${\mathcal R}_{i-1}$, and we set $t := t+1$.
    Note that, by construction, $Y_t^i \neq \varnothing$ because it contains vertex $z_t$.
    Furthermore, $Y_t^i =  \bigcap\{B_i(x) \cap V_{i \pmod 2}: x \in X_t^i\}.$
    Overall, by maximality of the selected vertices $z_t$, all subsets in ${\mathcal Q}_i$ must be disjoint.
    Therefore, the above algorithm for computing ${\mathcal P}_i,{\mathcal Q}_i$ is correct.
    Its implementation requires to store all vertices $z \in \bigcup_{j=1}^{\ell_{i-1}} Z_j^{i-1}$ in a max-heap, with their respective priorities being the number of subsets in which they appear in ${\mathcal R}_{i-1}$.
    Since all priorities are integer between $1$ and $\ell_i$, we can use for our implementation of max-heap a bucket queue~\cite{Di}.
    Then, all operations on the heap run in $\cO(\delta_p)$ time, where $\delta_p$ denotes the difference between the old and new values of the maximum priority of the elements in the queue.
    Since there are $\cO(\sum_{j=1}^{\ell_{i-1}}|Z_j^{i-1}|) = \cO(m)$ operations that are performed on the queue, and the maximum priority monotonically decreases from $\ell_{i-1}$ to $1$, the running time of the algorithm is in $\cO(m + \ell_{i-1}) = \cO(m)$.
\end{proof}

Our second intermediate result, Lemma~\ref{lem:abs-rectract-2}, follows from Proposition~\ref{prop:abs-rectract-1} and Lemma~\ref{lem:abs-rectract-1}:

\begin{mylemma}\label{lem:abs-rectract-2}
    For  $v \in V$ and  any $X \subseteq V$,   we can compute in  $\cO(m)$ time the set $S:=\bigcap\{N_2(v) \cap I(v,x): x \in X\}$.
\end{mylemma}
\begin{proof}
    First, we reduce to the case $X \cap B_3(v) = \varnothing$:
    \begin{enumerate}[(1)]
    \item {Case 1: $X \cap B_1(v) \neq \varnothing$}. We return
      $S=\varnothing$.
        \item {Case 2: $X \cap N_2(v) \neq \varnothing$}. If $|X \cap N_2(v)| > 1$, then we return $S=\varnothing$. Otherwise, let $x$ be the unique vertex of $N_2(v) \cap X$. We return $S=\{x\}$ if $x \in \bigcap\{I(v,x'): x' \in X\}$ (that can be verified in $\cO(m)$ time by starting a BFS at $x$) and $S=\varnothing$ otherwise.
        \item {Case 3: $X \cap N_3(v) \neq \varnothing$}. Let $Y := \bigcap\{N_2(v) \cap N(x) : x \in N_3(v) \cap X\}$. Note that we can compute $Y$ in $\cO(m)$ time. If $X \subseteq N_3(v)$, then we return $S=Y$. Otherwise, we also compute $Y' := \bigcap\{ N_2(v) \cap I(v,x') : x' \in X \setminus B_3(v) \}$ and we return $S=Y \cap Y'$.
    \end{enumerate}
    Thus, from now on, we can assume that $X \cap B_3(v) = \varnothing$.
    By Theorem~\ref{absolute-retracts}, $G$ satisfies the interval-outergate property.
    Furthermore, by Lemma~\ref{lem:outergate}, we can compute in $\cO(m)$ total time, for every $x \in X$, an outergate $g_v(x)$ in $B_1(v)$.
    Let $A = \bigcap\{N(v) \cap N(g_v(x)): x \in X\} = \bigcap\{N(v) \cap I(v,x): x \in X\}$, which can also be computed  in $\cO(m)$ time.
    If $A = \varnothing$, then so must be $\bigcap\{N_2(v) \cap I(v,x) : x \in X\}$, and we return $S=\varnothing$.
    Therefore, we now assume that $A \neq \varnothing$. From the definition of the sets $A$ and $S$, we have the inclusion $S\subseteq N(A)$.

    \begin{myclaim}\label{claim:binS}
       If $A\ne\varnothing$, then  in  $\cO(m)$ time  we can either decide that $S=\varnothing$ or compute   a vertex $b \in S$ such that $A \subseteq N(b)$.
    \end{myclaim}

    Suppose without loss of generality that $v \in V_0$. To establish the claim, first assume that $S\ne \varnothing$. 
    Then the half-balls $B_1(a) \cap V_0$, for every $a \in A$, and $B_{d(v,x)-2}(x) \cap V_0$, for every $x \in X$, pairwise intersect.
    Since $G$ satisfies the Helly property for half-balls, there exists a vertex $b \in V_0$ such that $b \in \bigcap\{N(a) : a \in A\}$ and $b \in \bigcap\{ B_{d(v,x)-2}(x) : x \in X  \}$.
    Then, $A \subseteq N(b)$ and $b \in S$.
    Therefore $S$ is nonempty if and only if such a vertex $b$ exists. To test is such $b$ exists,  we pick an arbitrary vertex  $a \in A$.
    Let $r := \max\{ d(a,x) : x \in X \} \geq 3$.
    We define a profile $\pi'$ such that
$$\begin{cases}
    \pi'(x) = r/d(a,x) \ \text{for every} \ x \in X \\
    \pi'(y) = r/3 \ \text{for every} \ y \in A \\
    \pi'(z) = 0 \ \text{for every other} \ z.
\end{cases}$$
In particular, $r_{\pi'}(a) = r$.
If there exists a vertex $b$ as desired, then $r_{\pi'}(b) < r_{\pi'}(a)$.
Conversely, if there exists $b \in N(a)$ such that $r_{\pi'}(b) < r_{\pi'}(a)$, then necessarily $A \subseteq N(b)$ and $b \in \bigcap\{ N(a) \cap I(v,x) : x \in X \} \subseteq \bigcap\{ N_2(v) \cap I(v,x) : x \in X \}=S$.
Therefore, we are done by applying Proposition~\ref{prop:abs-rectract-1} to the vertex $a$ and profile $\pi'$. This concludes the proof of Claim \ref{claim:binS}.

In what follows, we assume given $b\in S$ such that $A \subseteq N(b)$. It remains to show how to compute the whole set $S$.
By Lemma~\ref{lem:outergate}, we can compute in total $\cO(m)$ time, for every $x \in X$, an outergate $g_b(x)$ in $B_1(b)$ (with possibly $g_b(x) = x$ if $d(v,x) = 4$).
We stress that $g_b(x) \in N_4(v) \cap I(v,x)$.
The following claim is the cornerstone of our proof.

\begin{myclaim}\label{claim:outerS}
    For every $x \in X$, $N_2(v) \cap I(v,x) \cap N(A) = N_2(v) \cap B_2(g_b(x)) \cap N(A)$.
\end{myclaim}
In one direction, pick any $u \in N_2(v) \cap B_2(g_b(x))$.
Then $d(v,u) + d(u,x) = 2 + d(u,x) \le 4 + d(g_b(x),x) = d(v,x)$.
Therefore, $N_2(v) \cap B_2(g_b(x)) \subseteq I(v,x)$.
Conversely, let $u \in N_2(v) \cap I(v,x)$ be having a neighbor in $A$.
Since $A \subseteq N(b)$, $d(u,b) = 2$.
By Theorem~\ref{absolute-retracts}, $G$ is modular.
Then, a median of triple $u,b,x$ must be contained in $N(u) \cap N(b) \cap N_{d(v,x)-3}(x) \subseteq N(g_b(x))$.
In particular, $N_2(v) \cap I(v,x) \cap N(A) \subseteq B_2(g_b(x))$.
Therefore, Claim \ref{claim:outerS} is proven.

Finally, let $\pi''$ be the 0-1-profile on $G$ such that $\pi''(w) = 1$ if and only if $w = g_b(x)$ for some $x \in X$.
To compute $U := \bigcap\{ B_2(g_b(x)) : x \in X \}$, we apply  Lemma~\ref{lem:abs-rectract-1} for $\pi''$ and $k=2$.
Then, we can return $S=U \cap N_2(v) \cap N(A)$. The correctness follows from Claims \ref{claim:binS} ,\ref{claim:outerS}  and the inclusion $S\subseteq N(A)$.
\end{proof}

Combining Proposition~\ref{prop:abs-rectract-1} and Lemmas~\ref{lem:abs-rectract-1},\ref{lem:abs-rectract-2}, we obtain the following result.

\begin{myproposition}\label{prop:abs-rectract-2}
    For $v \in V,$ in $\cO(m)$ time we can either output some $u \in B_2(v)$ such that $r_\pi(u) < r_\pi(v)$ or assert that $v \in C_\pi(G)$.
\end{myproposition}
\begin{proof}
    We may assume, by Proposition~\ref{prop:abs-rectract-1}, that $v$ is a local minimum for  $r_\pi$.
    By Proposition~\ref{bar-wp}, there exists $u \in N_2(v)$ such that $r_\pi(u) < r_\pi(v)$ or $v \in C_\pi(G)$. Analogously to the definition of the set $S$ from Lemma \ref{lem:abs-rectract-2},
    let $$S_1:= \bigcap\{ N_2(v)\cap I(v,y) : y \in F_\pi(v) \}.$$
    Pick any  $u \in N_2(v)$. We claim that a first necessary condition for  $r_\pi(u) < r_\pi(v)$ is that $u$ belongs to $S_1$.
    Indeed, suppose by contradiction the existence of some $y \in F_\pi(v)$ such that $u \notin I(v,y)$.
    Then $d(u,y) > d(v,y) - 2$.
    Since $G$ is bipartite, the distances $d(u,y)$ and $d(v,y)$ have the same parity.
    In particular, $d(u,y) \geq d(v,y)$, which implies $r_\pi(u) \geq r_\pi(v)$, giving a contradiction.
    Therefore, the assertion is proven. Applying Lemma~\ref{lem:abs-rectract-2} to $v$ and  $F_\pi(v)$, the set  $S_1$ can be computed in $\cO(m)$ time.

    From now on, we assume that $S_1$ is nonempty,  otherwise, $v \in C_\pi(G)$.
    Note that this implies $B_1(v) \cap F_\pi(v) = \varnothing$.
    Then, let $X = \{ x \in V \setminus F_\pi(v) : \pi(x)(d(v,x)+2) \ge r_\pi(v)\}$.  Let  also
    $$S_2= \{ u \in N_2(v): d(u,x) \le d(v,x) \ \ \forall x \in X\}.$$
    We claim that a second necessary condition for $u\in N_2(V)$ for having $r_\pi(u) < r_\pi(v)$ is that $u\in S_2$.
    Indeed, if $d(u,x) > d(v,x)$ for some $x \in X$, then $d(u,x) = d(v,x)+2$ because $G$ is bipartite. Hence, $r_\pi(u) \geq \pi(x)(d(v,x)+2) \ge r_\pi(v)$,  a contradiction.
    So, our assertion is proven.
    Let us further prove that the set $S_2$ 
    can also be computed in $\cO(m)$ time.
    If $v \in X$, then there is no solution; in particular, $v \in C_\pi(G)$.
    Moreover, any solution must be contained in $\bigcap\{ N(x) : x \in X \cap N(v) \}$, and this latter subset can be computed in $\cO(m)$ time.
    As a result, we assume from now on that $X \cap B_1(v) = \varnothing$.
    By Theorem~\ref{absolute-retracts}, $G$ satisfies the interval-outergate property.
    Furthermore, by Lemma~\ref{lem:outergate}, we can compute an outergate $g_v(x)$, for every $x \in X$, in $\cO(m)$ total time.
    Observe that all vertices in $\bigcap\{ N_2(v) \cap B_2(g_v(x)) : x \in X \}$ satisfy the condition $d(u,x) \le d(v,x)$ for any  $x \in X$, and thus belong to $S_2$. 
    Let us prove that there are no other solutions.
    For that, let $u \in S_2$  
    and let $x \in X$ be arbitrary.
    If $u \in I(v,x)$, then $u$ is adjacent to some vertex of $N(v) \cap I(v,x) \subseteq N(g_v(x))$, which implies $d(u,g_v(x)) \le 2$.
    Otherwise, $d(u,x) = d(v,x)$.
    Since $G$ is modular, there exists a median vertex $w$ for the triple $u,v,x$.
    In this situation, $w \in N(u) \cap N(v) \cap I(v,x) \subseteq N(u) \cap N(g_v(x))$, which again implies $d(u,g_v(x)) \le 2$.
    Let $\pi_X$ be the 0-1-profile on $G$ such that $\pi_X(w) = 1$ if and only if $w = g_v(x)$ for some $x \in X$.
    By Lemma~\ref{lem:abs-rectract-1} applied to $\pi_X$ and $k=2$, all vertices of $N_2(v)$ belonging to $S_2$ can also be computed in $\cO(m)$ time as the set of all vertices of $N_2(v)$ such that
     $r_{\pi_X}(u)\le 2$.

    In order to complete the proof, we are left to show that if an arbitrary $u \in N_2(v)$ belongs to the sets $S_1$ and $S_2$, then $r_\pi(u) < r_\pi(v)$.
    Suppose, by contradiction, that it is not the case for some $u$. Let $z \in F_\pi(u)$ be arbitrary. Since we assume $r_\pi(u) \geq r_\pi(v)$, $d(u,z) \geq d(v,z)$. Since $u\in S_1$, 
    $z \notin F_\pi(v)$. Furthermore, $d(u,z) > d(v,z)$, or else $r_\pi(u) < r_\pi(v)$. Since $G$ is bipartite, $d(u,z) = d(v,z) + 2$. Since $u\in S_2$, 
    $z \notin X$. However, it implies $r_\pi(u) = \pi(z)(d(v,z)+2) < r_\pi(v)$, providing a contradiction.
\end{proof}
Consequently, Algorithm~\ref{alg:improveeccentricity-bar} implements the {\sf ImproveEccentricity} procedure in the case of bipartite Helly graphs.
\begin{algorithm2e}[h]
    \label{alg:improveeccentricity-bar}
    \SetKwInOut{Input}{Input}
    \SetKwInOut{Output}{Output}
   \Input{A bipartite Helly graph $G=(V,E),$ a profile $\pi:V\rightarrow R^+,$ and a vertex $v\in V$}
   \Output{A vertex $u\in B_2(v)$ with $r_\pi(u)<r_\pi(v)$ if it exists and $v$ otherwise}
   \If{$v$ is not a local minimum of $r_\pi$ in $B_1(v)$}{\KwRet{$u\in B_1(v)$ such that $r_\pi(u)<r_\pi(v)$}\hfill \tcp{$\cO(m)$ by Proposition \ref{prop:abs-rectract-1}}}
   Compute the set $S_1=\bigcap \{N_2(v) \cap I(v,y) : y\in F_\pi(v)\}$ ; \hfill \tcp{$\cO(m)$ by Lemma \ref{lem:abs-rectract-2}}
   Let $X= \{ x \in V \setminus F_\pi(v) : \pi(x)(d(v,x)+2) \ge r_\pi(v)\}$\;
   Compute $g_v(x)$ for all $x\in X$ ; \hfill \tcp{$\cO(m)$ by Lemma \ref{lem:outergate}}
   Let $\pi_X(w)=1$ if and only if $w=g_v(x)$ for some $x\in X$ \;
   Compute the set $S_2 = \{u\in N_2(v): r_{\pi_X}(u)\le 2\}$ ; \hfill \tcp{$\cO(m)$ by Lemma \ref{lem:abs-rectract-1}}
   \eIf{$S_1\cap S_2 = \varnothing$}{\KwRet{v}}{\KwRet{$u\in S_1\cap S_2$}}

   \caption{{\sf ImproveEccentricity}$(G,\pi,v)$ for bipartite Helly graphs}
\end{algorithm2e}

%

The following main result of this section is an easy by-product of Proposition~\ref{prop:abs-rectract-2} and the descent algorithm presented in Section ~\ref{descent}. It establishes the statement of Theorem \ref{main} in the case of bipartite Helly graphs.
\begin{mytheorem}\label{thm:main-bar}
Let $G$ be a bipartite Helly graph.
    \begin{enumerate}[(1)]
        \item For any profile on $G$, a central vertex can be computed in $\ctO(m\sqrt{n})$ time with high probability.
        \item For any 0-1-profile on $G$, a central vertex can be computed in deterministic $\cO(m\sqrt{n})$ time. 
    \end{enumerate}
\end{mytheorem}




\begin{myremark} Bandelt et al. \cite{BaHeFa} established the following correspondences between Helly graphs and bipartite Helly graphs. Let $G$ be a reflexive graph (i.e., each vertex of $G$  has a loop). The graph $B(G)$ has two copies of each vertex
of $G$ and two vertices of $B(G)$ are adjacent if and only if the corresponding vertices of $G$ are adjacent. Then $G$ is a Helly graph if and only if $B(G)$ is a bipartite Helly graph \cite{BaHeFa}. Conversely, to each bipartite graph $G=(X\cup Y,E)$
Bandelt et al.  \cite{BaHeFa} associated two reflexive graphs $S_X$ and $S_Y$: $S_X$ has $X$ as the set of vertices and two vertices are adjacent in $S_X$ if and only if these two vertices are identical or have distance 2 in $G$; the graph $S_Y$ is
defined in an analogous way. Then a bipartite graph $G=(X\cup Y,E)$ is bipartite Helly if and only if the graphs $S_X$ and $S_Y$ are Helly graphs \cite{BaHeFa}.

One may think that the center problem in a bipartite Helly graph $G=(X\cup Y,E)$ can be reduced to two center problems in the Helly graphs $S_X$ and $S_Y$, to which the algorithm of Ducoffe~\cite{Du_Helly} can be applied. Unfortunately,
this is not the case: simple examples show that the central vertices of $S_X$ and $S_Y$ computed by the algorithm of \cite{Du_Helly} maybe different and even if they have the same eccentricity in $S_X$ and $S_Y$, they may have different eccentricities in $G$.
But most importantly, the graphs $S_X$ and $S_Y$ may have a quadratic number of edges with respect to the number of edges $|E|$ of $G$: each vertex of degree $d$ give raise to $\frac{d(d-1)}{2}$ edges in the graph $S_X$ or $S_Y$.
It was explained in~\cite{Du_abs} how these issues could be circumvented in the case of diameter computation on bipartite Helly graphs. However, evidence was also given in~\cite{Du_abs} that a similar approach could not be easily extended to the center problem.
\end{myremark}

\section{The center problem in cube-free median graphs}\label{sec:median}

In this section, we present an $\cO(n\log^2 n)$-time divide-and-conquer algorithm for computing a weighted center of a cube-free median graph $G$.
Our method generalizes the algorithm of \cite{KaHa}
for computing the center of a tree and uses similar ingredients: divide-and-conquer, median computation, and
unimodality of the radius function.

\subsection{Median graphs.} Recall that a \emph{median} of $u,v,w$ is a vertex  from the intersection $I(u,v)\cap I(v,w)\cap I(w,u)$.
A graph $G$ is \emph{median} if any triplet $u,v,w$ has a unique median. A graph $G$ is \emph{cube-free median} if $G$ does not contain the cube $Q_3$ as a subgraph.
Median graphs satisfy $\QC$, are modular and thus bipartite. In fact, the median graphs are exactly the $K_{2,3}$-free modular graphs.
A topological local-to-global characterization of median graphs as 1-skeleta of CAT(0) cube complexes was given in \cite{Ch_CAT}.

It is not immediately clear from the definition, but median graphs have a rich metric and cubical structure. For that, let us recall some important notations and terminology.
The \emph{star} $\St(v)$ of $v$ is the union of all cubes containing $v$. For an edge $uv$, let $H(u,v)=\{ x\in V(G): d(x,u)<d(x,v)\}$. In a bipartite graph $G$, and thus in a median graph, the sets
$H(u,v)$ and $H(v,u)$ define a bipartition of $V(G)$. For a gated set $M$ and  $z\in M$, let $\Fib_M(z)$ be the set of all $x\in V$
whose gate in $M$ is $z$ and call $\Fib_M(z)$  the \emph{fiber} of $z$. The \emph{boundary} $\Upsilon_M(z)$ of  $\Fib_M(z)$ is the subgraph induced by all vertices of
$\Fib_M(z)$ having neighbors in other fibers $\Fib_M(z'), z'\in M$.  A \emph{branch} of a tree $T$ rooted at a vertex $z$ is a path of $T$ with one end in $z$ and another end in a vertex of $T$.
The following properties of median graphs are well-known, see for example \cite{BaCh_survey,BeChChVa_med,ChLaRa,Mul}:

\begin{mytheorem} For a median graph $G=(V,E)$, we have:
    \begin{enumerate}[(1)]
        \item  The convex sets, the locally-convex sets,  and the gated sets of $G$ are the same;
        \item The stars $\St(v)$ and the sets $H(u,v),H(v,u)$ are convex, and thus gated;
        \item If $G$ has $n$ vertices and the largest cube has dimension $d$, then $G$ contains at most $dn$ edges. In particular,
        any cube-free median graph contains at most $2n$ edges;
      \item If $M$ is gated, each fiber $\Fib_M(u)$ is gated and
        $u'\in \Fib_M(u), v'\in \Fib_M(v)$ and $u'\sim v', u\ne v$
        imply $u\sim v$~\cite{Mul};
      \item For a cube-free median
        graph $G$, for each fiber
        $\Fib_{\St(v)}(z)$ of a star $\St(v)$, 
      its boundary $\Upsilon_{\St(v)}(z)$ is a tree with gated branches~\cite{ChLaRa};
      \item For any $u$ and $v$, $u$ and all neighbors of $u$ in $I(u,v)$
        belong to a hypercube of $I(u,v)$. If $G$ is cube-free, then
        $u$ has at most two neighbors in $I(u,v)$~\cite{Mul}.
    \end{enumerate}
\end{mytheorem}

Call the last property the \emph{downward cube property}  (\emph{DCP}).  For an edge $uv$ of a median graph $G$, the sets $H(u,v)$ and $H(v,u)$ are called \emph{halfspaces}.
The edges running between two complementary halfspaces $H(u,v)$ and $H(v,u)$ of $G$ are called \emph{$\Theta$-classes}. Notice that any two edges from the
same $\Theta$-class of $G$ define the same pair of halfspaces.

\begin{mylemma} \label{opposite-fibers} If $G$ is cube-free median, $S=uvwz$ is a square of $G$, and $x$ and $y$ are two vertices of $G$ belonging to opposite fibers of $S$, say $x\in \Fib_S(u)$ and $y\in \Fib_S(w)$, then
$d(x,y)=d(x,u)+2+d(w,y)$  and thus  $u,v,w,z \in I(x,y).$
\end{mylemma}

\begin{proof}
    Since $u\in I(x,w)$ and $v,z\in I(u,w)$, it suffices to prove that $w\in I(x,y)$. 
    If this is not the case, then we can find a neighbor $t$ of $w$ belonging to  $I(w,x)$ and $I(w,y)$. Since $v,z\in I(w,x)$, by DCP, $t$ coincides with  $v$ or $z$. But this is impossible since $w$ is the gate of $y$ in $S.$
    This contradiction shows that $u,v,w,t\in I(x,y)$ and $d(x,y)=d(x,u)+2+d(w,y)$.
\end{proof}

By Corollary \ref{wp-cube-free-median}, radius functions are $G^2$-unimodal in cube-free median graphs. The following example shows that for any positive integer $p$ there exists a median graph (in fact a hypercube), in which the radius function is not $p$-weakly peakless:

\begin{myexample} For a given integer $p>0$, let $Q_r$ be a hypercube of dimension $r>p$. Obviously, $Q_r$ is a median graph. Pick any pair $u,v$ of diametral vertices of $Q_r$ and consider the following profile $\pi$ on $Q_r$:
$\pi(u)=\pi(v)=0$ and $\pi(x)=1$ if $x\ne u,v$. Then $r_{\pi}(u)=r_{\pi}(v)=r-1$ and $r_{\pi}(x)=r$ for any $x\ne u,v$ (this is because for $x$ there exists an opposite vertex $x'$ with $\pi(x')=1$. Since $r>p$, $r_{\pi}$ is  not $p$-weakly peakless.
\end{myexample}

\subsection{Improving neighbors and improving second neighbors.} From now on,  $G$ is a cube-free median graph and $\pi$ is any profile.
A neighbor $z$ of $v\in V$ is
\emph{improving} (an \emph{i-neighbor}) if $r_{\pi}(z)<r_{\pi}(v)$. 
A vertex $z$ with $d(z,v)=2$ is an \emph{improving second neighbor} (an \emph{is-neighbor}) if $r_{\pi}(z)<r_{\pi}(v)$.
If  $z\in I(v,c)$ for some $c$, then $z$ is
\emph{improving  in direction $c$.} In this subsection we present several properties of i-neighbors and is-neighbors of vertices of $G$.


\begin{mylemma} \label{convex-path} If $P=(v,z,w)$ is a gated path of $G$, then $2r_{\pi}(z)\le r_{\pi}(v)+r_{\pi}(w)$. In particular, $z$ is an i-neighbor of $v$ if $r_{\pi}(v)>r_{\pi}(w)$. Furthermore, one cannot have $r_\pi(v)=r_\pi(z)=r_\pi(w)$.
\end{mylemma}

\begin{proof} 
    Let $x$ be a furthest from $z$ vertex, i.e., a vertex of $\pi$ maximizing $\pi(x)d(x,z)$. If $x$ belongs to the fiber $\Fib_P(z)$ of $z$, then $d(v,x)=d(w,x)=d(z,x)+1$, whence $r_{\pi}(v)\ge \pi(x)d(v,x)=\pi(x)d(z,x)+\pi(x)=r_{\pi}(z)+\pi(x)$. Analogously, $r_{\pi}(w)\ge \pi(x)d(w,x)=r_{\pi}(z)+\pi(x)$ and in this case we obtain $2r_{\pi}(z)< r_{\pi}(v)+r_{\pi}(w)$. Now, suppose that $x$ belongs to one of the fibers $\Fib_P(v)$ or $\Fib_P(w)$, say $x\in \Fib_P(w)$. Then $d(z,x)=d(w,x)+1$ and $d(v,x)=d(z,x)+1$, whence $r_{\pi}(w)\ge \pi(x)d(w,x)=r_{\pi}(z)-\pi(x)$ and $r_{\pi}(v)\ge \pi(x)d(v,x)=r_{\pi}(z)+\pi(x)$. This implies that  $2r_{\pi}(z)\le r_{\pi}(v)+r_{\pi}(w)$ also in this case.

    Now, suppose by way of contradiction that $r_\pi(v)=r_\pi(z)=r_\pi(w)$. Since $2r_{\pi}(z)< r_{\pi}(v)+r_{\pi}(w)$ if $x\in \Fib_P(z)$, this case is impossible. On the other hand, in the case when $x\in \Fib_P(w)$ we deduce that $r_{\pi}(v)\ge \pi(x)d(v,x)=r_{\pi}(z)+\pi(x)$. Therefore, we conclude that also the case when $x$ belongs to one of the fibers $\Fib_P(v)$ or $\Fib_P(w)$ is impossible. This establishes the second assertion.
\end{proof}

\begin{mylemma} \label{two-neighbors} Any vertex $v$  
    has at most two i-neighbors. If $v$ has two i-neighbors $z_1,z_2$, then $v,z_1,z_2$ are included
    in a square 
    of $G$.  
\end{mylemma}

\begin{proof} Suppose $v$ has three i-neighbors $z_1,z_2,z_3$. Let $x\in F_\pi(v)$. 
    Since $r_{\pi}(z_i)<r_{\pi}(v)$,  $z_1,z_2,z_3\in I(v,x)$,  contradicting the fact that $G$ is cube-free. Now, let $v$ have two i-neighbors
$z_1,z_2$ and   $x\in F_\pi(v)$. 
    Then $z_1,z_2\in I(v,x)$. By  $\QC$ there exists a common neighbor $w$ of $z_1,z_2$ one step closer to $x$.
    Consequently, $v,z_1,z_2,w$ induce a square $S$.
    %
\end{proof}

\begin{mylemma} \label{no-improving-neighbor-with-weights} If  $v\notin C_{\pi}(v)$ 
    has no i-neighbor in direction $c\in C_{\pi}(G)$, then $v$ has no i-neighbor at all  but
    has a unique is-neighbor $w$ in all directions  $c'\in C_\pi(G)$. 
\end{mylemma}

\begin{proof} Pick $c\in C_{\pi}(G)$ and suppose  $v$ has no i-neighbor in direction $c$. Since $v\notin C_\pi(G)$, $d(v,c)\ge 2$. 
    We assert that $v$ has an is-neighbor. This is so when $d(v,c)=2$ since $c$ is an is-neighbor of $v$. If $d(v,c)\ge 3$,   since $r_{\pi}$ is $2$-weakly peakless,
    there exists  $w\in I(v,c)\cap B_2(v)$ with $r_{\pi}(w)<r_{\pi}(v)$. Since $v$ has no i-neighbor in direction $c$, $d(v,w)=2$, \textit{i.e.}, $w$ is an is-neighbor of $v$.
    From now on, let $w$ be any is-neighbor of $v$.  We assert that $w$ is an is-neighbor in direction of all  $c'\in C_\pi(G)$. First, note that $w\in \St(v)$. Indeed, if $v$ and $w$
    have a unique common neighbor $z$,  by Lemma~\ref{convex-path}, $z$ is an i-neighbor of $v$, a contradiction.
    Thus $v$ and $w$ have two common neighbors $z_1,z_2$. Pick the square $S=vz_1wz_2$, which is gated. 
    For $u\in S,$ let $x_u\in \supp(\pi)$ such that $r_\pi(u)=\pi(x_u)d(u,x_u).$
    We assert that $x_{z_1}\in \Fib_S(z_2)$ and $x_{z_2}\in \Fib_S(z_1)$. 
    If $x_{z_1}\in \Fib_S(z_1)\cup \Fib_S(v)$, then $r_\pi(w) \ge \pi(x_{z_1})d(w,x_{z_1})> \pi(x_{z_1})d(z_1,x_{z_1}) = r_\pi(z_1),$ a contradiction.
    If $x_{z_1}\in \Fib_S(w)$, then $r_\pi(v) \ge \pi(x_{z_1})d(v,x_{z_1})> \pi(x_{z_1})d(z_1,x_{z_1})=r_\pi(z_1),$ again a contradiction.
    Hence, $x_{z_1}\in \Fib_S(z_2)$ and $x_{z_2}\in \Fib_S(z_1).$ By a similar proof, $x_v\in \Fib_S(w).$ Now we show $C_\pi(G)\subseteq \Fib_S(w).$ Suppose $c'\in C_\pi(G)\setminus \Fib_S(w).$ If $c'\in \Fib_S(v)$, since $x_v\in \Fib_S(w),$ by Lemma~\ref{opposite-fibers}, $v\in I(c',x_v).$ Hence, $r_\pi(c')\ge \pi(x_v)d(c',x_v)\ge \pi(x_v)d(v,x_v)= r_\pi(v),$ a contradiction with
$v\notin C_\pi(G)$.
    Similar arguments with $z_i$ instead of $v$ show that $c'$ cannot belong to $\Fib_S(z_1)\cup \Fib_S(z_2)$. Hence, $C_\pi(G)\subseteq \Fib_S(w)$ and $w\in I(v,c')$ for any $c'\in C_\pi(G)$.

    Now suppose  $v$ has an i-neighbor $z$. Since $v$ has no i-neighbor in direction $c$, $z\ne z_1,z_2$, thus $z\in \Fib_S(v)$, where $S=vz_1wz_2$. Since $x_v\in \Fib_S(w)$, by Lemma~\ref{opposite-fibers},
$d(z,x_v)=1+2+d(w,x_v)=1+d(v,x_v)$, thus $r_\pi(z)\ge \pi(x_v)d(z,x_v)>\pi(x_v)d(v,x_v)=r_\pi(v)$. Hence, $v$ has no i-neighbor.
    It remains to show that $w$ is unique. Suppose $w'$ is another is-neighbor of $v$. By what has been proven above, $w$ and $w'$ are is-neighbors for all centers,
    thus $w,w'\in I(v,c')$ for any $c'\in C_\pi(G)$. 
    Since $G$ is bipartite, $d(w,w')=2$ or $d(w,w')=4$. If $d(w,w')=2$, let $z$ be the median of $w,w',v$. Then $z\sim w,w',v$.  Suppose by contradiction there exists another common neighbor $z'$ of $w,w'$. Since $G$ is $K_{2,3}$-free, $z' \not\sim v$. Then, $w,w' \in I(v,z')$. However, since $w,w' \in \St(v)$, the latter implies the existence of at least three neighbors of $v$ in $I(v,z')$, and so, by DCP, of an induced $Q_3$ in $G$. Since $G$ is cube-free, we deduce that $z$ is the unique
    common neighbor of $w$ and $w'$. By Lemma~\ref{convex-path}, $2r_\pi(z)\le r_\pi(w)+r_\pi(w')<2r_\pi(v)$, thus $z$ is an i-neighbor of $v$ in direction $c$, a contradiction. Now, let $d(w,w')=4$.
    Since $w\in I(v,c)$ and $G$ is cube-free, $z_1$ and $z_2$ are the unique neighbors of $v$ in $I(v,c)$. Since  $w'\in I(v,c)$, $w'$ must be adjacent to
    one of $z_1,z_2$, thus $d(w,w')=2$, a contradiction.
\end{proof}

\subsection{The algorithm.}  Let $G$ be a cube-free median graph; then $G$ contains at most $2n$ edges.
We start with $G$ as the search region $R$. At all iterations,  the search region $R$ is convex and contains the center $C_\pi(G)$.
At each iteration, we compute a median/centroid $v$ of the current
region $R$ in $\cO(|R|)$ time \cite{BeChChVa_med}. We compute the star
$\St(v)$ of $v$ in $R$ and $r_\pi(z)$ for all vertices $z$ of $\St(v)$
(see Section~\ref{eccentricies-in-a-star}).  If $R=\{ v\}$ or
$r_\pi(v)\le r_\pi(z)$ for all $z\in \St(v)$, we return $v$ as a
central vertex of $G$. Otherwise, we show that $v$ has an i-neighbor
or an is-neighbor in $R$. Using this improving ($2$-)neighbor of $v$, we
either compute a center or a halfspace $H$ of $R$ \emph{separating} $C_\pi(G)$ from $v$,
i.e., $C_\pi(G)\subseteq H$ and $v\in R\setminus H$. Then the search is
continued in $H$.  Algorithm \ref{alg:cut-on-best-neighbor} presents the main routine of the divide-and-conquer algorithm.
Algorithm  \ref{alg:Reduce Region} presents the division step, which
computes a separating halfspace $H$ of $R$.  


\begin{algorithm2e}[h]
    \label{alg:cut-on-best-neighbor}
    \SetKwInOut{Input}{Input}
    \SetKwInOut{Output}{Output}
    \Input{A cube-free median graph $G=(V,E)$ and a profile $\pi:V\rightarrow R^+$}
    \Output{A vertex of $C_\pi(G)$}
    $R\leftarrow G$ \;
    \While{$|R|>1$}
               {
               	$R\leftarrow {\sf ReduceConvexRegion}(G,\pi,R)$
               }
   \KwRet{the unique vertex of $R$}\;

   \caption{{\sf CutOnBestNeighbor}$(G,\pi)$ for cube-free median graphs}
\end{algorithm2e}


\begin{algorithm2e}[h]
    \label{alg:Reduce Region}
    \SetKwInOut{Input}{Input}
    \SetKwInOut{Output}{Output}
    \Input{A cube-free median graph $G=(V,E),$ 
    a profile $\pi:V\rightarrow R^+$, a convex region $R$ of $G$ containing $C_\pi(G)$}
    \Output{A proper convex subregion of $R$ containing $C_\pi(G)$ or a central vertex of $C_\pi(G)$}
    Compute a median vertex $v$ of $R$ using the algorithm of \cite{BeChChVa_med}\;
    Compute the star $\St(v)$\;
    Compute $r_\pi(z)$ for all $z\in \St(v)$ using the algorithm from Subsection \ref{eccentricies-in-a-star}\;
    \If{$v$ is a local minimum of $r_\pi$ in $\St(v)$}{\KwRet{$\{v\}$} \hfill \tcp{Case 0}}
    \If{$v$ has an is-neighbor $w$ but no improving neighbor}{\KwRet{$R\cap H(z,v)$ where $z\sim v,w$} \hfill \tcp{Case 1}}
    Let $z$ be a i-neighbor \;
    Construct the fiber $\Fib_{\St(v)}(z)$ and its boundary $\Upsilon_{\St(v)}(z)$ \;
    Compute a local minimum $u$ of the radius function $r_\pi$ on $\Upsilon_{\St(v)}(z)$ using the algorithm from Subsection \ref{ss:local-min-upsilon}\;
    \If{$u$ is a local minimum in $\St(u)$}
        {\KwRet{$\{u\}$ ; \hfill \tcp{Case 2}}}
    \If{$u$ has a unique i-neighbor $t$}
        {\KwRet{$R\cap H(t,u)$ ; \hfill \tcp{Case 4}}}
    \If{$u$ has an is-neighbor $w$ but no i-neighbor}
        {\KwRet{$R\cap H(t,u)$ where $t\sim u,w$ and $t\notin (z,u)$-path of $\Upsilon_{\St(v)}(z)$ ; \hfill \tcp{Case 3}}}

    \KwRet{$R\cap H(z,v)$ ; \hfill \tcp{Case 5}}

   \caption{{\sf ReduceConvexRegion}$(G,\pi,R)$}
\end{algorithm2e}
%

\subsection{Correctness of {\sf ReduceConvexRegion}.} Let $R$ be the current search region containing $C_\pi(G)$ and let $v$ be a median vertex of $R$.
We show that {\sf ReduceConvexRegion} correctly
computes a halfspace $H$ of $R$  separating $C_\pi(G)$ from $v$ if such a halfspace $H$ exists (otherwise, $v$ is returned as a central vertex).
We distinguish several cases.

\medskip\noindent
{\bf Case 0:} {\it $v$ does not have i- and is-neighbors in $R$.} Then  $v\in C_\pi(G)$. Indeed, let $c \in C_\pi(G) \subseteq R$. Since $r_\pi$
is 2-weakly peakless in $G$, if $d(v,c)\ge 3$, then $v$ has an i- or is-neighbor $w$ in $G$ in direction $c$. Since $R$ is convex, $w\in I(v,c)\subseteq R$, thus $w\in R$, a contradiction.
Thus $d(v,c)=2$ and $r_\pi(c)<r_\pi(v)$. Then $c\notin \St(v)$, i.e., $v$ and $c$ have a unique
common neighbor $z$. By Lemma~\ref{convex-path},
$r_\pi$ is convex on $(v,z,c)$, yielding $r_\pi(z)< r_\pi(v)$. Hence $z$ is an i-neighbor, a contradiction. 

\medskip\noindent
{\bf Case 1:} {\it $v$ does not have i-neighbor in $R$ but has an is-neighbor $w\in R$.} We set $H:=H(z,v)$ in $R$, where
$z\sim v,w$. We assert that $v\notin H$ and $C_\pi(G)\subseteq  H$. Indeed, since $R$ is convex,
$I(v,c)\subseteq R$ for any $c\in C_\pi(G)$. By Lemma~\ref{no-improving-neighbor-with-weights}, $w\in I(v,c)$.
Since $w\in I(v,c)$ and $z\in I(v,w)$, we have $c\in H(z,v)$ and $v\in H(v,z)$.

\medskip Now suppose that $v$ has an i-neighbor $z\in R$. We construct
the fiber $\Fib_{\St(v)}(z)$ of $z$ and its boundary
$\Upsilon_{\St(v)}(z)$, which is a tree with gated branches. By
Lemma~\ref{convex-path}, $r_\pi$ is convex on each branch of
$\Upsilon_{\St(v)}(z)$, however $r_\pi$ is not unimodal on
$\Upsilon_{\St(v)}(z)$. Nevertheless, we can find in $\cO(n\log n)$ time a
\emph{local minimum} $u$ of $r_\pi$ on $\Upsilon_{\St(v)}(z)$ (see
Section~\ref{ss:local-min-upsilon}). 
Let $P=(u_0=v,u_1=z,\ldots,u_{k-1},u_k=u)$ be the path obtained by adding $v$ to the gated branch
$I(z,u)\subseteq \Upsilon_{\St(v)}(z)$. Since $\St(v)$ is gated, $P$
is locally-convex and thus gated.  
Then compute $r_\pi(t)$ for all
vertices $t$ of the star $\St(u)$.

\medskip\noindent
{\bf Case 2:} {\it $u$ does not have i- and is-neighbors in $R$.} Then we return $u$ as a center; the proof is as in Case~0.

\medskip\noindent
{\bf Case 3:} {\it $u$ does not have any i-neighbor in $R$ but has an is-neighbor $w\in R$.}  By Lemma~\ref{convex-path}, $u$ and $w$ have two common neighbors $t,t'$. At least one of them, say $t$,
does not belong to the path $P$ (i.e., $t\ne u_{k-1}$). Then as $H$ we take the halfspace $H(t,u)$ of $R$.
The proof that $C_\pi(G)\subseteq H=H(t,u)$ is the same as in Case~1. Since the path $P$ is gated and $t\notin P$, $d(v,t)=d(v,u)+1$,
which implies $v\in H(u,t)$, and so $v\notin H$.

\medskip\noindent
{\bf Case 4:} {\it $u$ has a unique i-neighbor $t$ in $R$.} We set $H:=H(t,u)$. By Lemma~\ref{no-improving-neighbor-with-weights}, $t\in I(u,c)$ for any $c\in C_\pi(G)$, thus $C_\pi(G)\subseteq H(t,u)=H$.
Since $r_\pi(u_{k-1})\ge r_\pi(u)$ and $r_\pi(t)<r_\pi(u)$, $t\ne u_{k-1}$. Since $P$ is convex, $t\notin P$, thus $d(v,t)=d(v,u)+1$ and  $v\in H(u,t)$.

\medskip\noindent
{\bf Case 5:} {\it $u$ has two i-neighbor $t$ and $t'$ in $R$.} We distinguish two subcases.

\medskip\noindent
{\bf Subcase 1:} {\it $t$ or $t'$, say $t$, does not belongs to $\Fib_{\St(v)}(z)$.} This case is impossible. Indeed, then we should have $t\in \Fib_{\St(v)}(w)$ with $d(v,w)=2$
and $v,w$ belong to a square $S'$. Since $u\sim t$, we would get
$z\sim w$, and so $z\in S'$, i.e., $S'=vzwz'$.  By Lemma~\ref{two-neighbors},  $u,t,t'$ are included in a square $S=utyt'$. Since the fibers are gated,
$y\notin \Fib_{\St(v)}(z)$ and $t'\notin \Fib_{\St(v)}(w)$.
If $t'\in \Fib_{\St(v)}(z)$, then $t'\in \Upsilon_{\St(v)}(z)$ because $t'\sim y$. Since $r_\pi(t')<r_\pi(u)$ and $u$ is a local minimum of $r_\pi$ in $\Upsilon_{\St(v)}(z)$,
this is impossible. Thus $t'\in \Fib_{\St(v)}(w')$, where $d(v,w')=2, w'\ne w,$ and $v,w'$ are included in a square $S''$.
Since $u\sim t'$, we get $z\sim w'$, and so $S''=vzw'z''$. Consider the fiber  $\Fib_{\St(v)}(q)$ of $\St(v)$ containing $y$. Since $y\sim t,t'$, either $q\in \{ w,w'\}$ or $q\sim w,w'$. If say $q=w$,
then $w'\sim w$ and we obtain a  $K_3=\{ z,w,w'\}$. Thus $q\ne w,w'$ and $q\sim w,w'$. If $q\in \{ z',z''\}$, say $q=z'$, then $w'\sim z'$ and $v,w',z,z',z''$ induce a forbidden $K_{2,3}$.
Hence $q\ne z',z''$. The median of $z',z'',y$ is  $p\sim z',z'',q$. But then $v,z',z'',p,z,w,w',q$ induce a forbidden 3-cube.
%

\medskip\noindent
{\bf Subcase 2:} {\it $t,t'\in \Fib_{\St(v)}(z)$.} We set $H:=H(z,v)$. Clearly, $v\in H(v,z)$, thus $v\notin H$. We assert that $C_\pi(G)\subseteq H$. Pick any $c\in C_\pi(G)$. Since $u$ has  i-neighbors,
by Lemma~\ref{no-improving-neighbor-with-weights}, $u$ has an i-neighbor in direction $c$, say $t$. Let $c'$ be the gate of $c$ in $P$. If $c'\ne v$, then $z\in I(v,c')\subset I(v,c)$, thus $c\in H(z,v)=H$
and we are done. Now we show that $c'=v$ is impossible. Let $p$ be the median of $t,v,c$. Since $t,v\in I(u,c)$, necessarily $p\ne v$. Let $z'$ be a neighbor of $v$ in $I(v,p)$.
Then $z'\in I(v,p)\subseteq I(v,t)$. Since $P$ is convex and $t\notin P$, also
$z\in I(v,t)$. By (QC), there exists $w\sim z,z'$ one step closer to $t$ than $z,z'$. Since the square $vzwz'$ belongs to $\St(v)$ and $w\in I(z,t)$, we get a contradiction with  $t\in \Fib_{\St(v)}(z)$.

\begin{figure}
    \begin{center}
        \includegraphics[scale=0.7,page=4]{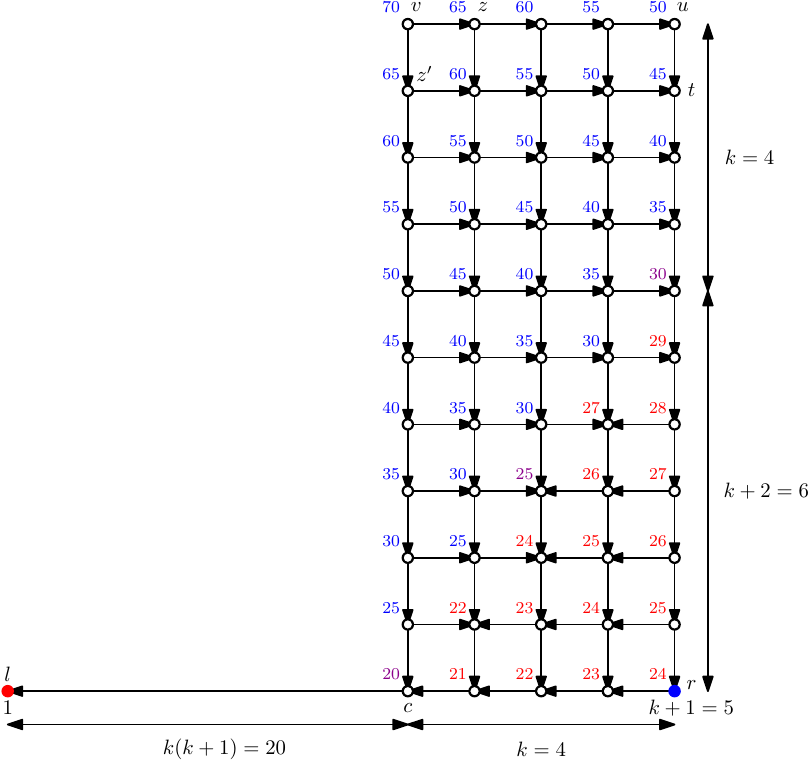}
    \end{center}
    \caption{An example showing that the halfspace $H(z,v)$ for an
      i-neighbor $z$ of $v$ does not necessarily intersect
      $C_{\pi}(G)$.}
    \label{cex-mediansSansCubes-grille}
\end{figure}

\begin{myexample}
  In the case of 0-1-profiles, for any i-neighbor $z$ of the current
  median vertex $v$ of $R$, the halfspace $H=H(z,v)$ separates $v$
  from some central vertex of $G$. The example from
  Figure~\ref{cex-mediansSansCubes-grille} shows that this is not
  longer true for weighted profiles. The cube-free median graph $G$
  consists of a rectangular grid of size $k\times (2k+2)$ and a
  horizontal path of length $k(k+1)$ glued to the bottom leftmost
  vertex $c$ of the grid (in the figure, $k=4$).  The profile consist
  of two vertices: the bottom leftmost vertex $l$ of $G$ of weight 1
  and the bottom rightmost vertex $r$ of $G$ of weight $k+1$. There is
  a unique central vertex, which is the vertex $c$ along which the
  grid and the path are glued;
  $r_\pi(c)=k(k+1)=1\cdot d(c,l) = (k+1)\cdot d(c,r)$. Now, let $v$ be
  the upper leftmost vertex of the grid. Then
  $r_\pi(v)=(k+1)(3k+2) = (k+1) \cdot d(v,r)$. The vertex $v$ has two
  neighbors in $G$, the horizontal neighbor $z$ and the vertical
  neighbor $z'$. Both $z$ and $z'$ are i-neighbors of $v$, namely
  $r_\pi(z)=r_\pi(z')=k(3k+2) = (k+1) \cdot (d(v,r)-1)
  <r_\pi(v)$. However the halfspace $H(z,v)$ does not separate $v$ and
  $c$. Therefore, in this case we need to compute a local minimum $u$
  of $r_\pi$ in $\Upsilon_{\St(v)}(z)$. In this example,
  $\Upsilon_{\St(v)}(z)=\Fib_{\St(v)}(z)$ and it coincides with the
  horizontal path with one end in $v$ and passing via $z$. Then $u$ is
  the other end of this path and
  $r_\pi(u)=(k+1)(2k+2) = (k+1)\cdot d(u,r)$. In this case, $u$ has a
  unique improving neighbor $t$ and
  $r_\pi(t) = (k+1)(2k+1) = (k+1)\cdot d(t,r)$, thus we are in Case~4
  and $H$ is the halfspace $H(t,u)$, which indeed separates $v$ and
  $c$.
\end{myexample}

\subsection{Computing a local minimum of $r_\pi$ in the tree $\Upsilon_{\St(v)}(z)$.\label{ss:local-min-upsilon}} We use divide-and-conquer and we start with $\Upsilon_{\St(v)}(z)$
as the search region $T$; $T$  will always contain a local minimum of $r_\pi$. At each iteration, we compute a centroid $u$ of $T$, and $r_\pi(u')$  for each $u'\in N[u]$ in $T$.
For that, since $N[u]\subset \St(u)$, we simply compute in $\cO(n)$ time  $r_\pi(u')$ for all $u'\in \St(u)$ (see the next Section~\ref{eccentricies-in-a-star} for details).
If $|T|=1$ or $r_\pi(u)\le r_\pi(u')$ for any  $u'\in N[u]$, then $u$ is a local minimum of $r_\pi$ in $\Upsilon_{\St(v)}(z)$ and we return $u$. Otherwise, if $r_\pi(u')<r_\pi(u)$ for some neighbor $u'$ of $u$ in $T$,
then clearly the subtree $T'$ of $T$ containing $u'$ and not containing $u$ will contain a local minimum of $r_\pi$ in $\Upsilon_{\St(v)}(z)$. Then we continue with $T'$ as the search region. Since $u$ is a centroid of $T$,
$|T'|\le |T|/2$, the algorithm has $\log n'$ iterations, where $n'$ is the number of vertices of $\Upsilon_{\St(v)}(z)$. Since each iteration is performed in $\cO(n)$ time and $n'\le n$, the algorithm has complexity $\cO(n\log n)$.
%

\subsection{Computing all eccentricities in a star $\St(v)$.} \label{eccentricies-in-a-star}  In this subsection, we show how to compute $r_\pi$ for all vertices of a star $\St(v)$.
This is used at each iteration of the algorithm in order to find the improving neighbors or the improving second neighbors of $v$.
Berg\'e et al. \cite{BeDuHa1} presented an algorithm which computes all eccentricities of a star in $\cO(n\log n)$ time. For cube-free median graphs, we present a simpler algorithm with complexity $\cO(n)$.
Our algorithm constructs the gated set $\St(v)$ and performs a BFS on $G$ to construct a partition of $V(G)$ into the fibers $\Fib_{\St(v)}(u): u\in \St(v)$. This can be done in linear $\cO(n)$ time as explained in \cite{ChLaRa}. 
The fibers of $G$ are divided into fibers of neighbors $u$ of $v$ and fibers of second neighbors $w$ of $v$ in $\St(v)$. We denote the set of first neighbors of $v$ by $N(v)$ and the set of
second neighbors of $v$ from $\St(v)$ by $N_2(v)$. We suppose that $\St(v)$ has $n'$ vertices. During the BFS that computes the partition of $V(G)$ into the fibers of $\St(v)$, our algorithm also computes in $\cO(n)$ time the values $k_i(z):=\max\{\pi(x)(d(x,z)+i):x\in \Fib_{\St(v)}(z)\}$ for all vertices $z\in \St(v)$ and all $i=0,1,\dots,4.$
%
The eccentricity of $v$ can be computed in $\cO(n')$ time by the formula
$r_{\pi}(v)=\max\{\{k_1(u): u\in N(v)\} \cup\{ k_2(w): w\in N_2(v)\}\}.$
The eccentricity of each neighbor $u$ of $v$ can be computed by the formula
\begin{align*}
    r_{\pi}(u)=\max & \{\{\pi(v),k_0(u)\} \cup \{ k_2(u'): u'\in N(v)\setminus \{ u\}\} \cup \{ k_1(w): w\in N_2(v)\cap N(u)\}\ \cup \\
    & \{ k_3(w): w\in N_2(v)\setminus N(u)\}\}.
\end{align*}
The first term of this formula can be computed in $\cO(1)$ time.
The second term can be computed in $\cO(1)$ time, after a preprocessing in $\cO(\deg(v))$ time in order to keep track of two neighbors of $v$ with maximum $k_2$-value.
The third term can be computed in $\cO(\deg(u))$ time by simply scanning all neighbors of $u$.
Finally, applying Lemma~\ref{lem:max-nonneighbor} to the subgraph induced by $\St(v)$ with $\kappa(w)=k_3(w)$ if $w \in N_2(v)$ and $0$ otherwise, we deduce that all fourth terms requested to compute the eccentricity of every $u\in N(v)$ can be computed in total $\cO(n')$ time.
It remains to compute the eccentricities $r_{\pi}(w)$ of vertices $w\in N_2(v)$. Let $u_1$ and $u_2$ be the two common neighbors of $w$ and $v$. Then $r_{w}$ can be computed by the following formula:
\begin{align*}
    r_{\pi}(w)=\max & \{\{2\pi(v),k_0(w), k_1(u_1), k_1(u_2)\} \cup \{ k_2(w'): w'\in N_2(v)\cap (N(u_1)\cup N(u_2))\}\ \cup \\
    & \{ k_3(u): u\in N(v)\setminus \{ u_1,u_2\}\} \cup \{ k_4(w'): w'\in N_2(v)\setminus (N(u_1)\cup N(u_2))\}\}.
\end{align*}
The first term can be computed in $\cO(1)$ time. To compute the second term, for every $u \in N(v)$, we compute $\max\{ k_2(w'): w'\in N_2(v)\cap N(u)\}$.
Then, for every $w \in N_2(v)$, we keep the maximum value computed for its neighbors $u_1,u_2$.
It takes total $\cO(n)$ time.
To compute the third term, we apply Lemma~\ref{lem:max-nonneighbor} to $\St(v)$ with $\kappa(u) = k_3(u)$ if $u \in N(v)$ and $0$ otherwise.
Finally, to compute the fourth term, 
we proceed in three steps:
\begin{enumerate}[(1)]
    \item First notice that $N(w) \cap N(v) \neq N(w') \cap N(v)$ for every distinct vertices $w,w' \in N_2(v)$. Indeed, if $w$ and $w'$ have the same common neighbors $u_1,u_2$ with $v$, then the vertices $v,w,w',u_1,u_2$ induce a $K_{2,3}$, which
    is forbidden in median graphs.
    \item Then, for every  $u \in N(v)$, we compute a vertex $w_u \in N_2(v) \setminus N(u)$ maximizing $k_4(w_u)$.
    For that, we apply Lemma~\ref{lem:max-nonneighbor} to $\St(v)$ with $\kappa(w) = k_4(w)$ if $w \in N_2(v)$ and $0$ otherwise.

    Observe that the fourth term of any $w \in N_2(u)$ must be equal to $k_4(w_{u_1})$ (or to $k_4(w_{u_2})$, respectively), unless $w,w_{u_1}$ ($w,w_{u_2}$, respectively) have a common neighbor in $\St(v)$, which can be verified in $\cO(1)$ time.
    If $w$ has no common neighbor with at least one of  $w_{u_1},w_{u_2}$, then we call it \emph{decided} (its fourth term has been already computed), and \emph{undecided} otherwise. Notice that $w$ is undecided if and only if
    $w_{u_1}$ is adjacent to $u_2$ and $w_{u_2}$ is adjacent to $u_1$. Furthermore,  every first neighbor $u \in N(v)$ has at most two undecided neighbors.
    Indeed, if $u$ has three undecided neighbors $w_1, w_2, w_3,$ then each $w_i,$ $i=1,2,3$ has a common neighbor $z_i$ with $v$ that is distinct from $u.$ Since there exists at most one $w \in N_2(v)$ such that $N(v) \cap N(w) = \{u,u'\},$ the vertices $z_1, z_2, z_3$ are pairwise distinct and they are all adjacent to $w_u.$ Consequently, $v,z_1, z_2, z_3, w_u$ induce a forbidden $K_{2,3}.$

    \item Let $K$ be obtained from  $\St(v)$ by making each undecided vertex $w\in N_2(v)$ adjacent to all vertices of $N(u_1)\cup N(u_2)$. There are at most $2\sum\{\deg(u) : u \in N(v)\} = \cO(n')$ additional edges in $K$.
    Therefore, $K$ contains at most $\cO(n)$ edges and can be computed in $\cO(n)$ time.
    We end up computing the fourth terms of undecided vertices of $N_2(v)$, by applying Lemma~\ref{lem:max-nonneighbor} to  $K$ with $\kappa(w) = k_4(w)$ if $w \in N_2(v)$ and $0$ otherwise.
\end{enumerate}

\subsection{Complexity.}
At each step, given the region $R$, we compute in $\cO(|R|)$ time a median vertex $v$ of $R$
using  \cite{BeChChVa_med}. Then we compute in $\cO(n)$ time $r_\pi(z)$ for all $z\in \St(v)$. We either return $v$ as a center (Case~0) or
$v$ has an i-neighbor or/and is-neighbor. If $v$ has no i-neighbor, then in $\cO(n)$ time we compute a separating halfspace $H$ of $R$ (Case~1). Finally,
if $v$ has an i-neighbor $z$, we compute in $\cO(n\log n)$ time a local minimum $u$ of $r_\pi$ in the tree $\Upsilon_{\St(v)}(z)$ and $r_\pi(t)$ for all  $t\in \St(u)$.
Then we either return $u$ as a center (Case~2) or we compute in $\cO(n)$ time a separating halfspace $H$ of $R$ (Cases~3-5). Namely, $H=R\cap H(a,b)$, where $ab$ is an edge of $R$
such that $ab=zv$ in Cases~1,5 and $ab=tu$ in Cases~3,4. To compute  $R\cap H(a,b)$, first we compute in $\cO(n)$ time the $\Theta$-classes of $G$ using \cite{BeChChVa_med}, then we traverse the edges of the $\Theta$-class $\Theta_i$ of $ab$. 
For each such edge we select the end belonging to $H(a,b)$. Then we perform a BFS in $R$ starting from the set of all such
vertices and which does not involve the edges of $\Theta_i$. The vertices of $R$ which are reached by this BFS
are exactly the vertices of $R\cap H(a,b)$, and it requires $\cO(n)$ time to compute them. 
Since we have $\log n$ iterations and the complexity of each iteration is in $\cO(n\log n)$, we obtain an algorithm with complexity $\cO(n\log^2 n)$.
Consequently, we have proven the following main result of this section (which is  Theorem \ref{thml:cube-free-median}):

\begin{mytheorem} \label{thm:cube-free-median}
    If $G$ is a cube-free median graph with arbitrary weights, then a central vertex of $G$ can be computed in deterministic time $\cO(n\log^2 n)$.
\end{mytheorem}

\section{$G^p$-unimodality of the radius functions of $\delta$-hyperbolic and $\whp$-coarse Helly graphs} \label{sec:hyperb-cHelly} In this section, we prove that the radius functions of $\delta$-hyperbolic and $\whp$-coarse Helly graphs are $G^p$-unimodal for some constant $p$ depending respectively only on $\delta$ and $\whp.$ 
We further refine our algorithmic results for $\delta$-hyperbolic (weakly) bridged or bipartite Helly graphs.

\subsection{$\delta$-Hyperbolic graphs.}
Gromov's hyperbolicity is an important notion in metric geometry, which revolutionized the geometric group theory. A graph $G=(V,E)$ is  $\delta$-\emph{hyperbolic} \cite{Gr} for $\delta\ge 0$ if  for any $u,v,x,y\in V$, the two largest of the sums
$d(u,v)+d(x,y)$, $d(u,x)+d(v,y)$, $d(u,y)+d(v,x)$ differ by at most
$2\delta \geq 0$. Gromov hyperbolic graphs
constitute a very important class of graphs in metric graph theory. The unweighted center problem for this class of graphs and the relationship between the radius, the diameter, and the
$\delta$-hyperbolicity were studied in \cite{ChDrEsHaVa}. Almost unimodality properties of the  classical radius function in
$\delta$-hyperbolic graphs (i.e., when $\pi(v)=1$ for all $v\in V$) were investigated earlier in \cite{AlRDr,DrGu_hyp,locality-dhg,locality-alpha-delta}. 
Here we consider arbitrary profiles.

The following result establishes the statement of Theorem \ref{Gpunimodality} related to $\delta$-hyperbolic graphs:

\begin{myproposition} \label{hyperbolic-wp} 
    All radius functions $r_\pi$ of a $\delta$-hyperbolic graph $G$ are  $(4\delta+1)$-weakly peakless.
\end{myproposition}

\begin{proof} 
    %
    Let $u,v$ be any pair of vertices of $G$ with $d(u,v)\ge 4\delta+2$. Pick any geodesic $\gamma(u,v)$ between $u$ and $v$ and let $w$ be a vertex of $\gamma(u,v)$ such that $d(w,u)\ge 2\delta+1$ and $d(w,v)\ge 2\delta+1$. This implies that $d(u,v)>\max(d(u,w),d(v,w))+2\delta.$ We assert that $r_{\pi}(w)<\max\{ r_{\pi}(u),r_{\pi}(v)\}$. Indeed, let $x$ be a furthest from
$w$ vertex of $\supp(\pi)$, i.e., $x$ is a vertex from the profile such that $r_\pi(w)=\pi(x)d(x,w)$.
    As
$d(u,v)+d(w,x)\le \max\{ d(u,x)+d(v,w),d(v,x)+d(u,w)\}+2\delta$ because $G$ is $\delta$-hyperbolic, and $d(u,v) > \max\{ d(u,w),d(w,v)\}+ 2\delta$, we deduce  that $d(w,x) <\max\{ d(u,x),d(v,x)\}$.   
    Consequently, $r_\pi(w)<\max\{ r_\pi(u),r_\pi(v)\}$, establishing that the radius function $r_\pi$ is strictly $(4\delta+1)$-weakly peakless.
\end{proof}


Let the $\delta$-hyperbolicity of a graph $G$ be the smallest $\delta$ for which $G$ is $\delta$-hyperbolic.
Theorem 9.6 of \cite{CCHO} (which refines Proposition 10 of \cite{ChDrEsHaVa}) characterizes (up to a constant factor) the $\delta$-hyperbolicity of a weakly modular graph $G$ by the largest size $\mu$ of a metric triangle of $G$ and the largest size $\kappa$ of an isometric square grid of $G$: $\delta=\Theta(\max\{ \mu,\kappa\})$.
Similarly to \cite[Theorem 9.6]{CCHO} about the hyperbolicity of weakly modular graphs, and using Lemma~\ref{lem:cb-triangle}, one can prove a similar result about the hyperbolicity of CB-graphs. From these two results we can conclude the following:

\begin{myproposition} \label{hypCBgraph} If $G$ is weakly modular  or a CB-graph, then $G$ is $\cO(\sqrt{n})$-hyperbolic.
\end{myproposition}

The $\delta$-hyperbolicity of median graphs, (weakly) bridged, and (bipartite) Helly graphs can be characterized by forbidding more structured patterns.  By Corollary 4 of \cite{ChDrEsHaVa} and Lemma 9.7 of \cite{CCHO},
the hyperbolicity of a median graph $G$ is characterized by the size of the largest isometric square grid of $G$. A similar result can be proved
for bipartite Helly graphs. Similarly, the hyperbolicity of a Helly graph $G$ is characterized by the size of the largest isometric king grid of $G$ (the king grid is obtained from the square grid by adding the two
diagonals of each square); see \cite[Proposition 4.7.]{CCGHO} and \cite{DrGu_hellyhyp}. Finally, by Corollary 4 of \cite{ChDrEsHaVa}, the hyperbolicity $\delta$ of a bridged graph $G$ is characterized by the size $\nu$ of a largest
isometric deltoid of $G$: $\delta=\Theta(\nu)$. A deltoid of size $\nu$ is a subgraph of $G$ that is isometric to the equilateral triangle with side $\nu$ of the triangular grid (the tessellation of the
plane into equilateral triangles with side 1). As a graph, the triangular grid can be obtained from the square grid by adding one diagonal to each square such that all such added diagonals lie on parallel lines. The deltoid of size
$\nu$ contains a triangular grid with side $\frac{\nu}{2}\times \frac{\nu}{2}$ (this  grid has the form of a lozenge and is obtained by gluing two deltoids with side  $\frac{\nu}{2}$ along a common side).
Therefore, the hyperbolicity of bridged graphs can be again characterized by the largest size of a triangular grid (largest lozenge). This characterization can be extended to all weakly bridged graphs.

The results of our paper and of \cite{Du_Helly}  can be combined with the result of \cite{ChDrEsHaVa} about the approximation of the
radius of $\delta$-hyperbolic graphs to get a fast deterministic algorithm for the center problem with 0-1-profiles in $\delta$-hyperbolic (weakly) bridged graphs, Helly, and bipartite Helly graphs. We call this method
\emph{FPscan-descent}.  We proceed as follows. Let $G$ be any $\delta$-hyperbolic graph from one of these classes of graphs and let $\pi$ be a 0-1-profile on $G$. Let $u,w$ be the pair of vertices returned (in $\cO(m)$ time) after two FP scans:
pick any vertex $v$, compute any farthest from $v$ vertex $u$ of the profile and then compute a furthest from $u$ vertex $w$ from the profile. Let $c$ be any vertex in the middle of a shortest $(u,w)$-path $P$
(if $d(u,w)$ is even, then $c$ is unique and if $d(u,v)$ is odd, then $P$ contains two middle vertices and $c$ is one of them). By \cite[Proposition 6]{ChDrEsHaVa}, the center of $\pi$ is contained in the ball $B_{5\delta}(c)$
of radius $5\delta$ centered at $c$ (then clearly that $r_{\pi}(c)\le r_{\pi}+5\delta$). Now, we set $u_0:=c$ and we apply the descent algorithm to the radius function $r_\pi$. Namely, at step $i$ given the current vertex $u_i$, we apply  {\sf ImproveEccentricity}$(G, \pi,u_i)$,
specific to each class of graphs. Let $(u_0,u_1,\ldots,u_\ell)$ be the improving path of $G^2$ (or of $G$ if $G$ is Helly) constructed by the descent algorithm. Since $u_\ell$ is a minimum of the radius function
$r_\pi$ in $B_2(u_\ell)$ and $r_\pi$ is $G^2$-unimodal, $u_\ell$ is a central vertex of $G$. Since $r_\pi(u_0)>r_\pi(u_1)>\cdots>r_\pi(u_\ell)$, $\pi$ is a 0-1-profile, and $u_0=c$ has distance $\le 5\delta$
from the center $u_\ell$, the length  $\ell$ of the improving path is at most $5\delta$. Since each call of {\sf ImproveEccentricity} is in $\cO(m)$ time, we obtain an algorithm with complexity $\cO(\delta m)$. If $\delta$ is constant, this
is a linear-time algorithm. Since by Proposition \ref{hypCBgraph} the hyperbolicity of $G$ is at most $\sqrt{n}$, this is a deterministic algorithm with complexity $\cO(\sqrt{n}m)$. The same holds for graphs with convex balls, but in this case
the final vertex $u_\ell$ is either central or is at distance at most 2 from the center. Summarizing, we get the following result (which is the assertion of Theorem \ref{main} for $\delta$-hyperbolic graphs):

\begin{mytheorem} \label{center-hyp} Let $G=(V,E)$ be a graph with $n$ vertices and $m$ edges and $\pi$ be a 0-1-profile on $G$.
\begin{enumerate}[(1)]
\item If $G$ is a $\delta$-hyperbolic bridged, weakly bridged, or bipartite Helly graph (respectively, CB-graph), then a central vertex of $G$ (respectively, a vertex  at distance at most 2 from the center)
can be computed in deterministic $\cO(\delta m)$ time via FPscan-descent method. If $\delta=\cO(1)$, then this is a linear-time algorithm for the center problem.
\item If $G$ is a  bridged, weakly bridged, or bipartite Helly graph (respectively, CB-graph) with a 0-1-profile, then a central vertex of $G$ (respectively, a vertex  at distance at most 2 from the center)
can be computed in deterministic $\cO(\sqrt{n} m)$ time via FPscan-descent method.
\end{enumerate}
\end{mytheorem}

Theorem \ref{center-hyp} has several consequences, when the hyperbolicity of $G$ is constant. For example, the hyperbolicity of chordal graphs is at most 1, thus we get a linear-time
algorithm for computing a central vertex for a chordal graph. As we noted above, such an algorithm already exists \cite{ChDr}. In fact, the FPscan-descent method can be viewed as a
generalization of the method used in \cite{ChDr}. Another class of graphs with constant hyperbolicity are the \emph{chordal bipartite graphs}, i.e., the bipartite graphs in which all
induced cycles have length 4. These graphs are hereditary modular and thus they are bipartite Helly graphs \cite{Ba_hmg}.  Consequently, we obtain the following
consequence of Theorem \ref{center-hyp}:

\begin{mycorollary}
    If $G$ is a chordal bipartite graph, then for any 0-1-profile on $G$, a central vertex can be computed in $\cO(m)$ time.
\end{mycorollary}

\begin{myremark}
  Previously, it was only known that a central vertex of a chordal
  bipartite graph can be computed in linear time for the classical
  center function~\cite{Du_abs}, i.e., when $\pi(v)= 1$ for any
  vertex $v$.
\end{myremark}

Since the size of a largest square grid, king grid, or triangular grid is roughly equivalent to hyperbolicity of bipartite Helly, Helly, and (weakly) bridged graphs,  we also obtain the
following consequence of Theorem \ref{center-hyp}:

\begin{mycorollary}
    If $G$ is a (1) bipartite Helly graph in which each square grid has size $\cO(1)$, (2) a Helly graph in which each king grid has size $\cO(1)$, or (3) a (weakly) bridged graph in which
    each triangular grid has size $\cO(1)$, then for any 0-1-profile on $G$, a central vertex can be computed in $\cO(m)$ time using FPscan-descent.
\end{mycorollary}

\begin{myremark} A linear-time algorithm for the center problem with 0-1-profiles for weakly bridged graphs not containing the triangular grid of size 2,
was recently and independently of us obtained by S. Seif \cite{Se}, using a different approach.
\end{myremark}

\subsection{Coarse Helly graphs.}
\newcommand{\calF}{\mathcal{F}}
%
A graph $G$ is called {\em $\alpha$-coarse Helly} if for any
family $\mathcal B=\{ B_{r_i}(v_i): i\in I\}$  of pairwise intersecting balls, inflating each ball $B_{r_i}(v_i)$ of $\mathcal B$ by a constant $\alpha$ leads to a family of balls with a nonempty common intersection,
i.e., $\bigcap_{i\in I} B_{r_i+\alpha}(v_i)\ne \varnothing$ holds.
This generalized Helly property was introduced in \cite{ChEs}, where it was shown that $\delta$-hyperbolic graphs are $2\delta$-coarse Helly,
and further was investigated in \cite{CCGHO} and \cite{DrGu} (and papers in geometric group theory, e.g., \cite{HaHoPe,OsVa}).
Clearly, any finite graph $G$ is $\alpha$-coarse Helly for  $\alpha \ge \lfloor diam(G) / 2 \rfloor$. The minimum~$\alpha$ for which a graph~$G$ is $\alpha$-coarse
Helly is called the \hg{} of~$G$, denoted by $\whp(G)$. Helly graphs are the graphs with the Helly gap $\whp(G)=0$. Several other graph families are known to have a \hg{} bounded by a constant \cite{DrGu}. 
Interestingly, even earlier, Lenhart et al.~\cite[Lemma 9]{LePoSaSeShSuTouWhYa} established that the balls of simple polygons endowed with link-metric
satisfy a Helly-type property which implies coarseness 1 (the link graph of a simple polygon $P$ has the points of $P$ as vertices and pairs of points $x,y\in P$ such that $[x,y]\subseteq P$ as edges).
Notice also that bipartite Helly graphs $G=(V_0\cup V_1,E)$  are $1$-coarse Helly. Indeed, if $\mathcal B=\{ B_{r_i}(v_i): i\in I\}$ is a collection
of pairwise intersecting balls of $G$, then the collection  ${\mathcal B}'=\{ B_{r_i+1}(v_i): i\in I\}$ of inflated balls pairwise intersect in $V_0$ (and also in $V_1$). Therefore, the inflated half-balls  $B_{r_i+1}(v_i)\cap V_0, i\in I$ pairwise intersect in $V_0$. By the Helly property for half-balls we deduce that there exists a vertex of $V_0$ belonging to $\bigcap_{i\in I} B_{r_i+1}(v_i)\cap V_0\subseteq \bigcap_{i\in I} B_{r_i}(v_i)$, and we are done.


In this subsection, we prove $G^{\cO(\alpha)}$-unimodality of all 0-1-radius functions $r_\pi$ of an $\whp$-coarse Helly graph.
Our starting point is the  interesting result of \cite{DrGu}: in any $\whp$-coarse Helly graph $G=(V,E)$ with
a 0-1-profile $\pi$, for any $v \in V$ such that  $r_\pi(v) >\rad_\pi(G) + \whp$, there is a vertex $v' \in B_{2\whp + 1}(v)$ with $r_\pi(v') < r_\pi(v)$.
For $\ell\ge 0$, let $C_\pi^\ell(G)=\{ v\in V: r_\pi(v)\le \rad_\pi(G)+\ell\}$. Clearly, $C_\pi(G)\subseteq C_\pi^\ell(G)$.
%
%
First, we  strengthen this result of \cite{DrGu} as follows:

\begin{mylemma}\label{lem:cHelly-almost-unimodal}
    If $G$ is $\whp$-coarse Helly graph, $\pi$ is a 0-1-profile, and $v$ an arbitrary vertex with $r_\pi(v)=\rad_\pi(G)+\alpha+k$ for some $k\ge 1$, then,
    \begin{enumerate}[(1)]
        \item either $C^{\whp+k}_\pi(G)\subset B_{2\alpha+1}(v)$ $($whence $C_\pi(G)\subset B_{2\alpha+1}(v))$,
        \item or for every $u\in C^{\whp+k}_\pi(G)$ with $d(v,u)\ge 2\whp +2$, there is a vertex $v'\in B_{2\alpha+1}(v)$ such that $r_\pi(v') < r_\pi(v)$ and $d(v',u)< d(v,u)$.
    \end{enumerate}
\end{mylemma}

\begin{proof}
    Let $M=\{v\in V : \pi(v)=1\} = \supp(\pi)$. Assume that $C^{\whp+k}_\pi(G)\setminus  B_{2\alpha+1}(v)\neq\varnothing$ and pick any $u\in C^{\whp+k}_\pi(G)$ with $d(v,u)\ge 2\whp +2$.
    Consider in~$G$ a system of balls $\mathcal F = \{B_{\rho(x)}(x) : x\in M\cup \{v,u \}\}$ with
    radii defined as follows: $\rho(v):= \alpha+1$, $\rho(u):= d(u,v)-\alpha-1\ge \alpha+1$ and $\rho(x):=r_\pi(v)-\alpha-1\ge \rad_\pi(G)$ for all  $x\in M$.
    We assert that all balls of $\mathcal F$ pairwise intersect.
    Pick any $x$ in  $M$. Balls $B_{\rho(x)}(x)$ and $B_{\rho(v)}(v)$ intersect as $d(x,v) \leq r_\pi(v)=\rho(x)+\rho(v)$.
    Balls $B_{\rho(x)}(x)$ and $B_{\rho(u)}(u)$ intersect as $d(x,u) \leq r_\pi(u) \le r_\pi(v)=(r_\pi(v)-\alpha-1)+(\alpha+1)\le \rho(x)+\rho(u)$. For each $y\in M$, balls $B_{\rho(x)}(x)$ and $B_{\rho(y)}(y)$ intersect as $d(x,y) \leq 2\rad_\pi(G)\le  \rho(x)+\rho(y)$. Finally, balls $B_{\rho(u)}(u)$ and $B_{\rho(v)}(v)$ intersect as $d(u,v) = \rho(u)+\rho(v)$. Since $G$ is $\whp$-coarse Helly, the system $\mathcal F$ of
    pairwise intersecting balls has a common intersection when radii of all balls are extended by $\whp$.
    Therefore, there is  $v'$ such that $d(v',v) \le \rho(v)+\alpha= 2\whp + 1$, $d(v',u) \leq \rho(u)+\alpha=d(v,u) - 1$, and $d(v',x) \leq \rho(x)+\alpha =r_\pi(v) - 1$ for all $x \in M$.
    The latter shows that $r_\pi(v') \le r_\pi(v) - 1$ and we are done. 
\end{proof}

The following result establishes the statement of Theorem \ref{Gpunimodality} in the case of $\alpha$-coarse Helly graphs:

\begin{myproposition}\label{cor:cHelly-almost-unimodal} For any unweighted radius function $r_\pi$ of an  $\whp$-coarse Helly graph $G$ and any vertex $v$, if $r_\pi(v)>\rad_\pi(G)+\alpha$, then $v$ is not
    a local minimum of $r_\pi$ in $G^{2\alpha+1}$.
\end{myproposition}

\begin{proof} Let $p=2\alpha+1$. If $d(v,C_{\pi}(G))\le p$, then in $G^p$ the vertex $v$ has a neighbor $v'\in C_\pi(G)$.  Then $r_{\pi}(v')\le r_\pi(v)$ and thus
    either $v$ is a central vertex or $r_{\pi}(v')<r_\pi(v)$ and thus $v$ is not a local minimum of $r_\pi$ in $G^p$. Now suppose that $d(v,C_\pi(G))>p$. By Lemma~\ref{lem:cHelly-almost-unimodal}, for any vertex $u\in C_\pi(G)$,
    there exists a vertex $v'\in B_{2\alpha+1}(v)$ such that $r_\pi(v') < r_\pi(v)$ and $d(v',u)<d(v,u)$. Thus, $v$ is not a local minimum in $G^p$ of the radius function $r_\pi$.
\end{proof}

At the difference to other classes of graphs, for which we prove $G^p$-unimodality via $p$-weakly peaklessness, we do not have such a proof for coarse Helly graphs. Even worse, we cannot prove that if a vertex $v$ is far away from the center $C_\pi(G)$, then $e_\pi(v)$ is much larger than $\rad_\pi(G)$. Furthermore, no good characterizations or polynomial recognition
algorithms are known for $\alpha$-coarse Helly graphs, even in the case $\alpha=1$. Even more, we are inclined to believe that already the recognition of 1-coarse Helly graphs is coNP-hard. On the other hand, Dragan and Guarnera  \cite{DrGu} proved the
following relation between the radius and the diameter of an $\alpha$-coarse Helly graph, which generalizes similar results for Helly graphs \cite{Dr_thesis,Dr_Helly} and for $\delta$-hyperbolic graphs \cite{ChDrEsHaVa}:

\begin{myproposition}[\!\!\cite{DrGu}]\label{prop:cHelly-Diam2Rad}
    For any subset $M\subseteq V$ of an $\alpha$-coarse Helly graph, $2rad(M) \ge diam(M) \geq 2rad(M)-2\whp(G)-1$.
\end{myproposition}

Dragan and Guarnera \cite{DrGu} asked \emph{whether
  $2rad(M) \ge diam(M) \geq 2rad(M)-2\whp(G)-1$ for all $M\subseteq V$
  implies that $G$ is $\alpha$-coarse Helly?}  A positive answer even
of the relaxed version when $G$ would be $\cO(\alpha)$-coarse Helly
would indicate that the graphs, for which the exact or approximate
computation of the center is done by searching a region around the
middle of a shortest path returned by a few FP scans, are all
$\cO(\alpha)$-coarse Helly.  As proved in \cite{CCGHO}, the Helly gap
$\alpha(G)$ is a maximal distance from a point in the Hellyfication of
$G$ to a closest vertex of $G$; however this result cannot be used to
compute $\alpha(G)$ since the Hellyfication of $G$ may contain an
exponential number of vertices.  Therefore it will be interesting to
investigate whether the Hellyfication of $G$ contains specific points
which are furthest from $V(G)$. Each point of the Hellyfication
corresponds to a set of radii of the vertices of $G$, which give
pairwise intersecting balls. The question of \cite{DrGu} can be
reformulated in this language as follows:

\begin{myquestion}
Is it true that $G$ is $\alpha$-coarse Helly iff for any $r$ and any family of pairwise intersecting
balls $\{ B_{r}(x_i): i\in I\}$ of radius $r$ the intersection $\cap_{i=1}^n B_{r+\alpha}(x_i)$ of the inflated balls is nonempty?
\end{myquestion}

The second question about coarse Helly graphs is on the existence of subquadratic-time algorithms for the center problem. This problem seems nontrivial already for 1-coarse Helly graphs, because it must generalize our result for bipartite Helly graphs:

\begin{myquestion}
Is it possible to compute a central vertex of an $\alpha$-coarse Helly graph $G$ in subquadratic time? What can be said in case of $\alpha=1$ and in case of 0-1-profiles?
\end{myquestion}

\section{Recognition and lower bounds} \label{sec:lower-recogn}  In this section we show on one hand that the graphs with $G^p$-unimodal radius functions can be recognized in polynomial time.
On the other hand, under the Hitting Set Conjecture, we show that deciding whether a given vertex is a local minimum of the radius function (and thus {\sf ImproveEccentricity}) in a graph  with  $G^p$-unimodal radius
functions cannot be solved in subquadratic time, even if $p=2$. 
\subsection{Recognition.}

In this subsection, we present a polynomial-time recognition algorithm for all graphs with $G^p$-unimodal radius functions 
(for median function, see \cite{BaCh_med} for $p=1$ and \cite{BeChChVa_G2} for any $p$). This proves Theorem \ref{th:recognition}.

\begin{mytheorem}
    \label{thm:recognition} Given a graph $G$ with $n$ vertices and a positive integer $p$, one can decide in $\cO(n^4)$ time whether all radius functions $r_\pi$ of $G$ are $G^p$-unimodal.
\end{mytheorem}

%
%
\begin{proof}
It suffices to check if $G$ admits a non $G^p$-unimodal radius function.
For that, for each vertex $u$, we check if there exists a profile $\pi$ such that $u$ is a local minimum of $r_\pi$ in $G^p$, however $u \notin C_\pi(G)$.
We consider each $v \in V$ with $d(u,v)>p$ and we search for  $\pi$ for which $r_\pi(v) < r_\pi(u)$.
We claim that a necessary and sufficient condition for existence of such $\pi$ is $\UC(u,v)$:
\emph{for every $x \in B_p(u)$, there is a vertex $w_x$ such that $d(x,w_x) \ge d(u,w_x)$ and $d(x,w_x) > d(v,w_x)$.}
%
Indeed, let there exists  $\pi$ for which $u$ is a local minimum of $r_\pi$ in $G^p$ and $r_\pi(v) < r_\pi(u)$.
For every $x \in B_p(u)$, we pick a vertex $w_x$ such that $r_\pi(x) = \pi(w_x)d(x,w_x)$.
Then, $d(x,w_x) \ge d(u,w_x)$, otherwise $r_\pi(x) < \pi(w_x)d(u,w_x) \le r_\pi(u)$.
In the same way, $d(x,w_x) > d(v,w_x)$, otherwise $r_\pi(v) \ge r_\pi(x) \ge r_\pi(u)$.
This proves the necessity of $\UC(u,v)$. Conversely, let $\UC(u,v)$ holds.
Set $W = \{w_x : x \in B_p(u)\}$.
Let $\pi$ be the profile with support $W$ such that, for every $w \in W$, we have $\pi(w) = 1/\min\{d(w,x) : x \in B_p(u) \ \text{and} \ w = w_x\}$.
Then, for every $x \in B_p(u)$, $r_\pi(x) \ge \pi(w_x)d(x,w_x) \ge \frac 1 {d(x,w_x)} d(x,w_x) = 1$.
In particular, $r_\pi(u) \ge 1$.
For every $w \in W$, there  exists some $x \in B_p(u)$ such that $w = w_x$ and $\pi(w) = 1/d(x,w_x)$.
Then, $\pi(w)d(u,w) = \frac 1 {d(x,w_x)} d(u,w_x) \le \frac 1 {d(x,w_x)} d(x,w_x) = 1$.
Hence $r_\pi(u) = 1$. Thus $u$ is a local minimum for $r_\pi$ in $G^p$.
Finally, for every $w \in W$, let again $x \in B_p(u)$ be satisfying $w = w_x$ and $\pi(w) = 1/d(x,w_x)$.
Then, $\pi(w)d(v,w) = \frac 1 {d(x,w_x)} d(v,w_x) < \frac 1 {d(x,w_x)} d(x,w_x) = 1$.
Therefore, $r_\pi(v) < 1 = r_\pi(u)$.
Note that $\UC(u,v)$ can be checked in $\cO(|B_p(u)|n) = \cO(n^2)$ time, if the distance matrix of $G$ is given.
Hence, the running time of the algorithm is in $\cO(n^2 \times n^2 + nm) = \cO(n^4)$.
\end{proof}

\subsection{Lower bounds.}

In this subsection, we prove a conditional hardness result for the problem of deciding whether a given vertex is a local minimum of the radius function in $G^2$ (Theorem \ref{th:localminimum}): 

\begin{mytheorem}
    \label{thm:localminimum}
    Assuming the Hitting Set Conjecture, for any $\varepsilon > 0$, deciding if $v$ is a local minimum in $G^2$ of
    the radius function $r_\pi$ for a graph $G$  with $2$-weakly-peakless radius functions  cannot be solved in $\cO(n^{2-\varepsilon})$ time, even if $G$ has only $\ctO(n)$ edges, and every center function $r_\pi$  of $G$ is $2$-weakly-peakless.
\end{mytheorem}

%
\begin{proof}
Recall the {\em Hitting Set Conjecture}, introduced by Abboud et al.~\cite{AVW16}:
There is no $\varepsilon > 0$ such that for all $\gamma \ge 1$, there is an algorithm that given
two lists $X, Y$ of $n$ subsets of a universe $U$ of size at most $\gamma \log{n}$, can decide in $\cO(n^{2-\varepsilon})$ time if there is a set in the first list that intersects every set in the second list, i.e., a hitting set.
Let $\langle X,Y,U \rangle$ be an instance of the  Hitting Set problem.
The following graph $G\langle X,Y,U \rangle$ is a slight variation of the HSE graph introduced in~\cite{AVW16}.
The vertex set is $X \cup Y \cup U \cup \{a,b,c,v,w\}$, where $U$ is a clique.  For every $z \in X \cup Y$,
and every $u \in U$, $z \sim u$ iff $u \in z$. Additionally, $a \sim v$, $v \sim w$, $b \sim c$, for every $z \in X \cup U$, $a,b \sim z$, and for every $x \in X$, $v \sim x$.
Since  $U$ is of size at most $\gamma \log{n}$, for some constant $\gamma$, we can construct $G\langle X,Y,U \rangle$ in $\ctO(n)$ time.
We can suppose wlog that $U$ is covered both by the sets of $X$ and by the sets of $Y$; equivalently, each $u\in U$ has a neighbor both
in $X$ and in $Y$.  Notice also that the vertices of $Y$ are adjacent only to vertices of $U$, thus all vertices of $Y$ are simplicial (i.e., their neighborhoods are cliques).
Theorem~\ref{th:localminimum} 
follows by combining the following two lemmas. 

\begin{mylemma}\label{lem:hs-1}
    The vertex $v$ is a local minimum in $G:=G\langle X,Y,U \rangle$ (in $G^2$, respectively) of the classical radius
    function $r_\pi$ (i.e., where $\pi(z)=1$ for any $z\in X \cup Y \cup U \cup \{a,b,c,v,w\}$)
    iff there is no set in $X$ that intersects every set in $Y$.
\end{mylemma}
\begin{proof}
By construction of  $G$, $r_\pi(v) = 3$; namely, $v$ has distance 3 to $c$ and all vertices of $Y$.
Since $w$ is a pendant vertex adjacent to $v$, $r_\pi(w) = 4$. Consequently,  the radius of  $G$ is at least $2$.
Furthermore, every vertex of eccentricity $2$ must be contained in the ball $B_2(w) = X \cup \{a,v,w\}$.
Observe that $r_\pi(a) = d(a,c) = 3$. Thus, either the radius of $G$ is 3 and $v \in C_\pi(G)$ or there exists a vertex $x$ such that $r_\pi(x) = 2$.
In the second case, by definition of $G$,  we must have $x\in X \subseteq N(v)$.
By definition of  $G$, a vertex $x \in X$ has eccentricity  $2$ iff the set $x$ of $X$
intersects every set in $Y$. Consequently, $v$ is a local minimum of $r_\pi$ in $G$ (resp. in $G^2$) iff there is no set in $X$ that intersects every set in $Y$.
\end{proof}

\begin{mylemma}\label{lem:hs-2}
    Any radius function $r_\pi$  on $G := G\langle X,Y,U \rangle$ is $2$-weakly peakless.
\end{mylemma}

\begin{proof} We start with a useful remark: \emph{Let $r,s, r\ne s$ be vertices of a graph $G'$ such that $B_1(r) \subseteq B_1(s)$.
If $r_\pi(r) < r_\pi(s)$, then $C_\pi(G') = \{r\}$ and $r_\pi(s) = \pi(r)$.}
Indeed, let $t$ be a furthest from $s$ vertex.  
If $t \neq r$, then $r_\pi(r) \ge \pi(t) d(t,r) \ge \pi(t) d(t,s) = r_\pi(s)$, which is impossible. 
Therefore, $t = r$, which implies $r_\pi(s) = \pi(r)$.
Furthermore, for every  $z \neq r$ of $G'$, $r_\pi(z) \ge \pi(r) d(r,z) \ge \pi(r) = r_\pi(s) > r_\pi(r)$.
Hence, $C_\pi(G') = \{r\}$ and we are done.

Let $\pi$ be any  profile on $G$. Let $z,z'$ be such that $d(z,z') \ge 3$.
We assume in what follows that for every vertex $s$ of $I^\circ(z,z')$ the inequality $r_\pi(s)\ge \max\{r_\pi(z),r_\pi(z')\}$ holds.
By the construction of  $G$, either at least one of the vertices $z,z'$ belongs to $Y$ or one of the vertices $z,z'$ coincides with one of the vertices $w,c$.
Consequently, in both cases, at least one of the vertices $z,z'$, say $z$,  is simplicial. Pick any $s \in I^\circ(z,z')$ adjacent to $z$. Since $z$ is simplicial, we have $B_1(z)\subseteq B_1(s)$.
We cannot have $r_\pi(s) = \pi(z)$, because otherwise, $r_\pi(z') \ge 3\pi(z) > \pi(z) = r_\pi(s)$, contrary to our assumption about the pair $z,z'$.
Therefore, by the above claim, $r_\pi(s)\le r_\pi(z)$. Since $r_\pi(s)\ge \max\{r_\pi(z),r_\pi(z')\}$, we obtain that $r_\pi(s)=r_\pi(z)$.
Since $z$ is not a furthest from $s$ vertex, the equality  $r_\pi(s)=r_\pi(z)$ implies that $z$ cannot be of degree one, and so, that it is different from vertices $w$ and $c$.  
Consequently, $z \in Y$ and $s\in U$.
If furthermore $r_\pi(z') \ge r_\pi(z)$, then we are done because $r_\pi(s) = r_\pi(z)\le \max\{ r_\pi(z), r_\pi(z')\}$ and equality holds when $r_\pi(z)=r_\pi(z')$.
Thus, from now on we can suppose that  $r_\pi(z') < r_\pi(z)=r_\pi(s)$.

Assume further that $z'$ is not simplicial (otherwise, we conclude the proof by considering $z'$ instead of $z$ and $s' \in N(z') \cap I(z,z')$ instead of $s$).
Since $d(z,z') \ge 3$, necessarily $z' \in X \cup \{v\}$.
Let $t$ be a furthest from $s$ vertex.  
Since  $r_\pi(z) = r_\pi(s)$, we must have $d(z,t)\le d(s,t)$. This implies that $t$ must be different from the vertices  $a,b,c,v,w,z'$.
On the other hand, since  $r_\pi(z') < r_\pi(s)$, we must have $d(z',t)<d(s,t)$. This implies that $t$ cannot belong to  $U \cup Y$, as we prove it now. Indeed,
for any vertex $u\in U$ we have $d(s,u)\le 1$ (since $s\in U$) and $d(z',u)\ge 1$ (since $z'\notin U$). Analogously, for any $y\in Y$ we have $d(s,y)\le 2$
(since $s\in U$ and $U$ is a clique)
and  $d(z',y)\ge 2$ (since $z'\in X\cup \{ v\}$).
If $z' \in X$, then furthermore $t$  cannot belong to $X \setminus z'$, because $d(z',x)=2$ and $d(s,x)\le 2$
for any $x\in X \setminus z'$. This leads us to a contradiction.
Therefore, $z' = v$. Consequently, in this case, $a \in I^\circ(z,z')$.
Suppose $r_\pi(a) \ge r_\pi(s) = r_\pi(z)>r_\pi(z')$ (otherwise, we are done), and let $r$ be a furthest from $a$ vertex.
Since $r_\pi(v)=r_\pi(z') < r_\pi(a)$, we must have $d(v,r)<d(a,r)$. This implies that $r$ must coincide with one of the vertices $v,w$.
But then, $a \in I^\circ(z,r)$, which implies $r_\pi(z) > r_\pi(a)$, a contradiction.
Consequently, any radius function $r_\pi$ is $2$-weakly peakless.
\end{proof}
This concludes the proof of Theorem \ref{thm:localminimum}.
\end{proof}

\section{Perspectives} \label{sec:perspectives}

In this paper, using $G^2$-unimodality of the radius functions, we presented local search subquadratic algorithms (randomized or deterministic) for the weighted and unweighted center problems on several important
classes of graphs:
bridged and weakly bridged graphs, graphs with convex balls, and bipartite Helly graphs. We also presented a divide-and-conquer algorithm with deterministic complexity $\cO(n\log^2 n)$
for the weighted center problem on cube-free median graphs (a subclass of bipartite Helly graphs). Prior to our work, it was only known how to compute in linear time the 0-1-centers of chordal graphs (a well-known class of graphs at the intersection of  bridged and hyperbolic graphs)
and a subquadratic-time algorithm for the  0-1-centers of median graphs.

We do not have analogous results for the weighted centers of $\delta$-hyperbolic graphs (and, more generally, of $\delta$-coarse Helly graphs)
and we believe that they do not exist. This is because at small scale (in balls of radius $\delta$),
$\delta$-hyperbolic graphs may be arbitrary. Therefore, the  algorithm of \cite{ChDrEsHaVa} for approximating the radius and
returning an approximate center of a $\delta$-hyperbolic graph 
seems to be the best one can get for the center problem. The results of \cite{ChDrEsHaVa} are for 0-1-profiles, thus generalizing them to arbitrary profiles is open:

\begin{myquestion} Is it possible to compute for any profile $\pi$ on a $\delta$-hyperbolic graph $G=(V,E)$ in $\mathcal{O}(|E|)$ time  (or even in subquadratic time) a vertex $v$ that has distance at most $\mathcal{O}(\delta)$ from the center $C_\pi(G)$?
\end{myquestion}


There are also several more general  perspectives on the center problem (and our paper provides partial answers to each of them). The first is about the graphs for which the center problem can be solved efficiently:

\begin{myquestion} Characterize the graphs for which the center problem can be solved in subquadratic time for all profiles.
\end{myquestion}

In the same vein, it will be important to investigate the metric properties of graphs, which can be useful in the efficient computation of radius, center, and diameter. The 0-1-center problem in chordal
graphs can be solved in linear time,  however assuming the Orthogonal Vectors Conjecture,  the diameter problem cannot be solved in subquadratic time. It will be interesting to investigate the center problem for classes of graphs for which the diameter problem
cannot be solved efficiently. This is related to the question how to compute the radius of a graph without computing the eccentricities of all its vertices.
It is also important to understand the power and the limits of local search  and $G^p$-unimodality for the center problem in graphs.

More generally, it will be important to investigate  general $G^p$-unimodal functions (and their weakly peakless variants)  in graphs from the perspectives of discrete convex analysis \cite{Hi,Mu,MuSh1,MuSh2} and the relationships between the complexity of local search and the geometry of graphs \cite{BrChRe}. Last but not least, it will be important and challenging to create an algorithmic theory for the classes of graphs
studied in Metric Graph Theory. This question is intimately related to the notion of metric (fat) minor of a graph, independently introduced in papers \cite{ChDrNeRaVa}  and  \cite{FuPa}  and the  coarse and metric graph theory (an interesting research program on coarse graph theory based on the notion of metric/fat minor
was suggested in recent paper  \cite{GePa}).

\subsection*{Acknowledgements} This work
has been supported in part by the PHC Br\^{a}ncu\c{s}i LoSST  and the ANR project DISTANCIA (ANR-17-CE40-0015).

\newpage

\section{Appendix: The center problem in CB-graphs}\label{sec:cb-alg}
In this section, we first recall the definition of CB-graphs together with characterizations and properties of these graphs. They are useful to prove that their radius functions are $G^2$-unimodal. Then, we consider the algorithmic implications of this result
and of the descent algorithms from Section~\ref{descent}. 
This part is similar to that for weakly bridged graphs (see Section~\ref{s:bridged}), but it is technically more involved, and the correctness proof requires additional distance conditions to hold for the center. Hence, we need to complete the algorithm with a different, and costlier, search procedure, which ensures that it is always correct even if such distance conditions do not hold.


\subsection{CB-graphs}  A graph $G$ is called a \emph{graph with convex balls} (abbreviately, a \emph{CB-graph}) if all balls $B_r(v)$ of $G$
are convex~\cite{SoCh,FaJa}. 
A topological characterization of CB-graphs via their triangle-pentagon complexes was given in \cite{ChChGi}. 

As (weakly) bridged graphs, CB-graphs satisfy the \emph{Interval
  Neighborhood Condition} $\INC$, i.e., for any $u,v\in V, u\ne v$,
the neighbors of $v$ in $I(u,v)$ form a clique. CB-graphs do not
satisfy the triangle condition, but they satisfy the
\emph{Triangle-Pentagon Condition} $\TPC$, i.e., for any $v,x,y\in V$
such that $d(v,x)=d(v,y)=k$ and $x\sim y$, either there exists
$z \in B_{k-1}(v)$ such that $z \sim x,y$, or there exists
$z \in B_{k-2}(v)$, $w \sim x,z$, and $w' \sim y,z$ such that $xwzw'y$
is a pentagon.

Recall now the metric characterizations of CB-graphs.

\begin{mytheorem}\label{thm: CB-graphs} For a graph $G$, the following are equivalent:
  \begin{enumerate}[(1)]
  \item $G$ is a CB-graph
  \item\label{th-cb-1} $G$ satisfies $\INC$ and isometric cycles have
    length 3 or 5~\emph{\cite{SoCh,FaJa}}.
  \item\label{th-cb-2}  $G$ satisfies the
    conditions $\INC$ and $\TPC$~\emph{\cite{ChChGi}}.
  \end{enumerate}
\end{mytheorem}


As in (weakly) bridged graphs, the $1$-balls in CB-graphs are
outergated.

\begin{mylemma}[\!\!\cite{ChChGi}]\label{lem:cb-outergate}
    Any  CB-graph $G$ satisfies the interval-outergate property.
\end{mylemma}

We will also need  the following result about the metric triangles:

\begin{mylemma}[\!\!\cite{ChChGi}]\label{lem:cb-triangle}
  In a CB-graph $G$, any metric triangle $uvw$ is strongly equilateral
  or of type $(2,2,1)$, i.e., $d(u,v) = d(u,w) = 2$ and $d(v,w) = 1$
  (assuming that $d(u,v) \geq d(u,w) \geq d(v,w)$).
\end{mylemma}



%


In CB-graphs, the property of Lemma~\ref{lem:wm-wx} holds but only for
vertices at distance at least $3$.

\begin{mylemma}\label{lem:cb-wx} If $G$ is a CB-graph and $u,v,x\in V$
  such that $d(x,u)<d(x,v)$ and $d(u,v) \ge 3$, then there exists
  $w_x\in N(v)\cap I(v,u)$ such that $w_x\in I(v,x)$.
\end{mylemma}

\begin{proof}
  Let $u'v'x'$ be a quasi-median (metric triangle) of the triplet
  $u,v,x$ and assume that $d(u,v)=d(u,u')+d(u',v')+d(v',v)$,
  $d(u,x)=d(u,u')+d(u',x')+d(x',x)$, and
  $d(v,x)=d(v,v')+d(v',x')+d(x',x)$. If $v'\ne v$, then any neighbor
  of $v$ in $I(v,v')\subset I(v,u)\cap I(v,x)$ can be selected as the
  desired $w_x$. Now suppose that $v'=v$.  Since $d(x,u)<d(x,v)$, the
  vertices $u',v',x'$ are pairwise distinct and the metric triangle
  $u'v'x'$ cannot be equilateral. Therefore, by
  Lemma~\ref{lem:cb-triangle}, the metric triangle $u'v'x'$ must have
  type $(2,2,1)$, and we conclude that $d(x',u')=1$,
  $d(x',v)=d(u',v) = 2$. Since
  $d(u,x) = d(u,u')+1+d(x',x) < d(v,x) = 2+d(x',x)$, we must have
  $u=u'$. However, this contradicts the assumption that
  $d(u,v) \geq 3$.
\end{proof}

\subsection{$G^2$-unimodality of the radius function.}

We show that as in (weakly) bridged graphs, in a CB-graph, any radius
function is $2$-weakly peakless.

\begin{myproposition}\label{thm:cb}
  All radius functions $r_\pi$ of a CB-graph $G$ are $2$-weakly
  peakless.
\end{myproposition}

\begin{proof}
  Let $u,v$ be any two vertices of $G=(V,E)$ with $d(u,v)\ge 3$ and
  suppose that $r_\pi(u)\le r_\pi(v)$.  If $r_\pi(u)=r_\pi(v)$, then
  for any vertex $x\in \supp(\pi)$ and any vertex
  $w\in I^{\circ}(u,v)$, by convexity of balls we have
  $\pi(x)d(x,w)\le \max\{ \pi(x)d(x,u),\pi(x)d(x,v)\}\le
  r_\pi(u)=r_\pi(v),$ yielding $r_{\pi}(w)\le
  r_\pi(u)=r_\pi(v)$. Thus, suppose that $r_\pi(u) < r_\pi(v)$.  By
  Lemma~\ref{lem:cb-outergate}, there exists $w \in I(v,u)$ such that
  $d(v,w)=2$ and $N(v)\cap I(v,u)\subseteq N(w)$. We will prove that
  $r_\pi(w) < r_\pi(v)$.

  Recall that
  $F_\pi(v) = \{ x \in \supp(\pi): \pi(x)d(x,v)=r_\pi(v) \}$. Since
  $G$ is a graph with convex balls, for any $z \in I^{\circ}(v,u)$ and
  any $y \in \supp(\pi) \setminus F_\pi(v)$, the following inequality
  holds:
  $\pi(y) \cdot d(z,y) \leq \max\{\pi(y) \cdot d(v,y), \pi(y) \cdot
  d(u,y)\} \leq \max\{\pi(y) \cdot d(v,y), r_\pi(u)\} < r_\pi(v)$.
  Therefore, in order to prove $r_\pi(w) < r_\pi(v)$, it suffices to
  prove that $d(w,x) < d(v,x)$ for every $x \in F_\pi(v)$.  By
  Lemma~\ref{lem:cb-wx}, for every $x \in F_\pi(v)$, there is some
  $w_x \in N(v) \cap I(u,v)$ such that $d(w_x,x) < d(v,x)$.  Set
  $W = \{w_x: x \in F_\pi(v)\}$. By definition
  $W \subseteq N(v) \cap I(u,v) \subseteq N(w)$.  Finally, for any
  $x \in F_\pi(v)$ we have $w_x,u\in B_{d(x,v)-1}(x)$.  Since
  $w\in I(w_x,u)$, the convexity of $B_{d(x,v)-1}(x)$ yields
  $d(w,x) \leq \max\{d(w_x,x),d(u,x)\}<d(x,v)$.
\end{proof}

\begin{mycorollary}\label{cor:cb}
  Let $G=(V,E)$ be a CB-graph and let $\pi$ be any profile.  Then for
  any $v\in V$, either $v\in C_\pi(G)$ or there exists $w \in B_2(v)$
  such that $r_\pi(w) < r_\pi(v)$.
\end{mycorollary}

\begin{proof}
  Suppose that $v\notin C_\pi(G)$ and let $c \in C_\pi(G)$. Then
  $r_\pi(c)<r_\pi(v)$. If $d(v,c)\le 2$, we are done. So, let
  $d(c,v) \geq 3$.  By Theorem~\ref{thm:cb}, for any vertex
  $u \in I(c,v)$ such that $d(u,v) = 2$ and
  $N(v) \cap I(v,u) \subseteq N(u)$ we have $r_\pi(u) < r_\pi(v)$.
\end{proof}

\subsection{Outergated and 2-outergated sets.}

Given a set $S$ and a vertex $z \in V \setminus S$, recall that $z^*$
is an outergate of $z$ with respect to $S$ if $d(z,z^*) = d(z,S)-1$
and $z^*$ is adjacent to $\proj_z(S)$, the set of all vertices of $S$
at distance $d(z,S)$ from $z$.  More generally, for an integer $i$ we
say that a vertex $z^*$ is an \emph{$i$-outergate} of a vertex $z$ if
$d(z,z^*)=d(z,S)-i$ and $d(z^*,u)=i$ for any $u\in \proj_z(S)$. Then
the set $S$ is called {\em $i$-outergated} if every vertex of
$V \setminus S$ has an $i'$-outergate for some $i'\le i$.  The
outergates and outergated sets correspond to the case $i=1$.

By Lemma~\ref{lem:cb-outergate}, in a CB-graph $G$ all unit balls
$B_1(v)$ are outergated sets. The cliques of $G$ are not always
outergated. However, they are 2-outergated:

\begin{mylemma}\label{lem:cb-clique-1-cb}
    Let $K$ be a clique of a CB-graph $G$. Then $K$ is 2-outergated.  More precisely, for any  $z\in V\setminus K$, either
    $z$ has an outergate or $\proj_z(K)=K$ and $z$ has a 2-outergate $z'$; as $z'$ one can select any vertex such that $d(z,z')=d(z,K)-2$ and $z'$ is
    included in a pentagon $z'xuvy$ with $u,v\in K$ not included in a 5-wheel.
\end{mylemma}

\begin{proof} Let $d(z,K)=k$. If there exists $u\in K\setminus \proj_z(K)$, then $\proj_z(K)\subseteq I(u,z)\cap N(u)$ and the fact that $z$ has an outergate follows from Lemma~\ref{lem:cb-outergate}.
    Thus assume  $\proj_z(K)=K$. First suppose that $\TC$ applies to every triplet $z,u,v$ with $u,v\in K$. We assert that then $z$ has an outergate. Indeed, let
$z^*$ be a vertex of $B_{k-1}(z)$ adjacent to a maximum number of vertices of $K$. Let $K'=K\cap N(z^*)$. By $\TC$, $|K'|\ge 2$ and let $u,v\in K'$. Suppose that there exists  $w\in K\setminus K'$.
    By $\TC$ there exists  $z'$ adjacent to $u$ and $w$ and having distance $k-1$ to $z$. By $\INC$ the vertices $z^*,z'\in I(u,z)$ are adjacent. By maximality of $z^*$, $K'$ contains a vertex, say $v$, not adjacent to
$z'$. But then  $z^*,z',w,v$ induce a forbidden 4-cycle. This contradiction shows that $K'=K$ and thus $z^*$ is an outergate. Thus suppose that $K$ contains two
    vertices $u,v$ such that to the triplet $z,u,v$ only $\PC$ applies. Thus there exists a vertex $z^*$ such that $d(z,z^*)=k-2$ and $d(z^*,u)=d(z^*,v)=2$. Since $\TC$ does not apply to the triplet $z,u,v$ and
$z^* \in I(u,z) \cap I(v,z)$, $z^*$ has no outergate. By the first part of the proof we conclude that $\proj_{z^*}(K)=K$ and thus $d(z^*,w)=2$ for any $w\in K$. Consequently, $z^*$ is a 2-outergate of $z$.
\end{proof}

\begin{mylemma}\label{lem:cb-clique-2} Let $K$ be a clique in a CB-graph $G=(V,E)$, let $z \in V \setminus B_1(K)$, and let $z^* \in B_1(K) \cap B_{d(z,K)-1}(z)$ be maximizing $|N(z^*) \cap K|$.
    Then $z^*$ is an outergate of $z$ with respect to $K$ if and only if $N(z')\cap B_1(K) \subseteq B_1(z^*)$ for every $z' \in N(z^*) \cap I(z,z^*)$.
\end{mylemma}

\begin{proof} Let $K'=N(z^*)\cap K$ and $d(z,K)=k$. Obviously, $K'\ne \varnothing$. First, suppose that $z^*$ is an outergate of $z$ in $K$, i.e., $K'=\proj_z(K)$. Pick any vertex $z'\in N(z^*)\cap I(z,z^*)$ and let  $y \in N(z') \cap B_1(K)$.
    We assert that $y\sim z^*$.  Since $y \in N(z') \cap B_1(K)$ and $y\notin K$, $y$ is adjacent to a vertex $u$ of $K$.  Obviously, $u\in \proj_{z'}(K)\subseteq \proj_z(K)=K'$. Hence, $z^*$ is adjacent to $u$. Since $z'\nsim u$ from the choice of $z'$,
    we obtain a forbidden 4-cycle induced by the vertices $z^*,z',y,u$. Hence, $z^*\sim y$.
    Conversely, suppose that $N(z')\cap B_1(K) \subseteq B_1(z^*)$
    holds for every $z' \in N(z^*) \cap I(z,z^*)$, however $z^*$ is
    not an outergate of $z$ in $K$. Then there exists a vertex
    $w\in \proj_z(K)\setminus K'$. Let $u \in K'$ and first suppose
    that $\TC$ applies to the triplet $z,u,w$. By $\TC$ there exists
    a vertex $t\sim u,w$ at distance $k-1$ to $z$. Since
    $z^*,t\in N(u)\cap I(u,z)$, by $\INC$ $z^*\sim t$.  Pick any
    $v\in K'$. Since $z^*\nsim w$ and the 4-cycle $(z^*,t,w,v)$ is not
    induced, we conclude that $t\sim v$. Consequently, $t$ is adjacent
    to $w$ and to all vertices of $K'$, contradicting the maximality
    choice of the vertex $z^*$. Thus, we can assume that only $\PC$
    applies to the triplet $z,u,w$.  By \cite[Theorem 3.3]{ChChGi},
    there exists a vertex $z'$ at distance $k-2$ from $z$ and a common
    neighbor $t$ of $z'$ and $w$ such that $w,u,z^*,z',t$ induce a
    pentagon. But then $t\in N(z')\cap B_1(K) \setminus B_1(z^*)$,
    contrary to our assumption.  This shows that $K'=\proj_z(K)$,
    i.e., that $z^*$ is an outergate of $z$.
\end{proof}

\subsection{Computing $r_\pi$ for the vertices of a clique.}
From Lemmas~\ref{lem:cb-clique-1-cb} and~\ref{lem:cb-clique-2} we
derive the following algorithmic procedures:

\begin{mylemma}\label{lem:alg-clique}
    Let $G$ be a CB-graph and $K$ be a clique of $G$. Then in $\cO(m)$ time we can compute the set of all vertices of $G$ without an outergate with respect to $K$.
\end{mylemma}

\begin{proof} Applying Lemma~\ref{lem:outergate} for $S=K$, in total $\cO(m)$ time we can map  every vertex $z \in V \setminus K$ to some $z^* \in B_{d(z,K)-1}(z) \cap B_1(K)$.
    Then, our initial problem reduces to that of deciding for every $z \in V \setminus K$ separately whether $z^*$ is an outergate of $z$ with respect to $K$.
    For that, we apply the following marking process.
    We consider every $z \in V \setminus B_1(K)$ sequentially, by nondecreasing distance to $K$. If $d(z,K)=2$,
    we mark vertex $z$ if and only if $N(z)\cap B_1(K) \not\subseteq B_1(z^*)$. Otherwise, if $d(z,K) \geq 3$, then we mark vertex $z$ iff there exists some $x \sim z$
    such that $d(x,K)=d(z,K)-1$, $|N(z^*) \cap K|=|N(x^*) \cap K|$, and $x$ is already marked.
    This can be done in $\cO(m)$ time, while we execute a BFS with start subset $K$.

    Now, we prove that this algorithm is correct. Pick any $z \in V \setminus B_1(K)$. First, assume that $z$ has no outergate with respect to $K$.
    By Lemma~\ref{lem:cb-clique-2} there exist  vertices $z' \in N(z^*) \cap I(z^*,z)$ and   $y\in N(z') \cap B_1(K)\setminus B_1(z^*)$.
    Let $u$ be any neighbor of $z^*$ in $K$ and $v$ be any neighbor of $y$ in $K$. Since $u\sim v, z^*\nsim y$ and $G$ does not contain induced 4-cycles,
    necessarily $z'yvuz^*$ is a pentagon. If $z'$ is marked, then every vertex of $I(z',z)$ must be also marked and we are done.
    So, suppose by way of contradiction that $z'$ is not marked.  This implies that $(z')^* \neq z^*$ and that $N(z')\cap B_1(K)\subseteq B_1((z')^*)$.
    By Lemma~\ref{lem:cb-clique-2} $(z')^*$ is an outergate of $z'$.  Since $d(z',v)=d(z',u)=2=d(z',K)$, the vertices $u$ and $v$ belong to $\proj_{z'}(K)$, thus
    they are both adjacent to $(z')^*$. Since $d(z,z')=d(z,K)-2$ and $d(z',K)=2$, necessarily $v\in \proj_{z'}(K)\subseteq \proj_z(K)$.  Since $u$ is an
    arbitrary vertex of $K$ adjacent to $z^*$ and $v$ is not adjacent to $z^*$, we obtain a contradiction with the maximality choice of $z^*$
    in Lemma~\ref{lem:outergate}. This contradiction shows that the vertices $z'$ and $z$ are marked.
    %

    Conversely, assume that the vertex $z$ is marked. We assert that $z$ has no outergate with respect to $K$. We proceed by induction on $d(z,K)$. If $d(z,K)=2$, by the marking algorithm we have
    $N(z)\cap B_1(K) \not\subseteq B_1(z^*)$. By Lemma~\ref{lem:cb-clique-2} and the maximality choice of $z^*$, $z$ has no outergate in $K$. Now assume that $d(z,K)>3$. By the marking algorithm,
    there exists a marked vertex $z'$ on a shortest path between $z$ and some vertex of $\proj_z(K)$ such that $d(z',K)=2$ and $|N((z')^*)\cap K| = |N(z^*) \cap K|$. By the induction hypothesis, $z'$ has no outergate in $K$.
    Suppose by way of contradiction that $z$ has an outergate with respect to $K$.  Equivalently, $z^*$ is an outergate of $z$, i.e., $N(z^*) \cap K=\proj_z(K)$. Since $d(z,K)=d(z,z')+d(z',K)$, we have $\proj_{z'}(K)\subseteq \proj_z(K)$.
    Since $|N((z')^*)\cap K|=|N(z^*) \cap K|$ and $N((z')^*)\cap K\subseteq \proj_{z'}(K)$, this is possible only if $N((z')^*)\cap K=\proj_{z'}(K)$, contradicting our assumption that $z'$ has no outergate with respect to $K$.
    Consequently, $z$ has no outergate with respect to $K$, establishing the correctness of the marking algorithm.
    %
\end{proof}

\begin{myproposition}\label{thm:cb-cliques}
    Let $G$ be a CB-graph and  $\pi$ be a profile on $G$.
    Then for every clique $K$, in $\cO(m)$ time, we can compute $r_\pi(w)$ for every $w \in K$.
\end{myproposition}

\begin{proof}
    We partition $V$ in $K,X,Y$, where $X$ contains the vertices of $V \setminus K$ with no outergate in $K$, and $Y = V \setminus (X \cup K)$.
    By Lemma~\ref{lem:alg-clique}, we can compute this partition in $\cO(m)$ time.
    Furthermore, by Lemma~\ref{lem:outergate}, in total $\cO(m)$ time we can compute an outergate $g_K(y)$ for every vertex $y \in Y$.
    Then, for every $u \in N(K)$, we compute the weights $\alpha(u) = \max\{ \pi(y)d(y,K): y \in Y \ \text{and} \ g_K(y) = u  \}$ and $\beta(u) = \max\{ \pi(y)(d(y,K)+1): y \in Y \ \text{and} \ g_K(y) = u  \}$.
    This can be done in additional $\cO(m)$ time. Pick any $w \in K$.
    Recall that according to Lemma~\ref{lem:cb-clique-1-cb} we have $d(x,w) = d(x,K)$ for any $x \in X$.
    Consequently,
    \begin{flalign*}
        r_\pi(w) = \max& \{\{ \pi(w'): w' \in K \setminus \{w\} \} \cup \{ \pi(x)d(x,K): x \in X \} \\
        \cup& \{ \alpha(u): u \in N(K) \cap N(w) \} \cup \{ \beta(u): u \in N(K) \setminus N(w) \}\}.
    \end{flalign*}
    Note that we need to compute $\max\{ \pi(x)d(x,K): x \in X \}$ only once, which can be done in $\cO(m)$ time.
    Furthermore, we claim that we can also compute in $\cO(m)$ time the values $\max\{ \beta(u): u \in N(K) \setminus N(w) \}$, for every vertex $w \in K$.
    For that, it suffices to apply Lemma~\ref{lem:max-nonneighbor} to the subgraph induced by $N[K]$ and to the cost function $\kappa$ such that: $\kappa(w) = 0$ if $w \in K$; and $\kappa(u) = \beta(u)$ if $u \in N(K)$.
    Doing so, for every $w \in K$, we can compute $r_\pi(w)$ in additional $\cO(\deg(w))$ time. Overall, the total runtime is in $\cO(m)$.
\end{proof}

\subsection{Local minima of $r_\pi$ in $G$.} We show how, given a
vertex $v$ of a graph $G$ with convex balls, to compute in $\cO(m)$
time a vertex $u$ minimizing $r_\pi$ on $B_1(v)$.  If
$r_\pi(v)=r_\pi(u)$, then $v$ is a local minimum of
$r_\pi$. 
This proposition is a generalization of
Proposition~\ref{thm:local-min} to CB-graphs.

\begin{myproposition}\label{thm:local-min-cb}
    Let $G$ be a CB-graph and $\pi$ be a profile on $G$.
    For any $v \in V$, in $\cO(m)$ time we can compute a vertex minimizing $r_\pi$ on $B_1(v)$. In particular, we can either output some $w \in N(v)$ with $r_\pi(w) < r_\pi(v)$ or assert that $v$ is a local minimum of $r_\pi$.
\end{myproposition}

\begin{proof}
    We may assume that $F_\pi(v) \cap N(v) = \varnothing$, otherwise, either $|F_\pi(v) \cap N(v)| > 1$ and $v$ must be a local minimum of $r_\pi$
    or $F_\pi(v) \cap N(v) = \{u\}$ and we are left outputting a vertex amongst $v,u$ with minimum $r_\pi$.
    By Lemma~\ref{lem:cb-outergate}, every $z \in F_\pi(v)$ has an outergate $g_v(z)$ with respect to $B_1(v)$.
    Furthermore, by Lemma~\ref{lem:outergate} we can compute $g_v(z)$, for any $z \in F_\pi(v)$, in $\cO(m)$ time.
    Let $K = \bigcap\{ N(v) \cap I(v,z): z \in F_\pi(v) \}$.
    Note that $K= \bigcap\{N(v) \cap N(g_v(z)): z \in F_\pi(v)\}$, thus we can compute $K$ in $\cO(m)$ time.
    For any $w\in N(v)\setminus K$ there exists $z\in F_\pi(v)$ such that
    $w\notin I(v,z)$. This implies that $d(w,z)\ge d(v,z)$, yielding $r_\pi(w)\ge r_\pi(v)$.  Therefore, if $K=\varnothing$, then $v$
    is a local minimum of $r_\pi$. If $K \neq \varnothing$,  since $K$ is the intersection of cliques $N(v) \cap I(v,z)$ with $z \in F_\pi(v)$,
    $K$ is a clique. By Proposition~\ref{thm:cb-cliques}, we can compute a vertex $u$ minimizing $r_\pi$ within $K$ in $\cO(m)$ time.
    Since $r_\pi(w)\ge r_\pi(v)$ for any $w\in N(v)\setminus K$, either $v$ or $u$ is a local minimum of $r_\pi$ in $B_1(v)$.
\end{proof}

\subsection{Local minima of $r_\pi$ in $G^2$ (under distance conditions).} In this part, we describe a
local-search technique, allowing to find an improving $2$-neighbor of $v$ (if there exists one).
First, we show that Lemma~\ref{lem:cb-wx} can be generalized in the
following way to obtain a lemma similar to Lemma~\ref{lem:wm-wX-2}, but only for
vertices at distance at least $3$.

\begin{mylemma}\label{lem:cb-wX-2}
  Let $G$ be a CB-graph, $X\subset V$ be nonempty, and $u,v,w \in V$
  be such that:
  \begin{enumerate}[(1)]
  \item $d(u,v) \geq 3$;
  \item for every $x \in X$, $d(u,x) < d(v,x)$;
  \item $w \in \bigcap\{N(v) \cap I(v,x):  x \in X\}$.
  \end{enumerate}
  Then there exists a vertex $w_X \in N(v) \cap I(v,u)$ such that
  $w_X \in \bigcap\{N(v) \cap I(v,x): x \in
  X\}$. 
\end{mylemma}

\begin{proof}
    Assume that $w \notin I(u,v)$, otherwise we can set $w_X = w$.  If $d(u,w)>d(u,v)$, then $v \in I(u,w)$ and $w,u\in B_{d(v,x)-1}(x)$ for any $x \in X$,
    contradicting the convexity of the ball $B_{d(v,x)-1}(x)$. Thus $d(u,w) = d(u,v)$.
    Pick any $y \in X$. By Lemma~\ref{lem:cb-wx}, there exists  $w_y \in N(v) \cap I(v,u)$ such that $d(y,w_y) < d(y,v)$.
    Since $w,w_y\in N(v)\cap I(v,y)$ and the ball $B_{d(y,v)-1}(y)$ is convex, the vertices $w_y,w$ are adjacent.
    Consequently, $w_y \in I(w,u)$. Since $u,w\in B_{d(x,v)-1}(x)$ for any $x\in X$, the convexity of the ball $B_{d(x,v)-1}(x)$ implies that  $d(x,w_y) \leq d(x,v)-1$.
    Thus  $w_y \in \bigcap\{N(v) \cap I(v,x):  x \in X\}$ and we can set $w_X = w_y$.
\end{proof}

%

The following result is the main technical result of this
subsection. It is in the same spirit as
Proposition~\ref{thm:loc-search} and the proof is essentially the
same: the only difference being that when $v$ is a local minimum, we
ask that $v$ is at distance at least $3$ from a central vertex. This
is because we use Lemmas~\ref{lem:cb-wx} and~\ref{lem:cb-wX-2} in the
proof instead of Lemmas~\ref{lem:wm-wx} and~\ref{lem:cb-wX-2}.

\begin{myproposition}\label{thm:loc-search-cb}
  Let $v$ be an arbitrary vertex of a CB-graph $G=(V,E)$ and let $\pi$
  be some profile on $G$.  If there exists some vertex $v'$ such that
  $r_\pi(v') < r_\pi(v)$ and $d(v,v') \neq 2$, then in $\cO(m)$ time
  we can output a vertex $u' \in B_2(v)$ such that
  $r_\pi(u') < r_\pi(v)$.
\end{myproposition}

\begin{proof}
  If $v'\sim v$, then Proposition~\ref{thm:local-min-cb} will return a
  vertex $u\in B_1(v)$ such that $r_\pi(u)\le r_\pi(v')<r_\pi(v)$ and
  we are done. So, assume that $v$ is a local minimum of $r_\pi$. Let
  $F_\pi(v) = \{z \in V: \pi(z) d(v,z) = r_\pi(v)\}$, i.e.,
  $F_\pi(v) $ is the set of vertices that are the furthest from $v$.
  Since $v$ is a local minimum but not a global minimum of $r_\pi$, we
  can suppose that $d(v,z)>1$ for any $z\in F_\pi(v)$.  For every
  neighbor $w \in N(v)$ of $v$, let
  $\Psi_\pi(w) = \{ z \in F_\pi(v): w\in N(v) \cap I(v,z) \}$.  We
  define an equivalence relation $\equiv$ on $N(v)$ as follows:
  $w \equiv w' $ if and only if $\Psi_\pi(w) = \Psi_\pi(w')$.  Then,
  we partition the vertices of $N(v)$ according to the relation
  $\equiv$. For that, we use outergates with respect to $N(v)$, which
  by Lemma~\ref{lem:cb-outergate} exist for every $z \in F_\pi(v)$.
  By Lemma~\ref{lem:outergate} we can compute such an outergate
  $g_v(z)$ for every $z \in F_\pi(v)$ in total $\cO(m)$ time.  Then,
  for every $w,w' \in N(v)$, we have $w \equiv w'$ if and only if
  $N(w) \cap \{g_v(z): z \in F_\pi(v)\} = N(w') \cap \{g_v(z) : z \in
  F_\pi(v)\}.$ Consequently, we can compute in $\cO(m)$ time the
  equivalence classes on $N(v)$ with respect to $\equiv$ by using
  partition refinement techniques.  Next, in $\cO(n)$ time we compute,
  for every vertex $u \in B_2(v) \setminus B_1(v)$, the number of
  vertices $z \in F_\pi(v)$ such that $g_v(z) = u$.  Doing so, in
  total $\cO(m)$ time, we can also compute $|\Psi_\pi(W_i)|$ for every
  equivalence class $W_i$ of $\equiv$ (where we set
  $\Psi_\pi(W_i) = \Psi_\pi(w)$ for arbitrary $w \in W_i$).  Let $W_0$
  be an equivalence class maximizing $|\Psi_\pi(W_0)|$.  If
  $|\Psi_\pi(W_0)| = |F_\pi(v)|$, then we abort.  From now on, we
  assume that $|\Psi_\pi(W_0)| < |F_\pi(v)|$.  Let
  $z \in F_\pi(v) \setminus \Psi_\pi(W_0)$ be arbitrary.  Then in
  $\cO(m)$ time we compute $N(v) \cap I(v,z) = N(v) \cap N(g_v(z))$.
  Let $w_{\max} \in N(v) \cap I(v,z)$ be a vertex maximizing
  $ |N(w_{\max}) \cap W_0 |$.  Finally, we apply
  Proposition~\ref{thm:local-min-cb} in order to compute a vertex $v^+$
  minimizing $\pi$ within $ N(w_{\max}) $ and the algorithm returns
  this vertex $v^+$.

    Now, we prove the correctness of the algorithm in the case of CB-graphs. Recall that we assume that $v$ is a local minimum of $r_\pi$ and that there exists a vertex $v'$ such that $d(v,v')\ne 2$ and
    $r_\pi(v')<r_\pi(v)$. This implies that $d(v,v')\ge 3$.  By Lemma~\ref{lem:cb-outergate}, there exists a vertex $u \in I(v,v')$ such that $d(u,v) = 2$ and $N(v) \cap I(v,v') \subseteq N(u)$.
    Furthermore, by the proof of Theorem~\ref{thm:cb}, $r_\pi(u) < r_\pi(v)$.  By Lemma~\ref{lem:cb-wX-2}, there exists a vertex $w' \in N(v) \cap I(v,v') \subseteq N(u)$ such that $\Psi_\pi(W_0) \subseteq \Psi_\pi(w')$.
    By maximality of $|\Psi_\pi(W_0)|$, necessarily  $w' \in W_0$. To prove the correctness of the algorithm, we will prove that (1) $|\Psi_\pi(W_0)| = |F_\pi(v)|$ is impossible and
    that (2) $u$ is adjacent to $w_{\max}$. This will show that the algorithm will return a vertex $v^+\in N(w_{\max})\subset B_2(v)$ with $r_\pi(v^+)\le r_\pi(u)<r_\pi(v)$.
    First, suppose that  $|\Psi_\pi(W_0)| = |F_\pi(v)|$, i.e., $\Psi_\pi(W_0)= F_\pi(v)$ holds. Since $w'\in W_0$ we have $\pi(y)d(w',y)<\pi(y)d(v,y)$ for any vertex $y\in \Psi_\pi(W_0)$.
    Since $v$ is a local
    minimum of $r_\pi$, the profile must contain a vertex $y'\notin \Psi_\pi(W_0)= F_\pi(v)$ such that
    $\pi(y')d(w',y') = r_\pi(w') \ge r_\pi(v)>\pi(y')d(v,y').$  This implies $d(w',y')=d(v,y')+1$. Since $\pi(y')d(u,y')\le r_\pi(u)<r_\pi(w')=\pi(y')(d(v,y')+1)$, we also conclude that
    $d(u,y')\le d(v,y')$. Consequently, the vertices $u$ and $v$ belongs to the ball $B_{d(v,y')}(y')$ and the vertex $w'\in I(u,v)$ does not belong to this ball, a contradiction with convexity of balls.  This establishes that
    $\Psi_\pi(W_0)$ is strictly included in $F_\pi(v)$ and proves that the vertex $z$ chosen by the algorithm exists.  It also proves that for any vertex $w\in W_0$ we have the equality $F_\pi(w) = F_\pi(v) \setminus \Psi_\pi(W_0)$.
    To prove that the vertex  $w_{\max}$ computed by the algorithm is adjacent to $u$, first we prove that $w_{\max}$ is adjacent to every vertex $w$ of $N(u)\cap W_0$ (this set is nonempty since it contains
    $w'$). 
    By Lemma~\ref{lem:cb-wx}, there exists a vertex
    $w_z\in N(v) \cap I(v,z) \cap I(v,v') \subseteq N(u)$.  Since
    $w,w_z \in I(v,u) \cap N(v)$, by $\INC$, the vertices $w,w_z$ are
    adjacent.  Similarly, since $w_z,w_{\max} \in N(v) \cap I(v,z)$,
    by $\INC$, $w_z$ and $w_{\max}$ are adjacent. Note that the sets
    $N(w_z) \cap W_0$ and $N(w_{\max})\cap W_0$ are comparable for
    inclusion (otherwise, we would get a forbidden induced $C_4$).  By
    our maximality choice of $w_{\max}$, we obtain
    $w\in N(w_z) \cap W_0\subseteq N(w_{\max})\cap W_0$, and thus
    $w_{\max}$ and $w$ are adjacent.  Then,
    $w_{\max}, u \in N(w) \cap I(w,z)$, and by $\INC$ $w_{\max}\sim
    u$. This concludes the proof.
    %
\end{proof}

\subsection{Getting very close to the center}
We show how to compute in $\ctO(m\sqrt{n})$ time a vertex $v^*$ of a
CB-graph $G$ which is either central or is located at distance 2 from any
central vertex of $G$.  Our result follows from a combination of
Proposition~\ref{thm:loc-search-cb} with random sampling and a
bootstrap approach (Lemma~\ref{lem:k-unimodal}).  For unweighted
profiles we derandomize this approach using
Proposition~\ref{number-of-steps}.
Using the sample-select-descent method in the weighted case and its deterministic version for 0-1-profiles (see Section \ref{descent}), the generic descent algorithm will construct a path $(v_0,v_1,v_2,\ldots v_\ell)$ of $G^2$ such that $r_\pi(v_0)>r_\pi(v_1)>r_\pi(v_\ell)$. By Lemma~\ref{lem:k-unimodal}, the length of this path is at most $\ctO(\sqrt{n})$. In the case of 0-1-profiles, the length of this path will be $\cO(\sqrt{n})$ (Proposition~\ref{number-of-steps}).
Let us call the vertex  $v^*:=v_\ell$ and the 2-ball $B_2(v^*)$ centered at $v^*$ \emph{terminal}. Consequently, a terminal vertex $v^*$ can be computed in $\ctO(m\sqrt{n})$ time (in $\cO(m\sqrt{n})$ time for 0-1-profiles).
From previous results it follows that either $v^*$ is a central vertex of $G$ or the whole center $C_\pi(G)$ is included in $B_2(v^*)\setminus B_1(v^*)$. Since  $C_\pi(G)$ is convex,
in the second case $C_\pi(G)$ must have diameter at most 3. In fact, if $C_\pi(G)\subseteq B_2(v^*)$ and $v^*$ is terminal, then $\diam(C_\pi(G))\le 2$. Indeed, let $x,y\in C_\pi(G)$ with $d(x,y)=3$.
Since $d(v^*,x)=d(v^*,y)=2$, any quasi-median of the triplet $x,y,v^*$ has type $(1,1,1)$. Thus $v^*$ will be adjacent to neighbors of $x$ and $y$ in $I(x,y)$. Since $I(x,y)\subset C_\pi(G)$,
$v^*$ cannot be a local minimum of $r_\pi$, contrary to our assumption that $v^*$ is a terminal vertex. Notice also that if $C_\pi(G)\subseteq B_2(v^*)$ and $\pi$ is a 0-1-profile, then $r_\pi(v^*)$ may differ from the radius $\rad_\pi(G)$ by at most 1.
Indeed, since $d(v^*,x)=2$ for any $x\in C_\pi(G)$, if $r_\pi(x)\le r_\pi(v^*)-2$, then for any common neighbor $w$ of $x$ and $v^*$ we will have $r_\pi(w)=r_\pi(x)+1<r_\pi(v*)$, contrary with the fact that $v^*$ is a terminal vertex.
Consequently, we have the following result:

\begin{mytheorem}\label{thm:center-cb-steps} For a CB-graph  $G$, the following holds:
  \begin{enumerate}[(1)]
  \item For any profile $\pi$, in $\ctO(m\sqrt{n})$ time with high probability one can
    compute a vertex $v^*$ such that, either
    $\diam(C_\pi(G))\ge 3$ and $v^*$ is a central vertex, or
    $C_\pi(G)\subseteq B_2(v^*) \setminus B_1(v^*)$.
  \item For any 0-1-profile $\pi$, in deterministic
    $\cO(m\sqrt{n})$ time one can compute a vertex $v^*$ with
    eccentricity $r_\pi(v^*)\le \rad_\pi(G)+1$.  Moreover, if
    $\diam(C_\pi(G))\ge 3$, then $v^*$ is a central vertex, otherwise
    $C_\pi(G)\subseteq B_2(v^*)$.
  \end{enumerate}
\end{mytheorem}

\subsection{Computing a central vertex of a CB-graph.} The goal of this subsection is to prove the following result:

\begin{mytheorem}\label{thm:center-cb}
  Given a CB-graph $G$ and a profile $\pi$, one can compute a central
  vertex $v \in C_\pi(G)$ in $\ctO(m^{3/2})$ time with high
  probability.
\end{mytheorem}

\begin{proof}
    The algorithm has two main steps.
    During its first step, we compute a terminal vertex $v^*$, as stated in Theorem~\ref{thm:center-cb-steps} (1).
    This can be done in $\ctO(m\sqrt{n})$ time with high probability.
    %
    %
    During the second step, we maintain a convex subset $X_i$ and two vertices $x_i,y_i$ such that:
    \begin{enumerate}[(1)]
        \item $x_i \in X_i$;
        \item for every $v \in V \setminus X_i$, $r_\pi(v) \geq r_\pi(y_i)$;
        \item either $y_i \in C_\pi(G)$ or $C_\pi(G) \subseteq B_2(x_i)$.
    \end{enumerate}
    Initially, $x_0 = y_0 = v^*$, and $X_0 = V$. Then, Properties 1
    and 2 trivially hold, while Property 3 follows from
    Theorem~\ref{thm:center-cb-steps} (1).  If at some point during the execution $X_i = \emptyset$, then
    we can assert by Property 2 that $y_i \in C_\pi(G)$.  Thus, in
    what follows, we always assume $X_i \neq \emptyset$.

    We first assume $|N(x_i) \cap X_i| \leq \sqrt{m}$.
    Then, we apply Proposition~\ref{thm:local-min-cb} to every vertex of $B_1(x_i) \cap X_i$, which results in the computation of a vertex $u_i$ that minimizes $r_\pi$ within $\bigcup\{ B_1(w_i) \mid w_i \in B_1(x_i) \cap X_i \}$.
    We claim that one of $u_i,y_i$ must be contained in $C_\pi(G)$.
    Indeed, assume that $y_i \notin C_\pi(G)$.
    Properties 2 and 3 imply that $C_\pi(G) \subseteq X_i \cap B_2(x_i)$.
    Since $x_i \in X_i$ (Property 1), the convexity of $X_i$ implies that $C_\pi(G) \subseteq \bigcup\{ B_1(w_i) \mid w_i \in B_1(x_i) \cap X_i \}$.
    In particular, $u_i \in C_\pi(G)$.
    Therefore, we are done outputting a vertex minimizing $r_\pi$ within $u_i,y_i$.

    From now on, we assume $|N(x_i) \cap X_i| > \sqrt{m}$.  Let $z_i$
    be such that $\pi(z_i) d(z_i,x_i) = r_\pi(x_i)$.  We set
    $X_{i+1} := X_i \cap B_{d(x_i,z_i)-1}(z_i)$.  Let also
    $x_{i+1} \in \proj_{x_i}(X_{i+1})$ (so as to satisfy Property 1).
    Finally, let $y_{i+1}$ be a vertex minimizing $r_\pi$ within
    $y_i,x_i$.  Note that for every vertex
    $v \in V \setminus X_{i+1}$, either $v \notin X_i$ and so
    $r_\pi(v) \geq r_\pi(y_i) \geq r_\pi(y_{i+1})$, or
    $d(v,z_i) \geq d(x_i,z_i)$ and so
    $r_\pi(v) \geq r_\pi(x_i) \geq r_\pi(y_{i+1})$ (Property 2).  We
    now claim that if $y_{i+1} \notin C_\pi(G)$, then
    $C_\pi(G) \subseteq B_2(x_{i+1})$ (Property 3).  For that, let
    $u_i \in C_\pi(G)$ be arbitrary. Since $y_{i+1} \notin C_\pi(G)$,
    we have $y_i, x_i \notin C_\pi(G)$. Note that $u_i \in X_{i+1}$
    because of Property 2 and our assumption that
    $y_{i+1} \notin C_\pi(G)$.  Furthermore,
    $d(x_i,x_{i+1}) \leq d(u_i,x_i) \leq 2$ because
    $x_{i+1} \in \proj_{x_i}(X_{i+1})$ and
    $u_i \in C_\pi(G) \subseteq B_2(x_i)$ by Property 3.  Suppose by
    contradiction $d(u_i,x_{i+1}) \geq 3$.  If $x_{i+1},x_i$ were
    adjacent, then we would obtain $d(u_i,x_{i+1}) = 3$ and
    $x_i \in N(x_{i+1}) \cap I(x_{i+1},u_i)$, thus contradicting the
    convexity of $X_{i+1}$.  Therefore, $d(x_i,x_{i+1}) = 2$. It also
    implies $d(x_i,u_i) = 2$. Note that the convexity of $X_{i+1}$
    also implies that $x_i \notin I(u_i,x_{i+1})$ and thus
    $d(u_i,x_{i+1}) = 3$. Consequently, there is no median vertex for
    $x_i,u_i,x_{i+1}$.  Consider a quasi-median $abc$ for
    $x_i,u_i,x_{i+1}$ such that:
    $d(x_i,u_i) = d(x_i,a)+d(a,b)+d(b,u_i)$,
    $d(u_i,x_{i+1})=d(u_i,b)+d(b,c)+d(c,x_{i+1})$ and
    $d(x_{i+1},x_i) = d(x_{i+1},c)+d(c,a)+d(a,x_i)$. If $abc$ is an
    equilateral triangle, then necessarily, $a=x_i$, $b \sim x_i,u_i$,
    and $c \sim x_i,x_{i+1},b$. Therefore, since
    $b,c \in I(u_i,x_{i+1}) \subseteq X_{i+1}$ as $X_{i+1}$ is convex,
    we have $d(x_i, X_{i+1}) = 1$, contradicting the definition of
    $x_{i+1}$. Consequently, by Lemma~\ref{lem:cb-triangle}, the
    triangle $abc$ has type $(2,2,1)$.
    If $a \ne x_i$, then $\max\{d(x_i,u_i),d(x_i,x_{i+1})\} \ge 1 + \max\{d(a,b),d(a,c)\} = 3$, which is impossible as $d(x_i,u_i) = d(x_i,x_{i+1}) = 2$. Therefore, $a = x_i$.
    As $X_{i+1}$ is convex, necessarily $b,c \in I(u_i,x_{i+1}) \subseteq X_{i+1}$.
    This implies $d(a,b) = d(a,c) = 2$.
    But then, $b = u_i$, $c = x_{i+1}$, and so $d(u_i,x_{i+1}) = 1 < 3$, a contradiction.
    Consequently,
    $d(u_i,x_{i+1}) \leq 2$ and Property 3 holds for
    $x_{i+1}, y_{i+1}$, and $X_{i+1}$.  We then repeat the step with
    $X_{i+1},x_{i+1},y_{i+1}$.

    Let us analyze the runtime of the second step.  Each iteration
    except the final one (when $|N(x_i) \cap X_i| \leq \sqrt{m}$) can
    be executed in $\cO(m)$ time and the final iteration can be done
    in $\cO(m\sqrt{m})$ time.  If we perform $k+1$ iterations, then
    the runtime of the algorithm is in $\ctO(km + m\sqrt{m})$.
    Vertices $x_0,x_1,\ldots,x_{k-1}$ must have degree $> \sqrt{m}$.
    Since $X_0 \supset X_1 \supset \ldots X_k$, and
    $x_i \in X_i \setminus X_{i+1}$ for every $i$ such that
    $0 \leq i < k$, these vertices are pairwise different.  As a
    result, $k = \cO(m/\sqrt{m}) = \cO(\sqrt{m})$.  Overall, the
    runtime is in $\ctO(m^{3/2})$.
\end{proof}

\end{document}